\documentclass[a4paper,12pt]{article}
\pdfoutput=1 

\usepackage{jheppub} 

\usepackage[T1]{fontenc} 
\usepackage{tensor}
\usepackage{mathrsfs}

\usepackage{mathtools}

\newcommand{\beq}{\begin{equation}}
\newcommand{\eeq}{\end{equation}}
\def\be {\begin{equation}}
\def\ee {\end{equation}}
\def\bs#1\es{\begin{split}#1\end{split}}
\def\ba#1\ea{\begin{align}#1\end{align}}
\def\baed#1\eaed{\begin{aligned}#1\end{aligned}}
\def\bged#1\eged{\begin{gathered}#1\end{gathered}}
\def\bea{\begin{eqnarray}}
\def\eea{\end{eqnarray}}

\newcommand{\sm}[1]{\text{{\tiny{$#1$}}}}

\def\\nonumber{\nonumber}

\def\la{\langle}

\def\d{\delta}

\def\e{\epsilon}

\def\G{\Gamma}

\def\k{\kappa}
\def\l{\lambda}

\def\o{\omega}

\def\nn{\nonumber}
\def\lM{\ell_{\rm M}}
\def\pd{\partial}

\def\bls{\bigg [}
\def\brs{\bigg ]}

\def\tbo{{\scriptscriptstyle{(1)}}}

\def\R{\text{Re}}

\newcommand{\cT}{\mathcal{T}}
\newcommand{\cZ}{\mathcal{Z}}

\newcommand{\cC}{\mathcal{C}}

\newcommand{\cK}{\mathcal{K}}

\newcommand{\cR}{\mathcal{R}}

\def\cD{\mathcal{D}}

\def\cA{{{\mathcal A}}}

\def\cM{\mathcal{M}} 
\def\cN{\mathcal{N}}
\def\cV{\mathcal{V}}

\def\al{\alpha}

\def\Re{\text{Re}}
\def\Im{\text{Im}}
\def\Tr{\text{Tr}}

\def\pa{\partial}
\def\na{\nabla}
\def\fr{\frac}
\def\tfr{\tfrac}

\def\we{\wedge}

\def\lra{\leftrightarrow}

\def\tbzero{{\text{\tiny{(0)}}}}

\def\tbtwo{{\text{\tiny{(2)}}}}

\let\foo\bar 
\renewcommand{\bar}[1]{ {\foo{  #1} }{} }

\newlength{\dhatheight}

\preprint{ IPMU19-0004}

\title{{\boldmath  F-theory Vacua  and   $\alpha'$\,--\,Corrections }}

\author{Matthias Weissenbacher}

\affiliation{Kavli Institute for the Physics and Mathematics of the Universe, University of Tokyo,\newline Kashiwa-no-ha 5-1-5, 277-8583, Japan}

\emailAdd{matthias.weissenbacher@ipmu.jp}

\abstract{In this work we analyze  F-theory and Type IIB orientifold compactifications to study $\alpha '$-corrections to the four-dimensional, $\cN = 1$ effective actions.  
In  particular, we obtain corrections to the K\"ahlermoduli space metric and its complex structure  for generic dimension originating from eight-derivative corrections to eleven-dimensional supergravity.    We propose a completion of the $G^ 2 R^3$ and $(\nabla G)^2R^2$-sector in eleven-dimensions  relevant in Calabi--Yau fourfold reductions. We suggest that the three-dimensional, $ \cN=2$  K\"ahler coordinates may be expressed as topological integrals  depending on the first, second, and third  Chern-forms of the divisors of the internal Calabi--Yau fourfold.

The divisor integral Ansatz for  the K\"ahler potential and K\"ahler coordinates may be lifted to four-dimensional, $\cN = 1$ F-theory vacua.  We identify a novel correction to the K\"ahler potential and coordinates at order  $ \alpha'^2$, which is leading compared to other known corrections in the literature.  At weak string coupling the correction arises from the intersection of $D7$-branes and $O7$-planes  with base divisors and the volume of self-intersection curves of divisors in the base.
In the presence of the conjectured novel $\alpha'$-correction resulting from the divisor interpretation the no-scale structure  may be  broken. Furthermore, we propose  a model independent scenario to achieve  non-supersymmetric AdS vacua for  Calabi-Yau orientifold backgrounds with negative Euler-characteristic.

}

\begin{document} 
\maketitle
\flushbottom

\section{Introduction}

Four-dimensional minimal super-gravity theories are of particular phenomenological interest. The effective actions are 
commonly derived by dimensionally reducing ten-dimensional supergravity actions arising in string theory with localized 
brane sources.
The stringy imprint  arises in the  form of $\alpha'$-corrections\footnote{Which is given by $\al' = l_S^2 $ with string length  $l_S$. The canonical convention for the definition of $\alpha'$ is w.r.t,\,\,the string tension $T$ as $T^{-1} = 2 \pi \alpha'$. } to the K\"ahler potential and coordinates  of the leading two-derivative action or in form of high-derivative couplings in four dimensions.
Such corrections have been shown to be crucial in determining the vacua of the effective theory in the process of moduli stabilization. 
However, to compute  $\alpha'$-corrections  in a truly minimal supersymmetric i.e.\,\,$\cN=1$ set-up has been a challenging endeavor.
A promising approach is to utilize F-theory which is a formulation of Type IIB string theory with  space-time filling  seven-branes at varying string coupling \cite{Vafa:1996xn}.   It captures the string coupling dependence in 
the geometry of an elliptically fibered higher-dimensional manifold.  The general effective actions of 
F-theory compactifications have been studied using the duality with M-theory \cite{Denef:2008wq,Grimm:2010ks}. 
A wide range of phenomenologically promising  geometric  F-theory backgrounds are known to giving rise non-Abelian gauge groups \cite{Blumenhagen:2006ci,Denef:2008wq,Weigand:2010wm}. 

The starting point of the M/F-theory duality is  the long wave length limit of M-theory, i.e.\;\;eleven-dimensional supergravity. Higher-derivative  or higher-order $l_{\rm M}$-corrections can then be followed through the  duality to give rise to $\alpha'$-corrections in the resulting four-dimensional $\cN=1$  theory. We first compactify eleven-dimensional supergravity including the  next to leading order eight-derivative or $l_{\rm M}^6$-couplings to three dimensions on a  supersymmetry preserving 8-dimensional background.  More precisely, we preform a classical Kaluza-Klein reduction 
of the purely gravitational M-theory $R^4$-terms   
 \cite{Green:1997di,Green:1997as,Kiritsis:1997em,Russo:1997mk,Antoniadis:1997eg,Tseytlin:2000sf} on
elliptically fibered Calabi--Yau fourfolds. Furthermore, one needs to consider the $G^2R^3$  and  $(\nabla G)^2R^2$-sector, where $G$ is the M-theory four-form field strength. One easily verifies that all those couplings carry eight derivatives.
We then implement the F-theory limit by
decompactifying the thee-dimensional theory to four space-time dimensions  and interpret the resulting $\alpha'$-corrections to the two-derivative effective theory. In particular, we study $l_{\rm M}^6$-corrections to the three-dimensional K\"ahler potential and K\"ahler coordinates of the  $\cN = 2$ theory, which then modify the four-dimensional  K\"ahler potential and K\"ahler coordinates in the F-theory limit. In particular, we identify a new leading order $\alpha'^2$-correction to the K\"ahler potential and coordinates which may break the no-scale structure.
It is then of interest to study its  effects in moduli stabilization scenarios.
\newline 

We start the discussion in section \ref{sec-copletionsnewGsectors} by reviewing the  $G^2R^3$  and  $(\nabla G)^2R^2$-sector. No super-symmetric completion of those sectors is known. In this work we propose a completion of the bosonic terms relevant for Calabi--Yau fourfold reductions. We start from a  general basis and fix the coefficients via comparison to controlled theories upon dimensional reduction. In particular, we compactify on Calabi--Yau threefolds and verify compatibility with $5d, \cN=2$ supergravity. Furthermore,  upon reduction on $S^1 \times K3$ we make use of the Heterotic/IIA-theory duality. 

This then allows us to fix the parameters such that we can perform a controlled dimensional reduction on Calabi--Yau fourfolds with a generic number of K\"ahler deformations in section \ref{sec-3daction}. Also in this section we review our previous results for the one-modulus case for which the integration in a three-dimensional K\"ahler potential and coordinates can be performed exactly.

In section \ref{sec-3dKaehlerGeneric} we suggest a proposal for the three-dimensional K\"ahler potential and coordinates for a generic number of K\"ahler moduli of the Calabi--Yau fourfold  background. The key new approach in contrast to our previous attempts \cite{Grimm:2015mua,Grimm:2014efa} is the formulation of the higher-derivative contributions as divisor integrals, analogous to  the discussion of the warp-factor  in \cite{Grimm:2015mua}.  We argue in \ref{secCYDivisorINt}  that the new formulation can indeed give rise to all relevant higher-derivative couplings in the reduction result  obtained in \ref{reductionCY4generic}. However, to match the reduction result  is beyond the aim of this work and we suggest that non-trivial identities  relating the higher-derivative objects are needed to perform this tasks. Let us stress that obtaining the correct building blocks from a K\"ahler potential and K\"ahler coordinates  is a big leap forward as this steps meets heavy obstacles as pointed  out in \cite{Grimm:2015mua}.
We then proceed  in \ref{secTopDivInt} by showing that the divisor integral K\"ahler coordinates can be re-expressed  as topological integrals. This is very intriguing as it will allow for a F-theory interpretation. Lastly in section \ref{sec-onemodcompatibiliyt}  we show compatibility with  to the one-modulus case where the K\"ahler potential and  coordinates could be fixed exactly \cite{Grimm:2017pid}.

In section \ref{sec_FtheoryMain} we discuss the F-theory uplift of  the three-dimensional $l_{\rm M}^6$-corrected K\"ahler potential and coordinates to four dimensions. The classical uplift of the topological integrals is well understood and can be performed rigorously. It is expected that the F-theory lift receives loop-corrections which result from integrating out Kaluza-Klein states on the $4d/3d$ circle at one-loop. As we encounter a   $l_{\rm M}^6$-correction to the K\"ahler coordinates with \cite{Grimm:2017pid} logarithmic dependence on the Calabi--Yau fourfold volume reminiscent of such a  loop correction we comment on a one-loop modification of the F-theory uplift. However, to present a complete analysis  of the F-theory uplift at one-loop is beyond the scope of this work.  
Due to this the resulting $\alpha'^2$-corrected four-dimensional K\"ahler potential  and coordinates thus carry free parameters we are not able to fix.  The  three-dimensional  K\"ahler coordinates generically lead to a breaking of the no-scale structure which may remain present in four-dimensions. This  breaking of the no-scale structure is also consistent with the one-modulus case \cite{Grimm:2017pid}.  However, we conclude that a better understanding of the F-theory uplift at one-loop is required before deciding the ultimate faith of the $\alpha'^2$-correction to the four-dimensional scalar potential.

To give an independent interpretation of the novel $\alpha^{\prime 2}$-correction 
we take the Type IIB weak string coupling limit \cite{Sen:1996vd}. The correction  is proportional to  the volume of the intersection curve  of $D7$-branes and
the $O7$-plane  with divisors in the K\"ahler base of the elliptically fibered Calabi--Yau fourfold. Moreover, it depends on the volume of the self-intersection curves of those divisors in the base. We also identify a second correction which survives the F-theory limit. However it vanishes due to conspiration of pre-factors. The latter correction is  proportional to the self-intersection of divisors in the base intersecting  the $D7$-branes and
the $O7$-plane. Both are expected to  arise from  tree-level string amplitudes involving oriented open strings with the topology of a disk a and non-orientable closed strings with the topology of a projective plane analogous to the $\al'^2$-correction encountered in \cite{Grimm:2013gma,Grimm:2013bha}. We also discuss the latter in this work.

In section \ref{moduli-stabilisation} we discuss the implications of the   $\alpha'^2$-corrections on moduli stabilization. 
We propose  a  scenario to achieve  non-supersymmetric AdS vacua for geometric backgrounds with negative Euler--characteristic  $\chi(B_3) < 0$, where $B_3$ is the base of the elliptically fibered Calabi-Yau fourfold in F-theory. In the IIB picture thus  for  Calabi--Yau oreintifold backgrounds with negative Euler-characteristic.  The  vacua are obtained due to an interplay of the Euler-Characteristic correction \cite{Becker:2002nn} and the  $\alpha'^2$-corrections  to the scalar potential.\footnote{ The form of the scalar potential due to the $\alpha'^2$-correction  obtained in \cite{Grimm:2013gma,Grimm:2013bha}  is similar to  the  one obtained at order $\alpha'^3$ in \cite{Ciupke:2015msa,Grimm:2017okk}.} We close by emphasizing that the discussion can be performed analogously for Calabi--Yau fourfolds with  $\chi(B_3) > 0$ which leads to  de Sitter extrema. We thus suggest that the scenarios may suffice  to construct an explicit counter example to the recent conjecture by \cite{Obied:2018sgi}. Let us emphasize that we do not study explicit geometric backgrounds in this work but derive constraints on the topological quantities such that vacua may be obtained. 

\section{Towards a completion of the $G^2R^3$ and $ (\nabla G)^2R^2$ sectors}
\label{sec-copletionsnewGsectors}

In section \ref{sec-HIghDerM} we review the known eleven-dimensional supergravity action at eight-derivatives. We consider the possibility of having additional $  G^2   \cR^3$ and $ (\nabla G)^2R^2$-terms in the eleven-dimensional action  in  section \ref{CY3checkM}, where $G$ denotes the M-theory four-form field strength, and $R$ is an abbreviation for the Riemann tensor. We propose a completion of these two sectors relevant for Calabi--Yau fourfold reductions.  Due to these potential novel terms one encounters an additional parameter freedom in the reduction result in section \ref{sec-3daction}.  However, as we make not use of this parameter freedom in the remaining work let us stress that this section stands independent. The reader more interested in the three and four-dimensional effective actions can thus safely skip the technical section  \ref{CY3checkM} and carry on with section \ref{sec-3daction}.

\subsection{Higher-derivative corrections in M-theory}
\label{sec-HIghDerM}
In this section we review the eleven-dimensional supergravity action 
including  the relevant eight-derivative terms.  Note that we comment on a completion of the  $  G^2   \cR^3$ and $ (\nabla G)^2R^2$-sector relevant for  a Calabi--Yau fourfold $CY_4$ reductions in the next section \ref{CY3checkM}. 
The bosonic part of the classical two-derivative $\cN=1$ action in eleven dimensions is given by
\be 
2 \k_{11}^2 \, S_{11}=\int_{M_{11}}   R \,   \ast \, 1-\frac{1}{2}   G \wedge   \ast \,   G-\frac{1}{6}   C \wedge   G \wedge   G\, .\label{eq:S0}
\ee
The purely gravitational sector is corrected at eight-derivatives by $  R^4$-terms  given by
\be
2 \k_{11}^2 \, S_{  R^4}=\int_{M_{11}}\big(  t_8   t_8-\tfrac{1}{24} \epsilon_{11} \epsilon_{11} \big)  \cR^4\,   \ast \, 1-3^2 2^{13}\,  C \wedge   X_8 \,\label{R4} \;\; .
\ee
First derived in \cite{Gross:1986iv,Duff:1995wd} these terms can be shown to be re related to the R-symmetry and conformal anomaly of the world-volume theory of a stack of $N$ M5-branes \cite{Tseytlin:2000sf}. 
Secondly the known contributions  \cite{Liu:2013dna} to  $  G^2   \cR^3$ and $ (\nabla G)^2R^2$-sector of the four-form field strength are given by
\be
2 \k_{11}^2 \, S_{  {\mathcal{G}}} =\int_{M_{11}}-\big(  t_8   t_8+\tfr{1}{96} \epsilon_{11} \epsilon_{11} \big)  G^2 \,   R^3\,   \ast \, 1+  s_{18} \, \big(  \nabla   G \big)^2  \,   R^2\,  \ast \, 1+256 \,   Z   G \we   \ast \,   G\, .\label{Gterms}
\ee 
The last term in \eqref{Gterms} was argued to be necessary to ensure Type IIA/M-theory duality when considering Calabi--Yau threefold compactifications \cite{Grimm:2017okk}. 
The precise definition of the higher-derivative terms in \eqref{R4} and \eqref{Gterms} can be found in the appendix  in \ref{section-basisG2R3}. The detailed index structure of the terms  $\big(   \nabla   G\big)^2   R^2$  in \eqref{Gterms} can be found in \ref{section-basisG2R3}.

\subsection{Checks on the $G^2R^3$ and $ (\nabla G)^2R^2$-sector}

\label{CY3checkM}

 It is well known that no supersymmetric completion of the eleven-dimensional  $  G^2   \cR^3$-sector  and $ (\nabla G)^2R^2$-sector  is known.  The eleven-dimensional eight-derivative terms involving two powers of the four-form field strength are lifted from the corresponding terms in the Type IIA  effective action. Those arise at the level of the five point-functions in the Type IIA superstring and partial indirect conclusions can be drawn at the level of the six-point function \cite{Liu:2013dna}. However,  let us stress that a conclusive study at the level of the  six-point function and especially at higher order n-point functions remains absent. In particular  a supersymmetric completion of the $  G^2   \cR^3$-sector  and $ (\nabla G)^2R^2$-sector employing the Noether coupling method would be of great interest.
It is thus desirable to discuss  possible extensions of the $  G^2   \cR^3$  and  $ (\nabla G)^2R^2$-sector beyond the known terms. In this section we accomplish this task and provide a complete maximal extension of the eleven-dimensional $G^2   \cR^3$ and  $ (\nabla G)^2R^2$-sector   relevant for Calabi--Yau fourfold reductions.\footnote{In other words due to the Calabi-Yau condition certain terms in the Ansatz yield zero upon reduction. Those coefficients can not be fixed by our arguments but constitute a complete description relevant for Calabi--Yau fourfold reductions. }

Instead of computing string amplitudes or employing the Noether coupling method we take a more pragmatic way here.  In \cite{Grimm:2017pid} a complete  basis of eight-derivative terms of the schematic form $  G^2   \cR^3$ was constructed. We then compliment this with a basis for the   $ (\nabla G)^2R^2$-sector given in appendix \eqref{section-basisG2R3}, both of which then upon dimensional reduction contribute to the kinetic terms of the three-dimensional vectors. 
To constrain the free parameters  we follow the same logic as in our previous work \cite{Grimm:2017okk,Grimm:2017pid}, namely deriving constraints on the parameter of the eleven-dimensional Ansatz by verifying  compatibility upon dimensional reduction with lower-dimensional supersymmetry. For example, as the $R^4$-sector is known to be complete one can fix certain lower-dimensional supersymmetry variables solely by deriving its dimensional reduction, which then can be compared to the ones derived from the $  G^2   \cR^3$ and  $ (\nabla G)^2R^2$-sector.

\hspace{0,4 cm}

Let us next discuss the general form of the relevant terms in the basis of $  G^2   \cR^3$ and  $(  \nabla   G)^2    \cR^2$.
The terms contributing to the three-dimensional effective action are those, which do not contain any Ricci tensors or scalars as these vanish trivially on a Calabi--Yau manifold. Taking into account the first Bianchi identity for the Riemann tensor a minimal basis of these terms is given in appendix \ref{section-basisG2R3}.  The general expansion of terms which may contribute in addition to \eqref{Gterms}  to the three-dimensional action is then
\be\label{newAnsatzG2R3}
2 \kappa_{11}^2 \,  S^{\rm{extra, \,gen}}=\alpha^2 \, \int_{M_{11}}\sum_{i=1}^{17}  C_{i} \, \mathcal{B}_i \,   \ast 1 \, + \sum_{i=1}^{24} C_{i+17} \, B_i \,   \ast 1 \, 
\ee
for some coefficients $C_i \in \mathbb{R}$. To restrict the parameters in  the Ansatz \eqref{newAnsatzG2R3}   we  first take a detour to Calabi--Yau threefold compactifications   and  furthermore  discuss the dimensional reduction on $K3 \times S^1$.
Thus in  particular, we provide the maximal complete extensions of the eleven-dimensional $  G^2   \cR^3$  and $ (\nabla G)^2R^2$-sector \eqref{newAnsatzG2R3}, which is compatible upon dimensional reduction with  five-dimensional, $\mathcal{N}=2$ supersymmetry, i.e.\,\,by  dimensional reduction on Calabi--Yau threefolds  to five dimension for a generic number of K\"ahler moduli. Moreover, we perform the dimensional reduction  on  $K3 \times S^1$ to six dimensions and employ the Heterotic - IIA duality to compare  the resulting four-derivative couplings to the  well known terms on the Heterotic side of the duality.
It turns out that these arguments are very restrictive and allow us to parametrize the  $  G^2   \cR^3$ basis with only five parameters \cite{Grimm:2017pid}. However when allowing for an interplay with the  $ (\nabla G)^2R^2$-sector the number of independent parameters reduces from forty-one to  thirteen.

Moreover the above analysis allows us to infer that the  $  G^2   \cR^3$  and $ (\nabla G)^2R^2$-terms are consistent with the partially known six-point function results \cite{Liu:2013dna}. Let us stress that it would be of great interest to study additional  constrains on this eleven-dimensional sector  by circular reduction to type  IIA effective supergravity. Any combination of novel terms need to be vanishing at the level of the five-point one-loop string scattering amplitude with two NS-NS two-form field and three graviton vertex operator insertions. We suggest that such a study will lead to fix the  remaining parameter freedom in the eleven-dimensional action.

By dimensionally reducing the  extension \eqref{newAnsatzG2R3} one  modifies the kinetic couplings of the three-dimensional vectors and introduces an additional parameter freedom. One may use to this to rewrite the reduction result in terms of $3d, \, \cN=2$ variables. 
In section  \ref{reductionCY4generic} we  perform the dimensional reduction of the $  G^2   \cR^3$ and $ (\nabla G)^2R^2$-extensions to three space-time dimensions on  Calabi--Yau fourfolds with arbitrary number of K\"ahler moduli.

 \paragraph{Calabi--Yau threefold checks to  $5d, \cN=2$.}  In the following we  derive constraints on the coefficients $C_{i}$ in   \eqref{newAnsatzG2R3}  by demanding compatibility with $\mathcal{N}=2$ supersymmetry in five dimensions upon compactification on a Calabi--Yau threefold. The $l_{\rm M}^6$-corrections give contributions to the  five-dimensional vector multiplets of the  $\mathcal{N}=2$ supergravity which   is expressed in terms of a real pre-potential $\mathcal{F}(X^I)$ and real special coordinates $X^I$. Note that  physical scalars in the vector multiplets obey 
\be\label{prepot5d}
\mathcal{F}(X^I)=\tfr{1}{3!} C_{IJK} \, X^I X^J X^K=1\, \;\;.
\ee
The  totally symmetric  and constant tensor $C_{IJK}$ is entirely  determined by the $U(1)$ Chern-Simons terms $\sim C_{IJK} \, A^I F^J F^K$, which  however do not receive $l_{\rm M}^{\, 6}$--corrections. One concludes that also the physical scalars $X^I$ remain uncorrected.

We  dimensionally reduce the action \eqref{newAnsatzG2R3} with general coefficients $C_i$ on a Calabi--Yau threefold $Y_3$ to five dimensions. As our focus is on the kinetic terms for the vectors  we  note that in order to dimensionally reduce one expands
\be
  G= F_{\rm 5D}^i \wedge \omega^{CY_3}_i\, ,
\ee
with the field strength of the  five-dimensional vectors  $F_{\rm 5D}^i$  and  the harmonic $(1,\,1)$-forms on the Calabi--Yau threefold $\omega_{i}^{CY_3}$,  $ i=1,\dots,  h^{1,1}(CY_3)$.
 The constraints imposed by supersymmetry are then inferred by making use of Shouten and total derivative identities on the internal space $CY_3$. The condition one encounters is that novel  terms \eqref{newAnsatzG2R3} may no contribute to the five-dimensional couplings, which is equivalent to the non-renormalisation of \eqref{prepot5d}. The computation is in principal straightforward (but tedious) and leads  us to impose the relations among the coefficients $C_1\,\dots,C_{41}$. Details can be found in the appendix \eqref{parameters}.

 \paragraph{Heterotic and type IIA duality.} 
In this section we compactify \eqref{newAnsatzG2R3} on $K3 \times S^1$.  We first circular reduce the basis of forty-one $ G^2  \cR^3$ and $ (\nabla G)^2R^2$-terms to ten dimensions  on $\mathbb{R}^{1,9} \times S^1$ to obtain a $l_{\rm M}^6$-modified  IIA supergravity theory.
The only terms relevant for us are the ones which arise from
\beq
  G_{11 M N O} =  e^{\frac{\phi}{3}}H_{M N O}  \;\; ,\;\;\;  M,N,O = 1,\dots,10 \;\;\; ,
\eeq
where $11$ denotes the direction along $S^1$ and with $H$ the field strength of the type IIA Kalb-Ramond tensor field.
We then check compatibility of the novel induced $H^2 R^3$-terms making use of the IIA - Heterotic duality by dimensional reduction on $K3$. Compactifying  type IIA on $K3$ is dual to the Heterotic string on $\mathbb{T}^4$. For our purpose it is enough to show that when compactifying the novel  $H^2 R^3$-terms on $K3$ those do not induce any $l_{\rm M}^6$-correction to the six-dimensional action.  In particular.  the absence of  four-derivative terms is imposed, which results in one further constraint on the parameters. The additional constraints  on the $C$'s arises from imposing the vanishing of the four-derivative terms such as e.g.
\beq
 \sim \chi(K3) \; H^{6D}{}^{\mu \nu \rho} \, H^{6D}{}_{\mu}{}^{ \nu_1 \rho_1} \, R^{6D}{}_{\mu \mu_1 \nu  \nu_1} \;\; ,
 \eeq
 with $\mu,\nu =1,\dots,6$.  One then infers the additional constraints on the parameters in \eqref{newAnsatzG2R3} to be
\beq\label{K3constraints}
C_2 = 0\;\; , \;\;\;  C_1 = - \tfr{1}{6} \big(8 C_3 + 2 C_{31} + C_{35} + 36 C_4 + 3 C_6 \big) \;\;\; .
\eeq
This concludes that by fixing the parameter \eqref{K3constraints} the proposed maximal extension of $  G^2   \cR^3$ and $ (\nabla G)^2R^2$-terms  in the M-theory effective action is fully consistent with the indirect six-point functions results discussed in  \cite{Liu:2013dna}.

\section{Three-dimensional effective actions revisited }
\label{sec-3daction}

F-theory can be viewed as a map of dualities which allows one to derive controlled IIB orientifold backgrounds at weak string coupling which incorporate for back-reacted $D7$ branes and $O7$-planes on the axio-dilaton \cite{Denef:2008wq,Vafa:1996xn,Grimm:2010ks}. The starting point of this journey is eleven-dimensional supergravity, which compactified on an appropriate eight-dimensional internal space gives a $3d,\,\cN=2$ supergravity theory which can then be related via the F-theory lift to a $4d,\,\cN=1$ supergravity theory.
The main objective of this section is the dimensional reduction of eleven-dimensional supergravity including the  novel eight-derivative couplings \eqref{newAnsatzG2R3}  on Calabi--Yau fourfolds for a generic number of K\"ahler moduli in section \ref{reductionCY4generic}. We start our discussion with a review of the generic properties of $3d,\, \cN =2$ supergravity theories in section \ref{3dN2sugra}. Finally, we conclude this section with a review of the one-modulus case in which the warp-factor as well as the higher-derivative couplings can be matched to the  $3d,\, \cN =2$  variables   \cite{Grimm:2017pid}.

\paragraph{Background solution.}  Let us set the stage by reviewing the fourfold solutions including eight-derivative terms studied in \cite{Becker:1996gj, Becker:2001pm,Grimm:2014xva}. The background solution is taken to be an expansion in terms of the dimensionful parameter \footnote{We follow the conventions of \cite{Tseytlin:2000sf}.}
\be \label{def-alpha}
\al^2=\fr{(4 \pi \, \k_{11}^2)^{\fr{2}{3}}}{(2\pi)^4 \, 3^2 \cdot 2^{13}}\, ,\qquad \qquad 2\k_{11}^2=(2\pi)^5 \, l^{\,9}_{\rm M} \, ,
\ee
which reduces to the ordinary direct product solution $\mathbb{R}^{1,2} \times CY_4$ without fluxes and warping to lowest order in $\al$.  At  order  $\al^2$  a warp-factor $W^\tbtwo = W^\tbtwo(z,\bar z)$ and fluxes are induced. The background solution is known \cite{Becker:2001pm, Grimm:2014xva} to then take the form
\begin{align}\label{background_metric}
\la d s^2 \rangle &= \e^{\al^2 \, \Phi^\tbtwo} \Big(\e^{-2 \al^2 \,W ^\tbtwo} \eta_{\mu \nu} \, d x^\mu  d x^\nu + 2\e^{\al^2 \, W^\tbtwo} \, g_{m \bar m}\,  d z^m  d \bar z^{\bar m}\Big)\, ,\\\label{background_metric0}
\la   G \rangle &= \al \, G^\tbo+ {\text{dvol}}_{\mathbb{R}^{1,2}} \wedge d \big(\e^{-3 \al^2 \, W^\tbtwo} \big)\, .
\end{align}
By solving the eleven-dimensional E.O.M.'s for the  metric $g_{m \bar m}$  of the internal space one encounters that it seizes to be Ricci flat i.e.\,Calabi--Yau \cite{Grimm:2013gma}. It receives a correction at order $\al^2$ as
\beq\label{background_metric2}
g_{m \bar m}=g^\tbzero_{m \bar m}+\al^2 \, g^\tbtwo_{m  \bar m}\, , \qquad g_{m \bar m}^\tbtwo \sim \pd_m \bar \pd_{\bar m}\,  \ast^\tbzero\big(J^\tbzero \wedge J^\tbzero \wedge F_4 \big)\, ,
\eeq
where $g^\tbzero$ is the lowest order, Ricci-flat Calabi--Yau metric and $J^\tbzero$ is its associated K\"ahler form and where  $F_4$ the non-harmonic part of the third Chern form. Latter  is however  irrelevant for the following discussion, as it only contributes couplings  to the effective action which are total derivatives \cite{Grimm:2013bha}. Furthermore, \eqref{background_metric2}  includes an overall Weyl factor $\Phi^\tbtwo= -\fr{512}{3}\ast^\tbzero \big(c_3^\tbzero \we J^\tbzero \big)$, which was first discussed in \cite{Grimm:2014xva} and a warp-factor $W^\tbtwo(z, \bar z)$ satisfying the warp-factor equation
\be\label{eq:warp}
\Delta^\tbzero \, \e^{3 \al^2  W^\tbtwo} \,  d \text{vol}^\tbzero_{Y_4}+\fr{1}{2} \al^2 \, G^\tbo \wedge G^\tbo-3^2 2^{13}\, \al^2 X^\tbzero_8=0\, .
\ee
The background value of the four-form field  strength \eqref{background_metric0} is given by the sum of the internal flux $G^\tbo \in H^4 (CY_4)$ and a warp-factor contribution. Due to lowest order supersymmetry constraints the flux  is to be self-dual with respect to the lowest order Calabi--Yau metric. Note that we do not discuss the corrections to the gravitino variations at order $l_{\rm M}^6$ here but refer the reader to \cite{Grimm:2014xva} for a detailed discussion.  Let us emphasize that the $l_{\rm M}^6$-gravitino variations are not known as a supersymmetric completion of eleven-dimensional supergravity at higher $l_{\rm M}$-order remains elusive. However, it is widely believed that  \eqref{background_metric}-\eqref{eq:warp} constitutes a supersymmetric background.

\subsection{Three-dimensional gauged $\cN=2$ supergravity}
\label{3dN2sugra}
In this section we briefly review $\mathcal{N}=2$ gauged supergravity in three dimensions where all shift symmetries are gauged. Shift symmetries corresponds to an isometry of the geometry of the scalar field space. Three-dimensional maximal and non-maximal supergravities are discussed in \cite{deWit:2004yr}.   For our purpose it is sufficient to consider three-dimensional $\cN=2$ supergravity coupled to chiral multiplets with  complex scalars  $N^a$, which are  gauged along the isometries $I^{ab}$  and subject to the constant embedding tensor $\Theta_{ab}$. One then infers the simply form of the $\mathcal{N}=2$ action to be
\begin{align}
S_{N=2}=\int_{M_3} \tfr{1}{2} R \, \ast   1-K_{a \bar b}\,  \nabla N^a \wedge \ast \, \nabla \bar N^{\bar b}-\tfr{1}{2} \Theta_{ab}\, A^a \wedge F^b-\big(V_{D}+V_{F} \big) \ast 1\, ,\label{chiralN=2}
\end{align}
where $K_{a \bar b}=\partial_{N^a} \partial_{\bar N^{\bar b}}K$ is a K\"ahler metric with K\"ahler potential $K$. The gauge covariant derivative $\nabla N^a$ is defined by $
\nabla N^a=d  N^a+\Theta_{bc}\, I^{ab} \, A^c\, $ .
The F-term scalar potential in \eqref{chiralN=2} is given by
\begin{align}
\label{ScPot1}
V_F&=\e^K \big(K^{a \bar b} D_i W \overline{D_b W}-4 \lvert W \rvert^2\big)\, ,
\end{align}
with $K^{a \bar b}=(K^{-1})^{a \bar b}$    the inverse of the  K\"ahler metric given by a hermitian matrix and $W$ a holomorphic super potential.
Furthermore, one finds that $V_{D} =K^{a \bar b} \, \partial_a D \partial_{\bar b} D - D^2\, $ where $D$ is a real function of the chiral fields $N^i$. Lastly, note that the vectors in the Chern-Simons term \eqref{chiralN=2}  are non-dynamical.

\paragraph{Dualization of the action. } One may now split the chiral fields as $N^a=(M^I, \, T_i)$ and dualizes the chiral multiplets  in \eqref{chiralN=2} with bosonic component $T_i$ into vector multiplets \cite{Grimm:2011tb}.  
Note that  dualization is in general not possible but requires $\Im T_i$ to admit a  shift symmetry. 
Upon Legendre dualization  the theory depends on the kinematic potential $\tilde K$ which is expressed in terms of the quantities of  the dual theory as 
\be
K(M, T)=\tilde K(M, L)-\Re T_i \, L^i\, \;\; ,\;\;\;  L^i = - \frac{\pd K}{\pd \Re T_ i}\;\; .\label{legendre}
\ee
One then derives the  dual action to take the form\footnote{One may choose a constant embedding tensor such that
\be
I^{ij}=-2 i \, d x^{ij}\, , \qquad I^{I  J}=\tilde I^{i \bar J}=0\,, \qquad I^{i J}
=0\, , \qquad \Theta_{IJ}=0\, .\nonumber
\ee }
\begin{align}
S_{\cN=2,\,  \rm {dual}}&=\int_{M_3} \tfr{1}{2}R \ast 1-\tilde K_{M^I \bar M^J} \, \cD M^I \we \ast \cD \bar M^{\bar J}+\tfr{1}{4} \tilde K_{L^i L^j} \,  d L^i \we\ast\, d L^j \nonumber\\
&+\int_{M_3}\tfr{1}{4} \tilde K_{L^i L^j} \, F^i \we\ast \,  F^j+\tfr{1}{2} \Theta_{ij} A^i \we F^j+F^i \we\Im \big[ \tilde K_{L^i M^I} \, \nabla M^I \big]\,\nonumber \\
&-\int_{M_3} \big(V_{D}+V_F \big) \ast \, 1\, ,\label{dualN=2}
\end{align}
 with  kinematic couplings given by
 \beq
 \tilde K_{L^i L^j}=\partial_{L^i} \partial_{L^j} \tilde K \;\; .
 \eeq
Note that the scalars $L^i$  belong to vector multiplets. One may furthermore infer from   \eqref{legendre}  that 
\be
K_{T_{i} \bar T_{j}}=-\tfr{1}{4} \tilde K^{L^i L^j}\, , \qquad \Re T_{i}=\tilde K_{L^j}\, , \qquad \fr{\partial L^i}{\partial T_j}=\tfr{1}{2} \tilde K^{L^i L^j}\, .
\ee
Left to discuss is the dualization of the scalar potential.\footnote{
The D-term results in
\begin{align}
V_D=\tilde K^{M^I \bar M^{\bar J}} \partial_{M^{\bar I}} \cT \,\partial_{\bar M^{\bar J}} \cT-\tilde K^{L^i L^j} \partial_{L^i}\cT \, \partial_{ L^j} D- D\, \;\;\; ,\;\;\; 
D =-\tfr{1}{2} L^i \, \Theta_{ij }\,  L^j\, .\label{Texpr}
\end{align}} 
The F-term scalar potential  in the vector multiplet language is then given by
\be\label{Ftermsimple}
V_F=\e^K \, \Big[ \tilde K^{M^I \bar M^{\bar J}} D_{M^I} W \, \overline{D_{M^J} W}-\big(4+ L^i \, \tilde K_{L^i L_j}\, L^j \big) \big| W \big| ^2\Big]\, .
\ee
where we have assumed that the superpotential does not depend on the scalars $L^i$ in the vector multiplet. This case is relevant when matching to the string theory reduction result in which the superpotential does not depend on the K\"ahler moduli, i.e.\;\;non-perturbative effects such as $M5$-brane instantons  are absent. For the discussion in this work this will be sufficient but one may choose to generalize \eqref{Ftermsimple} easily.

\subsection{Calabi--Yau fourfold reduction for generic $h^{1,1} $}
\label{reductionCY4generic}
In this section we discuss the reduction result of M-theory involving the eight-derivative action \eqref{eq:S0}-\eqref{Gterms} and \eqref{newAnsatzG2R3} on the warped background \eqref{background_metric}-\eqref{eq:warp} and allow for an arbitrary number of K\"ahler moduli of the internal manifold. Latter is achieved by deforming the background metric as
\beq\label{Kahlerdeformation}
g_{m \bar n} \to g_{m \bar n} + i \delta v^i  \omega^\tbzero_{i m \bar n} \;\; ,
\eeq
where $\delta v^i = \delta v^i (x)$ are infinitesimal scalar deformations and $\{\o^\tbzero_{i}\}$  are harmonic $(1,1)$-forms w.r.t\, the background Calabi--Yau metric $g^\tbzero$, with $ i=1,\dots,h^{1,1} (CY_4)$. The non-vanishing contribution for the dynamical three-dimensional vectors $A^i_\mu$ is derived by\footnote{Note that in the presence of $l_{\rm M}^6$-correction the deformations \eqref{Kahlerdeformation} and \eqref{Vectordeformation} may receive higher-order corrections as discussed in \cite{Grimm:2015mua,Grimm:2014efa}, none of which alter the dynamics of the resulting theory. We thus omit them from the present discussion.  }
\beq\label{Vectordeformation}
G_{\mu \nu m \bar n} = F^i_{\mu \nu} \o^\tbzero_{i m \bar n} \;\; ,\;\;\; F^i = dA^i \;\; .
\eeq
To enhance the  readability of the main text in the following we shift the  more technical steps to the appendix. To express the reduction result we need to introduce several higher-derivative building blocks.  Among them the familiar second and third Chern-forms  $c_2$ and $c_3$, respectively, and $Z, Z_{m \bar m}, Z_{m \bar m n \bar n}$ and $\mathcal{Y}_{ij }, \Omega_{ij }$. All higher-derivative objects are w.r.t.\,the zeroth $\al$-order Calabi--Yau metric. Their precise  definition can be found in appendix \ref{Conv_appendix}, in particular \eqref{def-Zmmmm}-\eqref{defOmega}. Here let us schematically note that\vspace{0.3cm}
\ba
Z, \,Z_{m \bar m}, \,Z_{m \bar m n \bar n} \sim \big(R \big)^3 \;\;\; , \;\; 
\mathcal{Y}_{ij }  \sim (\nabla \o_i)(\nabla \o_j)\big(R \big)^2 \; \;\; ,\;\; \Omega_{ ij } \sim ( \o_i)( \o_j) \big( R\big) \;\; . \\[-0.2cm]\nn
\ea 
where $R$ denotes the Riemann tensor on the internal manifold  and $\nabla$ is  the covariant derivative w.r.t.\,the Calabi--Yau metric.
The warp-factor dependence can be elegantly captured by introducing the warped volume and warped metric
\ba\label{def-cVW}
\cV_{\mathscr{W}} = \cV+ 3 \mathscr{W} \;\;\; , & \;\; \mathscr{W}= \int_{Y_4}  W^\tbtwo  *^\tbzero 1      \ , &
G^\mathscr{W}_{ij} =& \frac{1}{2 \cV_\mathscr{W}} \int_{Y_4} e^{3 \alpha^2 W^\tbtwo} \o^\tbzero_{i} \wedge *^\tbzero \o^\tbzero_{j} \, , 
\ea
which at zeroth order in $\alpha$ reduce to $\cV$ and 
$G_{ij} = \frac{1}{2 \cV} \int_{Y_4} \o^\tbzero_{i} \wedge *^\tbzero \o^\tbzero_{j} $. 
We also introduce 
\ba
  \cK_i^\mathscr{W} =& i \cV_\mathscr{W}  \, \omega^\tbzero_{i m}{}^m + \frac{9}{2} \alpha^2 \int_{Y_4} \pa_i W^\tbtwo| *^\tbzero 1\ ,
\ea
which at lowest order simply reduces to $ \cK^\tbzero_i = i \cV  \, \o^\tbzero_{i m}{}^m= \tfrac{1}{3!} \int_{Y_4}  \o^\tbzero_{i} \wedge  J^\tbzero \wedge J^\tbzero  \wedge J^\tbzero $. Note that we use the notation $ \cK^\tbzero_i $ to abbreviate the intersection number evaluated in the background, in contrast to the analogue quantities $\cK_i$ which may vary over the  K\"ahler moduli space. 
With these definitions we state that the action including the $l_{\rm M}^6$-corrections to the kinetic terms \cite{Grimm:2014efa,Grimm:2015mua} is given by
\ba \label{actionvs10}
S_\text{kin}  =&   \fr{1}{2 \k_{11}} \int_{\cM_3} \bls  R \ast 1 -   (G^\mathscr{W}_{ij} +\cV_\mathscr{W}^{-2} K_i^\mathscr{W}  K_j^\mathscr{W} )d \d v^i \we * d \d v^j  
    -  \cV^{2}_\mathscr{W} G_{ij}^\mathscr{W} F^i \we \ast F^j  & \nonumber \\[0.2cm]
     & - d \d v^i \we * d \d v^j    \frac{\alpha^2}{\cV_0} \int_{CY_4} \Big(768 Z  \o^\tbzero_{i m}{}^m\o^\tbzero_{j n}{}^n   - 3072 i Z_{m \bar n}  \o^\tbzero_{i}{}^{\bar n m} \o^\tbzero_{j s}{}^s \Big)*^\tbzero 1   \nonumber\\[0.2cm]
  &   +  d \d v^i \we * d \d v^j    \frac{\alpha^2}{\cV_0} \int_{CY_4} 3072  Z_{m \bar n r \bar s} \o_i^{\tbzero \bar n m} \o_j^{\tbzero  \bar s r }  *^\tbzero 1  &\nonumber\\[0.2cm]
&  -  F^i \we \ast  F^j   \alpha^2 \cV_0 \int_{CY_4} \Big( - 256  Z \o^\tbzero_{i m \bar n} \o^\tbzero_{j}{}^{ \bar n m } + 192 (7 - a_1)i Z_{m \bar n} \o^\tbzero_{i}{}^{\bar r m} \o^\tbzero_{j}{}^{\bar n}{}_{\bar r } \Big) *^\tbzero1   &\nonumber\\[0.2cm]
&  +  F^i \we \ast  F^j   \alpha^2 \cV_0 \int_{CY_4} 384 (1 + a_1)  Z_{m \bar n r \bar s} \o_i^{\tbzero \bar n m} \o_j^{\tbzero \bar s r}  *^\tbzero1 \;\;\; + \Theta_{ij} A^i \wedge F^i   \brs\  \; \;. &
\ea
The one parameter freedom $a_1$ arises from the uncertainty inherent in the $(\nabla G)^2 R^2$-sector. From the novel sector \cite{Grimm:2017okk}  we find\vspace{0.3 cm}
\beq\label{novelG2R3red}
\delta S_1 =  256 \;  F^i \we * F^j   \alpha^2 \cV  \int_{Y_4}  Z \omega^\tbzero_{i m \bar n} \o^\tbzero_{j}{}^{ \bar n m } \ast^\tbzero 1 \;\; .
\eeq 
Note that  novel eleven-dimensional terms \eqref{novelG2R3red} is precisely cancelled by the same structure in  \eqref{actionvs10}.
Lastly, one performs the dimensional reduction of \eqref{newAnsatzG2R3} to yield
\ba\label{new-reductioResult}
\delta S_2 = \;\; \;\; & F^i \we * F^j   \alpha^2 \cV \int_{Y_4} \Big( 8 i  (a_3 + a_4)  Z_{m \bar n}  \o^\tbzero_{i}{}^{\bar n m} \o^\tbzero_{j s}{}^s \ast 1 - 8 a_3 Z_{m \bar n r \bar s} \o_i^{\tbzero \bar n m} \o_j^{\tbzero  \bar s r } \Big)\ast^\tbzero 1 \nn  \\[0.2 cm] 
 +  \;& F^i \we * F^j   \alpha^2 \cV \int_{Y_4}  a_2\;  c_2 \wedge  \Omega_{ij}  \;\;\; ,
\ea
with the coefficients $a_3,a_4$ result from the unfixed eleven dimensional parameters, $a_3 = -C_{22} +4 C_3 $ and $a_4 = 18 C_4$.
Let us close this section with some remarks.
 Note that in \eqref{new-reductioResult}  one obtains a term proportional to the second Chern-form. 
In the limit  $h^{1,1} \to 1$, i.e.\;\;the one-modulus case we see that 
\beq
\delta S_2 \to a_4 \cZ \;\;,
\eeq 
as the term $\Omega_{ij}$ vanishes. For the physical arguments provided in \cite{Grimm:2017pid} where the one-modulus case is discussed  we  infer that  $\delta S_2 \to 0$ as it would change the physical interpretation else-wise. Hence in the remainder of this work we assume $C_{14} =0 $ and thus $a_4 = 0$.\footnote{Comparison to five point-scattering and six-point amplitudes  can in principle  fix the 11-dimensional coefficient of the basis, thus also $C_{14}$.}
Furthermore, note that the action \eqref{actionvs10} depends on the infinitesimal deformation $\d v^i $. To establish the connection to the full field space $ v^i$, i.e.\,the  coordinates on the K\"ahler moduli space we replace $\delta v^i \to v^i$ in the following.\footnote{Possible obstructions and subtleties to this step for  higher-derivative couplings of non-topological nature were discussed in \cite{Weissenbacher:2016gey}.} This will become relevant for the discussion in section \ref{sec-3dKaehlerGeneric}.

\subsection{Review  one-modulus  K\"ahler potential and coordinates}
\label{sec-reviewOneMod}

The  dimensional reduction of the eleven-dimensional supergravity action including higher-derivative terms on a warped  Calabi--Yau fourfold  background with one K\"ahler modulus, i.e.\;\;$h^{1,1} = 1$ case was discussed rigorously in \cite{Grimm:2017pid}. We devote this section to reviewing this discussion, in particular the derivation of the K\"ahler potential and coordinates of the $3d, \, \cN = 2$ theory. As a starting point  we may take the limit $h^{1,1} \to 1$ of the generic Calabi--Yau fourfold reduction result presented in  \eqref{actionvs10}-\eqref{new-reductioResult}.
One then infers  that the one-modulus $l_{\rm M}^6$-corrected action r  takes the standard form 
\be
S^{3d}=\int_{M_3}\fr{1}{2} R \ast 1+\fr{1}{4} \tilde{G}_{LL}(L)\, d L \wedge \ast \, d L+\fr{1}{4}\tilde{G}_{LL}(L)\,  \, F\wedge \ast \, F\, ,
\ee
with
\be\label{kinmet}
\tilde{G}_{LL}(L)=-\fr{4}{L^2} \Big(1-384 \, \al^2 \, \tilde \cZ  \, L \Big)=-\fr{4}{L^2}+1536 \, \al^2\, \tilde\cZ \, \fr{1}{L}\, .
\ee
and with the topological coupling depending on the third Chern-form given by
\beq
\cZ = (2 \pi^3)\,  \int_{CY_4} c_3 \wedge  J \;\;   \; ,\;\;\; \;\;  \cZ   = \cV^{\tfrac{1}{4}} \tilde\cZ \;\;\;,
\eeq
where we have used that $J = \omega_0 \cV^{\tfrac{1}{4}} $.
We can integrate the metric $\tilde G_{LL}$ to obtain the kinetic potential $\tilde K(L)$ and coordinate
\begin{align}
\tilde{K} & =4 \log L +1536\, \al^2 \,  \tilde\cZ\, L \, \big(\log (L)-1 \big)+4\, ,\label{Kpot}\\[0.2cm]
L & =\cV^{-\fr{3}{4}}-3 \, \al^2 \, \mathscr{W}\, \cV^{-\fr{7}{4}}\, , \label{resultL}
\end{align}
where we have chosen the integration constants in a convenient way. 

\paragraph{Determining the K\"ahler potential.}  One may next dualize the vector multiplet to a chiral multiplet, whose metric derives from a K\"ahler potential. As outlined in section \ref{3dN2sugra} this is achieved by a Legendre transformation of the kinetic potential 
\be
K=\tilde{K}-L \, \Re\, T\, , \qquad \qquad \Re \, T=\partial_L \tilde{K}\, .
\ee
One thus derives the K\"ahler potential  $K(T+\bar T)$ to be
\bea \label{Kpot1}
K &=& 4 \log L-1536 \, \al^2 \, \tilde \cZ \, L 
   =-3 \log \Big( \cV + \al^2 \big(4 \mathscr{W} + 512\, \cZ\big) \Big)\,,  
\eea
with corresponding coordinate
\bea
\Re T&=& \fr{4}{L}+1536 \al^2 \, \tilde\cZ \, \log L 
         = 4 \cV^{\fr{3}{4}}+12 \al^2 \, \cV^{-\fr{1}{4}}\mathscr{W}\,-1152\,  \al^2 \, \tilde \cZ \, \log \cV \ .\label{ReTcorrected}
\eea
Note that all quantities in the K\"ahler potential \eqref{Kpot1}  depend on the one-modulus $\cV$, i.e.\;\;the overall volume.

\paragraph{The no-scale condition and the scalar potential.}
We next argue that the $\lM^{\, 6}$-suppressed corrections to the K\"ahler potential in \eqref{Kpot1}  generically lead to a breaking of the no-scale condition and thus generate a $F$-term scalar potential.  One  straightforwardly computes that
\be
K_T \, K^{T \bar T} \, K_{\bar T} \; = \; \fr{K_T^2}{K_{T \bar T}} \;=\;4-1536 \,\, \frac{ \al^2 } {\cV} \cZ   \; .\label{NoScaleCond}
\ee
One may next infer the scalar potential originating from the breaking of the no-scale condition. It enters the effective action via the F-term scalar potential\footnote{Note that superpotential can not be renormalized perturbatively but may be subject to e.g.\,\,$M5$-instanton corrections which correspond to $D3$-instantons in the F-theory limit \cite{Blumenhagen:2010ja}.}  
\be
V_F=\e^K \big(K^{T \bar T}\,D_T W \overline{D_{ T}W}-4 \big| W \big|^2 \big)=-1536 \, \al^2 \, \fr{\big| W_0 \big|^2}{\cV^4} \cZ \;\;  \label{FScPot} \;\; .
\ee
Note that it exhibits a runaway direction for $\cV \to \infty$ if $\int_{Y_4} c_3 \wedge J < 0$ \footnote{An example with this property and $h^{1, 1}=1$ is the sextic fourfold. For the sextic one finds $\int_{Y_4} c_3 \wedge \omega =-420$.}. In  \eqref{FScPot} we assumed that the complex structure moduli  are stabilized by  the GVW superpotential \cite{Gukov:1999ya} given by
 \beq\label{GVWsuper}
W=\fr{1}{\lM^{\, 3}}\int_{Y_4}G^\tbo \wedge \Omega \, , \qquad \Omega \in H^{4,0}(Y_4) \, .
\eeq
 which in the vacuum then takes the  constant value $W_0$. A critical assessment of this two step procedure is discussed in \cite{Choi:2004sx,deAlwis:2005tf,Lust:2005dy}.
The runaway behavior of \eqref{FScPot} for large volume $\cV$ signals an instability of the solution for the case of a non-vanishing $W_0$  as recently examined  in \cite{Sethi:2017phn}. 

Let us conclude this section by emphasizing the importance of  the  one-modulus results in particular the integration into a K\"ahler potential and coordinates. In a following section we will  show compatibility with the generic moduli case which is exceedingly more complicated due to the appearance of  non-topological higher-derivative contributions to the K\"ahler metric.

\section{Three-dimensional  K\"ahler potential and coordinates}
\label{sec-3dKaehlerGeneric}

The eleven-dimensional higher-derivative corrections manifest themselves in terms of $l_{\rm M}^6$-modifications  of the  kinematic couplings of the two-derivative three-dimensional supergravity theory as discussed in the previous section \ref{reductionCY4generic}. The objective is to express these $l_{\rm M}^6$-modifications  to the  kinematic couplings  in the language of three-dimensional, $\cN=2$ supergravity. Namely these must result from a $l_{\rm M}^6$-correction to the K\"ahler potential and K\"ahler coordinates, i.e.\;\;fixing the complex structure on the K\"ahler moduli space. We reviewed this procedure for the one-modulus case, i.e.\;\;$h^{1,1} =1$ in \ref{sec-reviewOneMod}. In this section 
 we propose a novel description of the K\"ahler coordinates in terms of divisor integrals.
Due to these specific divisor integrals of the Calabi--Yau fourfold one manages to reproduce all high-derivative structures appearing  in the reduction result of  the  K\"ahler metric \eqref{actionvs10}-\eqref{new-reductioResult} which we discuss in section \ref{secCYDivisorINt}.
To motivate our Ansatz note that the K\"ahler coordinates 
are expected to linearise the action of M5-brane instantons on divisors $D_i$.\footnote{ In fact,
as discussed in \cite{Witten:1996bn} a holomorphic super-potential of the schematic form $W \propto e^{-T_i}$ can 
be induced by such instanton effects.} This implies that the $T_i$'s are expected to be integrals 
over divisors $D_i$.  In particular the Ansatz depends on the  first, second and third Chern-form of the Divisors $\tilde c_{1,2,3} = \tilde c_{1,2,3} (D_i)$.
Let us first recall further definitions 
\beq
\cZ_i = (2 \pi)^3\,  \int_{CY_4} c_3 \wedge \omega_i = (2 \pi)^3\,  \int_{D_i} c_3\;\;, \;\; \mathscr{W}_i = \int_{D_i} W^\tbtwo \ast 1 \;\;, \;\;  \mathscr{F}_i = 1536 \int_{D_i} F_6 \ast 1 \;\; .
\eeq
The Ansatz for the  K\"ahler potential and coordinates depends on the real parameters $\alpha_1, \dots,\alpha_9$ and $\kappa_{1},\dots, \kappa_{6} $. We assert the  K\"ahler potential  to take the form
\beq \label{Kahlerpot}
K = -3  \log{ \Big( \cV + \al^2 \big( 4 \mathscr{W}_i v^i+ \kappa_1 \cZ_i v^i +  \kappa_2 \cT_i v^i  \big) \Big)} \;\; ,
\eeq
and for the K\"ahler coordinates to be\footnote{We omit constants shifts such as $\cZ_i$ in the definition of the K\"ahler coordinates.}
\beq
\label{Kahlercord}
\Re T_i = \cK_i + \al^2\Big(  \mathscr{F}_i + 3 \mathscr{W}_i +\kappa_3  \frac{\cK_i }{\cV} \cZ_j v^j + \kappa_4 \cZ_i \log{\cV}  + \kappa_5 \frac{\cK_i }{\cV}  \cT_j v^j + \kappa_6 \cT_i  \Big) \;\; .
\eeq
Note that the warp-factor part of this Ansatz was fixed in \cite{Grimm:2014efa,Martucci:2014ska}.\footnote{Comparison of the warp-factor contribution of the one modulus K\"ahler coordinates  \eqref{ReTcorrected} and \eqref{Kahlercord} suggest that $\mathscr{F}_i \to  9\mathscr{W} \cV^{-1/4}$ in the one-modulus case.  }
In \eqref{Kahlercord} we introduce a novel divisor integral  higher-order correction 
\bea
\label{divisorInt}
\cT_i \;\;= \;\;&  \;\; \alpha_1 &  \int_{D_i} \tilde c_1 \wedge  \tilde c_1 \wedge  \tilde c_1 + \alpha_2    \int_{D_i}  \tilde c_1 \wedge  \tilde c_2 \ + \alpha_3   \int_{D_i} \tilde c_3 +  \alpha_4    \int_{D_i} \cC_1 \; \tilde c_1 \wedge  \tilde c_1 \wedge  \tilde J 
 \nonumber \\[0.2cm]
&+  \;\; \alpha_5 & \int_{D_i} \cC_1 ^2 \;  \tilde c_1 \wedge  \tilde J \wedge \tilde J
 +    \alpha_6  \int_{D_i} \cC_1  \;  \tilde  c_2 \wedge  \tilde J 
 +    \alpha_7  \int_{D_i}  \ast_6 ( \tilde c_1 \wedge  \tilde J ) \wedge   \tilde c_2  \\[0.2cm]
&+ \;\;  \alpha_8 & \int_{D_i}  \ast_6 ( \tilde c_1 \wedge  \tilde J ) \wedge \tilde c_1 \wedge \tilde c_1
+  \alpha_9  \int_{D_i} \cC_1 \, \tilde  c_1 \wedge    \ast_6 \; \tilde  c_1  \nonumber \;\;,
\eea
with $ \cC_1 =  \ast_6 (\tilde c_1 \wedge \tilde J^2) = 2 R_{m}{}^m{}_n{}^n$ and  $i= 1,...,h^{1,1}$ and where $D_i = PD(\omega_i)$ are the Poincare-dual divisors to the harmonic forms $\omega_i$ the Calabi--Yau fourfold. Furthermore, $ \tilde c_1,  \tilde c_2,  \tilde c_3$ are the corresponding Chern-forms of the Divisor and $\tilde J = i^*J$ the pull-back of the  K\"ahler form $i : D_j \to CY_4$. In the following $c_3$ is the third Chern-form of $CY_4$.  Note that although $c_1(CY_4) = 0$ the divisors i.e.\ sub-manifolds of complex co-dimension one generically have $c_1(D_i) := \tilde c_1 \neq 0$. Let us use the  notation
\beq
 \cZ = \cZ_i v^i \;\; ,
 \eeq
in the following. Furthermore, we choose the normalization
\beq\label{chooseal1}
\alpha_3 = 1 \;\;,
\eeq
which is argued for in section \ref{secCYDivisorINt}. Note that as  in the Ansatz \eqref{Kahlercord} we allow for additional pre-factors  \eqref{chooseal1}  can be imposed without loss of generality.

Let us next briefly outline the logic of this section.  In \ref{secCYDivisorINt} we  compute the variation of the Ansatz \eqref{divisorInt}  w.r.t.\,\,K\"ahler deformations of the Calabi--Yau fourfold and show the correlation with the higher-derivative structures encountered in the reduction result. We  will argue in section \ref{secTopDivInt} that the Ansatz \eqref{Kahlerpot}  and \eqref{Kahlercord} can be rewritten solely  in terms of topological quantities of the divisors. 
All the higher-derivative structures of the reduction result  \eqref{actionvs10}-\eqref{new-reductioResult} can be matched.  This steps fixes the relative factors $\alpha_1,\dots,\alpha_9$ with one remaining free parameter $\alpha_2$. 
 In section \ref{sec-onemodcompatibiliyt} we discuss the compatibility of this Ansatz with the one-modulus case which can be integrated exactly into a K\"ahler potential \cite{Grimm:2017pid} which induces certain relations among the $\kappa$'s in the Ansatz. However, a precise  determination  of the reminiang  $\kappa$-parameters  is beyond the aim of this work and we suggest that the matching of the reduction result is possible with the Ansatz \eqref{divisorInt}, \eqref{Kahlerpot}  and \eqref{Kahlercord}  which  then may fix all the parameters uniquely. 
Lastly, we provide further indirect evidence for this claim by comparison to the newly discovered structures \eqref{new-reductioResult} proportional to the second Chern form of the Calabi--Yau fourfold which may  also be reproduced  by the novel Ansatz. This insight however is not used in the direct line of arguments which precede through the following sections.

\subsection{K\"ahler coordinates as integrals on $CY_4$}
\label{secCYDivisorINt}

To write the  integrals \eqref{divisorInt} defined over Divisors $D_i = PD(\omega_i)$  as integrals over the Calabi--Yau fourfold we note that e.g.
\beq
\int_{D_i} \tilde c_1 \wedge  \tilde c_1 \wedge  \tilde c_1  = \int_{CY_4} \tilde c_1 \wedge  \tilde c_1 \wedge  \tilde c_1  \wedge \omega_i . 
\eeq
Note that it is crucial to maintain  $\tilde c_1$  instead of $c_1$ as latter  would vanish due to the Calabi--Yau condition. The induced metric on $D_i$ inherited from the ambient space is itself K\"ahler \cite{Kobayashi, Joyce} but generically not Calabi--Yau. Let us note that in previous work we considered the correction written in terms of topological quantity namely the third Chern-form of the Calabi--Yau fourfold. 
One may write the K\"ahler coordinates \eqref{divisorInt} in terms of a basis of well defined $CY_4$-integrals in terms the Calabi--Yau metric and covariant quantities thereof such as the Riemann tensors  if the parameters in  \eqref{divisorInt} obey the following relations
\bea\label{coeffIntegralB3toY4}
\alpha_5  &=&  - \tfrac{1}{8}\alpha_1 +  \tfrac{1}{24}+ \tfrac{1}{4} \alpha_4\; ,\nonumber \\
 \alpha_6  &=&  \tfrac{1}{2}\alpha_2 + \tfrac{1}{2}\; ,\nonumber \\
   \alpha_7  &=& \alpha_2 + 1 \; ,\nonumber \\
   \alpha_8  &=&  \tfrac{1}{2} \alpha_1 - \tfrac{1}{3} - \alpha_4 \; , \nonumber \\
    \alpha_9  &=& - \alpha_1 + \tfrac{1}{6}\; .
\eea
Thus in other words  by imposing \eqref{coeffIntegralB3toY4} we can rewrite the K\"ahler coordinates in terms of a higher-derivative density on the Calabi--Yau fourfold, which as we argue in appendix  \ref{secDivisorintegralsDetails} may take the form
\beq
\label{tiasCY4int2}
\cT_i = \int_{CY_4} \omega_i \wedge \mathcal{X}  \;\; , \;\;\; \mathcal{X} \sim R^3
\eeq
with  the higher-derivative $(3,3)$-form $\mathcal{X}$   defined in the appendix \eqref{definitionOfT}.
One can easily verify the property
\beq \label{propXY}
\mathcal{T}_i v^i =  \cZ   \;\; .
\eeq
To compute the  K\"ahler metric we need to  take derivatives of the K\"ahler potential w.r.t.\ to the K\"ahler coordinates as
\beq\label{KpotfromAnsatz}
K_{ij} = \frac{\partial^2 K }{\partial \Re T_i \partial \Re T_j} =  \frac{\partial^2 v^k}{\partial \Re T_i \partial \Re T_j}  \frac{\partial K}{\partial v^k} +  \frac{\partial v^k}{\partial \Re T_i } \frac{\partial v^l}{\partial \Re T_j}  \frac{\partial^2 K}{\partial v^k\partial v^l}  \;\; ,
\eeq
with 
 \beq
 \frac{\partial v^i}{\partial \Re T_j}  = \Big(  \frac{\partial \Re T_j }{\partial v^i}   \Big)^{-1} = \cK^{ij}  - \al^2 \kappa_5\cK^{ik} \Big( \frac{\partial}{\partial v^k} \cT_l  + \dots  \Big)\cK^{lj} \;\; ,
 \eeq
where $\cK^{ij}$ is the inverse intersection number of the Calabi--Yau fourfold defined in the appendix \eqref{intersectionids}.
The variation of $\cT_i$ w.r.t.\,to the K\"ahler moduli fields of the  Calabi--Yau fourfold constitutes the crucial new ingredient to generate and  match the higher-derivative structures in the reduction result \eqref{actionvs10}-\eqref{new-reductioResult} of the  K\"ahler metric. Let us next discuss it in more detail.

\paragraph{ Variational derivative  of  K\"ahler coordinates.}
The aim of this section is to argue that the Ansatz for the K\"ahler potential \eqref{Kahlerpot} and K\"ahler coordinates \eqref{Kahlercord} may reproduce  the  K\"ahler metric  in the Legendre dual variables which are in agreement with the reduction results. In other words we are able to encounter all relevant  higher-derivative structures  found in the reduction result \eqref{actionvs10}- \eqref{new-reductioResult}. However, let us stress that to  precisely match the factors in the reduction result is beyond the aim of this work. It is expected that additional  non-trivial identities relating the higher-derivative building blocks \eqref{variationTi} and \eqref{actionvs10} -\eqref{new-reductioResult}, and \eqref{higher-derYij} are required to perform this task.

Let us proceed with the main argument. It is straight forward to compute  derivatives of the previously encountered topological objects \cite{Grimm:2013bha} w.r.t.\,to the K\"ahler moduli fields as
\beq\label{variarionZs}
\frac{\partial}{\partial v^i} \cZ = \cZ_i \;\;\;, \;\;\;  \frac{\partial}{\partial v^j} \cZ_j = 0\;\; .
\eeq
Let us note that due to \eqref{variarionZs} no terms proportional to  the logarithm of the volume - $\bf \log{\cV}$ - appear in the  K\"ahler metric nor in the Legendre dual variables and thus \eqref{Kahlercord} and \eqref{Kahlerpot} are in  agreement with the reduction result in this regard.

Let us next compute the  variation of $\cT_i$ in \eqref{tiasCY4int2} w.r.t. to the K\"ahler moduli fields which gives

 \beq \label{variationTi}
 \frac{\partial}{\partial v^j} \cT_i =- \frac{3}{\cV}   \cK_{j} \cT_{i} + \frac{5}{\cV}   \cK_{i} \cT_{j} + 3\, \cT_{ij}+  \frac{3}{\, \cV}\cZ_i \cK_j  +  4 i  \int_{CY_4} Z_{m \bar n}  \omega_{i}{}^{\bar n s} \omega_{j s}{}^m  \ast 1  \;\; ,
 \eeq
 where
 \beq\label{defTsimple}
 \cT_{ij} =   \int_{CY_4} \ast_8 \big( \o_i \wedge \o_j \wedge J \big) \wedge \mathcal{X}  \;\; .
 \eeq
To compute \eqref{variationTi} we make extensive use of the compute Algebra package xTensor \cite{Nutma:2013zea}. We provide some more technical details in appendix \ref{variation}. There we also discuss  couplings of the K\"ahler metric proportional to the second Chern form of the Calabi--Yau fourfold. By using the relation 
 \ba\label{higher-derYij}
\mathcal{Y}_{ij}= - \fr16 \int_{Y_4} (  i Z_{m \bar n} \o_{i}{}^{\bar r m} \o_{j}{}^{\bar n}{}_{\bar r } + 2  Z_{m \bar n r \bar s} \o_i {}^{\bar n m} \o_j{}^{\bar s r} ) *1 \, , 
\ea
one infers that \eqref{variationTi} can be put in relation 
 to  $Z_{m \bar n r \bar s} \omega_i^{ \bar n m}\omega_i^{ \bar r s}$ and $\mathcal{Y}_{ij}$.
 Let us emphasize that establishing the relation of topological K\"ahler coordinates and the building blocks of the K\"ahler metric obtained  by dimensional reduction $ \sim Z_{m \bar n r \bar s} \omega_i^{ \bar n m}\omega_i^{ \bar r s}$ as well as $\sim Z_{m \bar n}  \omega_{i}{}^{\bar n s} \omega_{j s}{}^m$ has been a long standing problem posed in our previous work \cite{Grimm:2014efa,Grimm:2015mua}.

 Let us close this section by providing further arguments for the completeness of  higher-derivative building blocks in \eqref{variationTi}. By evaluating \eqref{KpotfromAnsatz} one obtains that  the  K\"ahler metric $K_{ij}$ contains $ \cV \cK_{kl}\cK_{ijk}\cT_j$ and  $\cK_{(i} \cT_{j)}$.\footnote{The precise form of the K\"ahler metric results from \eqref{KpotfromAnsatz} by inserting our  Ansatz \eqref{Kahlercord},\eqref{Kahlerpot} and by using the properties of the intersection numbers listed in equation \eqref{intersectionids}. Furthermore, one may use the relations on the higher-derivative building blocks \eqref{propXYApp}  and \eqref{propXYApp2}} Those structures arise  naturally from the variation of the K\"ahler coordinates \eqref{variationTi}, in particular $\cT_{ij} \sim \cK^{kl}\cK_{ijk} \cT_{i} + \dots$.  It has been argued for analogous  relations in \cite{ediss18713,Strominger:1985ks}.
Concludingly, the  divisor integral Ansatz  \eqref{Kahlercord} manages to reproduce all relevant higher-derivative building blocks which appear in the reduction result \eqref{actionvs10}-\eqref{new-reductioResult}. However, we also find  that we have one abundant object namely $\mathcal{Y}_{ij}$ which does not appear in the reduction result but would be generated by our Ansatz.  In \cite{Grimm:2015mua} we had argued for a relation in between the $\mathscr{F}$ and higher-derivative objects which in the light of this work most certainly needs a revision.
Let us close this section with remarks on the warp-factor in the K\"ahler potential  and coordinates and its potential  connection to the higher-derivative structures. In appendix \ref{variation} we review  the integration of the warp-factor into a K\"ahler potential in particular in \eqref{K-AnsatzApp} - \eqref{deriv-results} . From the definition  \eqref{def-Yij} one immediately infers that  
$\mathcal{Y}_{ij}  v^j = \mathcal{Y}_{ji} v^j =0$ and thus it takes special simplified role in the process of matching the reduction result.
 We thus suggest that a relation $\mathcal{Y}_{ij}  \sim \mathscr{F}_{ij}  $ might be established to proof the conjectured integration into a K\"ahler potential which revises the claims of \cite{Grimm:2015mua}.

\subsection{Topological divisor integrals  as K\"ahler coordinates}
\label{secTopDivInt}

In this section we argue that the  Ansatz for the K\"ahler coordinates \eqref{divisorInt} may  be rewritten in terms of "topological quantities" by fixing the coefficients in the Ansatz. The quotation marks refer to an abuse of the word as the integrands can be reduced to topological integrands by factorizing out K\"ahler moduli deformations, e.g.\,\,the intersection number of the Calabi--Yau fourfold $\cK_{ijkl}$ is a topological quantity, in contrast to the volume of a complex curve $\cK_{ijk}$  which is not as it depends on the position in moduli space. However one may write it in terms of the topological intersection numbers by factorizing out the K\"ahler moduli fields as $ \cK_{ijk} =\cK_{ijkl} v^l$.

To set the stage note that any closed form such as $\tilde c_1$ may be written in terms of its harmonic part plus a double exact contribution
\beq
\tilde c_1 = H\tilde c_1 + \partial\bar{\partial} \lambda \;\; ,
\eeq
where $\lambda$ is a function on the divisor. From the closure of $\tilde c_1$ and by using inferred relation thereof  in appendix \ref{appendixsecTopDivInt}   one may show that the Ansatz for the K\"ahler  coordinates  \eqref{divisorInt} can be rewritten as
\ba\label{newTi1}
\cT_i =  &\;\;\;  \alpha_1  \int_{D_i} \tilde c_1 \wedge  \tilde c_1 \wedge  \tilde c_1 + \alpha_2    \int_{D_i}  \tilde c_1 \wedge  \tilde c_2 \ + \alpha_3   \int_{D_i} \tilde c_3
 +  \frac{\alpha_4}{\cK_i}    \int_{D_i} \tilde c_1 \wedge \tilde J^2   \int_{D_i}  \; \tilde c_1 \wedge  \tilde c_1 \wedge  \tilde J   \nonumber  \\[0.2 cm]
 &+  \frac{\alpha_5}{\cK_i^2}   \int_{D_i}  \; \tilde c_1 \wedge   \tilde J ^2  \int_{D_i}  \; \tilde c_1 \wedge   \tilde J ^2 
 \int_{D_i}  \; \tilde c_1 \wedge   \tilde J ^2   + \frac{\alpha_6}{\cK_i}    \int_{D_i}  \; \tilde c_1 \wedge   \tilde J ^2    \int_{D_i}  \; \tilde c_2 \wedge   \tilde J  \nonumber  \\[0.2 cm] 
& + 2\alpha_6  \int_{D_i}  \; \tilde   \ast_6 ( H\tilde c_1  \wedge   \tilde J ) \wedge \tilde c_2
   -  \big( 2\alpha_4 + 8 \alpha_5 \big)  \int_{D_i}  \; \tilde \ast_6 (H\tilde c_1  \wedge \tilde J) \wedge  \tilde c_1  \wedge   \tilde c_1 \   \nonumber \\[0.2 cm]
   &-\frac{4\alpha_5}{\cK_i} \int_{D_i} \tilde c_1 \wedge \tilde J^2 \int_{D_i}  \tilde c_1 \wedge \ast_6 H \tilde c_1   \;\; ,
\ea
where $\cK_i$ denotes the volume of the divisor $D_i$.
Note that in order to obtain  \eqref{newTi1} one fixes the coefficients such that
\ba
\alpha_7  = 2 \alpha_6  \;\; , \;\;\; 
\alpha_8 =  2 \alpha_4 - 8  \alpha_5  \;\; , \;\;\;
\alpha_9 = - 4 \alpha_5 \;\;.
\ea
Additionally requiring that we can write $\cT_i$ as integrals on the Calabi--Yau fourfold one is led to additional constraints which in  combination with \eqref{coeffIntegralB3toY4} then impose 
\ba \vspace{0,3 cm}\label{finalconstraints}
\quad & \alpha_1 = \tfrac{1}{6} \;\; , \;\;\; \;\;\quad\quad
\quad  \alpha_3 = 1 \;\; , \;\;\; \;\;\quad\ \;\;\;\;\;
  \alpha_4 = -\tfrac{1}{12}  \;\; ,\;\;\;\; \;\;\;
  \alpha_5 = 0   \;\; , \;\;\;\; & \quad\nonumber \\[0.2cm]
 \quad &  \alpha_6 = \tfrac{1}{2} +\tfrac{1}{2} \alpha_2 \;\;\;,\;\;\;\;
     \alpha_7 = 1  +  \alpha_2  \;\; , \;\;\;\; \;\;\;\;
     \alpha_8 = -\tfrac{1}{6}  \;\;\; ,\;\;\;\;  \;\;\;
      \alpha_9 = 0 \;\;. & \;\; \vspace{0,3 cm}
\ea
One thus infers from \eqref{finalconstraints} the final form of the higher-derivative  K\"ahler coordinate divisor integral to be
\ba\label{TitopologicalFinal}
\cT_i &=     \tfrac{1}{6}  \int_{D_i} \tilde c_1 \wedge  \tilde c_1 \wedge  \tilde c_1 + \alpha_2    \int_{D_i}  \tilde c_1 \wedge  \tilde c_2 \ +   \int_{D_i} \tilde c_3
 -  \frac{1}{ 12\cK_i}    \int_{D_i} \tilde c_1 \wedge \tilde J^2   \int_{D_i}  \; \tilde c_1 \wedge  \tilde c_1 \wedge  \tilde J   \nonumber  \\[0.3 cm]
 &+\tfrac{1 }{2} \big(1+\alpha_2 \big) \frac{1}{\cK_i}    \int_{D_i}  \; \tilde c_1 \wedge   \tilde J ^2    \int_{D_i}  \; \tilde c_2 \wedge   \tilde J  
 + \big( 1 + \alpha_2\big) \int_{D_i}  \; \tilde   \ast_6 ( H\tilde c_1  \wedge   \tilde J ) \wedge \tilde c_2 \nonumber  \\[0.3 cm]
 &- \tfrac{1}{6} \int_{D_i}  \; \tilde \ast_6 (H\tilde c_1  \wedge \tilde J) \wedge  \tilde c_1  \wedge   \tilde c_1 \;\; .
\ea
Let us note that \eqref{TitopologicalFinal} is in indeed a sum of "topological integrals". 
In this sense after factorizing out K\"ahler moduli deformations one may vary the integrands of \eqref{TitopologicalFinal}  w.r.t.\,the  induced metric on the divisors  $D_i$ and find that the resulting variation constitutes a total derivative. This follows straightforwardly from the properties of $\tilde c_1,\tilde c_2,\tilde c_3 $ and $\tilde J$. The integrands involving the hodge star $\tilde \ast_6$ crucially have it act on only the harmonic part of the first Chern-form $H \tilde c_1$.

\subsection{One-modulus  compatibility }
\label{sec-onemodcompatibiliyt}
The one-modulus case can be integrated exactly into a K\"ahler potential as discussed in section \ref{sec-reviewOneMod}. Thus in this section we examine the limit $h^{1,1} \to 1$ of the generic moduli case  \ref{reductionCY4generic} to impose constraints on the $\kappa$-parameters in the Ansatz. We focus on the higher-derivative components and do not discuss the warp-factor contributions $\mathscr{W}$ and $\mathscr{F}$ here. Recall that
\beq
\cZ_i =  (2 \pi)^3 \int_{CY_4} c_3 \wedge \omega_i \;\; , \;\;\;\;\;\;\; \;\;  \cZ = \cZ_i v^i \;\; ,
\eeq
We made the Ansatz for the K\"ahler potential
\beq \label{Anstaz1}
K = -3  \log{ \Big( \cV + \al^2 \big( \kappa_1 \cZ_i v^i +  \kappa_2 \cT_i v^i  \big) \Big)} \;\; ,
\eeq
and for the K\"ahler coordinates
\beq
\label{Ansatz2}
\Re T_i = \cK_i + \al^2\Big(  \mathscr{F}_i + 3 \mathscr{W}_i +\kappa_3  \frac{\cK_i }{\cV} \cZ_j v^j + \kappa_4 \cZ_i \log{\cV} + \kappa_5 \frac{\cK_i }{\cV}  \cT_j v^j + \kappa_6 \cT_i  \Big) \;\; .
\eeq
Let us next analyse these expressions \eqref{Anstaz1} and \eqref{Ansatz2}
 in the case $h^{1,1} =1$. One finds that 
 \beq\label{toOneMod1}
\cK_i \to  4 \cV^{\tfrac{3}{4}}  \;\;\ ,\;\;\;\
\cK_{ij} \to  12 \cV^{\tfrac{1}{2}}  \;\;\ ,\;\;\;\
\cK_{ijk} \to   24 \cV^{\tfrac{1}{4}} \;\;\ ,\;\;\;\
\cK_{ijkl} \to  1   \;\;\ ,\;\;\;\; 
\cK^{ij} \to  \tfrac{1}{12} \cV^{-\tfrac{1}{2}}  
\eeq
and from the expression \eqref{tiasCY4int2} and \eqref{propXY} that
 in the one-modulus case
\beq\label{oneModT}
\cT_i  \to \tilde  \cZ  \;\;\; \;\;\text{with} \;\;\; \tilde \cZ =  (2\pi)^3 \int_{CY_4} c_3 \wedge \omega^0 \;\; . 
\eeq
The relation \eqref{oneModT} follows from \eqref{divisorInt} and \eqref{newTi1} due to the Calabi--Yau condition which leads to a vanishing of terms proportional to $ \tilde c_1$. One furthermore notes that  $J = \omega_0 \cV^{\tfrac{1}{4}}  $ and thus
\beq\label{toOneMod2}
 \cZ = \cV^{\tfrac{1}{4}} \tilde \cZ =   (2 \pi)^3 \int_{CY_4} c_3 \wedge J \;\; .
\eeq
Ones concludes that \eqref{Anstaz1} and \eqref{Ansatz2} for $h^{1,1} \to 1$ by using relations \eqref{toOneMod1}-\eqref{toOneMod2} become
\beq
K  \to  -3  \log{ \Big( \cV +\al^2 \,(\kappa_1 + \kappa_2)  \cV ^{\tfrac{1}{4}} \tilde \cZ  \Big)} \;\; .
\eeq
and
\beq
\Re T_i \to  4 \cV^{\tfrac{3}{4}}+ \al^2\Big(  (4\kappa_3 +4 \kappa_5 + \kappa_6 )  \tilde \cZ+  \kappa_4 \tilde \cZ \log{\cV}  \Big) \;\; ,
\eeq
Thus one infers by comparison to the one-modulus case  \eqref{Kpot1} \eqref{ReTcorrected} that
\beq\label{oneModConstraints}
\kappa_1 + \kappa_2 = 512\,  \;\;\; ,\;\;\;\;4\kappa_3 +4 \kappa_5 + \kappa_6  = 0 \;\;\;, \;\;\;   \kappa_4 =  -1152 \;\;\; .
\eeq
Additionally one aims to match  the Legendre dual coordinates to the one modulus case. To proceed one needs to specify the precise form of the K\"ahler coordinates in terms of Calabi--Yau fourfold integrals.  In section \ref{secCYDivisorINt} we emphasized that the match with the divisor integral form remains ambiguous. Let us proceed with a simple version given in \eqref{defTsimple} for the remainder of this section. One can then use
\beq
L^i  = -\frac{\pd K}{\pd T_i} = - \frac{\pd K}{\pd v^ji}\frac{\pd v^j}{\pd T_i} \;\; ,
\eeq
to find
\beq\label{Lcorrected}
L^i = \frac{v^i}{\cV}  +\frac{ \al^2 }{\cV}\Big( \kappa_6 \cK^{ij} \cT_ j + (3 \kappa_1 - 4 \kappa_3- 4 \kappa_5 - \kappa_6 )\cK^{ij}\cZ_j -  \frac{v^i}{3\cV}( 3 \kappa_1 -\kappa_3 - \kappa_5 + \kappa_4)  \cZ \Big) \;\;.
\eeq
To compute \eqref{Lcorrected} we only used the fact that $\big(\frac{\pd}{\pd v^j} \cT_ i \big) v^i = - \cT_ j  +\cZ_j$ which follows from $\cT_ i  v^i= \cZ$.  Lastly by imposing  \eqref{oneModConstraints}  one infers a match of \eqref{resultL}  with comparison of the one-modulus limit of \eqref{Lcorrected}, i.e.\,the  order $\al$-contributions vanishes in the limit. One can furthermore compute the scalar potential by evaluating \eqref{KpotfromAnsatz} which can be performed by  using \eqref{propXYApp2} and \eqref{intersectionids}  contracted with \eqref{Lcorrected}. One finds for a non vanishing flux-superpotential $W_0$ that
\beq\label{onemodScalarPot}
V_F =  \frac{|W_0|^2}{\cV^4} \frac{4 \,  \kappa_4}{3} \, \cZ \;\; ,
\eeq
which by imposing \eqref{oneModConstraints} matches the one modulus case given in \eqref{FScPot}. 
Moreover, note that from  \eqref{onemodScalarPot} one infers that for the Ansatz \eqref{Kahlerpot} and \eqref{Kahlercord} the no-scale structure is broken due to the imposed compatibility with the one-modulus case. 
\vspace{0.3cm}

 Let us close this section with a critical remark. In section \ref{secCYDivisorINt}  and \ref{secTopDivInt} we pointed out that the lift of the divisor integral expressions to integrals on the Calabi--Yau fourfold leaves certain parameters unfixed. In order to compute other quantities such as \eqref{onemodScalarPot} in full generality we suggest that a better understanding of the $\cT_i$ contribution is to be developed.

\section{F-theory uplift to  4d$, \, \bf \cN = 1$}
\label{sec_FtheoryMain}
In this section we utilize the duality between M-theory and F- theory to lift the $l_{\rm M}$-corrections in the three-dimensional theory obtained in the previous section  to $\al '$-corrections to the four-dimensional  effective theory arising from F-theory compactified on $CY_ 4$. 
This requires the Calabi--Yau manifold to be elliptically fibered over a three-dimensional K\"ahler base $B_3$.

In the following we consider the classical result of the F-theory uplift \cite{Grimm:2010ks}. One may parametrize the shrinking of the torus fiber by the parameter $\epsilon \to 0$. One then infers the scaling of the fields $v^0 \sim \epsilon$ and $v^\alpha \sim\epsilon^{-1/2}$. This leads to an identification of the $3d, \cN=2$ multiplet field $L^0 = \tfrac{v^0}{\cV} = \frac{1}{r^2}$ with $r$ the radius of the  $4d/3d$ circular reduction. To keep the base volume finite in the limit one finds
\beq
2 \pi v_b^\alpha = \sqrt{v^0}v^\alpha \;\; .
\eeq 
For simplicity, let us restrict to a smooth Weierstrass model, i.e.\;\;a geometry without non-Abelian singularities, that can be embedded in an ambient fibration with typical fibers being the weighted projective space $W \mathbb{P}_{231}$.  This implies having just two types of divisors $D_i, \, i = 1, . . . , h^{1,1}(CY_4)$. There is the horizontal divisor corresponding to the zero-section $D_0$, and the vertical divisors $D_\alpha, \, \alpha = 1, . . . , h^{1,1}(B_3)$, corresponding to elliptic fibrations over base divisors $D^b_\alpha $. Denoting the Poincare-dual two-forms to the divisors by $ \omega_i = (\omega_0 , \omega_\alpha )$, one expands  the K\"ahler form as
\beq \label{KaehlerFormCY4}
J=v^0 \omega_0 +v^\alpha \omega_\alpha \;\;  ,  
\eeq 
where $v^0$ is the volume of the elliptic fiber, and we choose the harmonic representatives of the class. 
We are now in a position to discuss the F-theory uplift of the individual terms in
\ba \label{correctionTbeforeUplift}
\cT_i &=     \tfrac{1}{6}  \int_{D_i} \tilde c_1 \wedge  \tilde c_1 \wedge  \tilde c_1 + \alpha_2    \int_{D_i}  \tilde c_1 \wedge  \tilde c_2 \ +   \int_{D_i} \tilde c_3
 -  \frac{1}{ 12\cK_i}    \int_{D_i} \tilde c_1 \wedge \tilde J^2   \int_{D_i}  \; \tilde c_1 \wedge  \tilde c_1 \wedge  \tilde J   \nonumber  \\[0.3 cm]
 &+\tfrac{1 }{2} \big(1+\alpha_2 \big) \frac{1}{\cK_i}    \int_{D_i}  \; \tilde c_1 \wedge   \tilde J ^2    \int_{D_i}  \; \tilde c_2 \wedge   \tilde J  
 + \big( 1 + \alpha_2\big) \int_{D_i}  \; \tilde   \ast_6 ( H\tilde c_1  \wedge   \tilde J ) \wedge \tilde c_2 \nonumber  \\[0.3 cm]
 &- \tfrac{1}{6} \int_{D_i}  \; \tilde \ast_6 (H\tilde c_1  \wedge \tilde J) \wedge  \tilde c_1  \wedge   \tilde c_1 \;\; .
\ea
and the correction
\beq \label{correctionZbeforeUplift}
\cZ_ \alpha =(2 \pi)^3 \int_{CY_4} c_3 \wedge \o_\al \;\ \; ,
\eeq
where $\cK_i$ is the volume of the divisor $D_i$.
Latter was discussed already   in \cite{Grimm:2013gma,Grimm:2013bha} however, we review these results in section \ref{topDivUplift}.
Note that the relation between the eleven-dimensional  Planck length $l_{\rm M}$ and the string length $l_s$ by the M/F-theory duality is obtained as
\beq
2 \pi l_s = \cV^{\tfrac{1}{2}}l_{\rm M} \;\; .
\eeq
As in the F-theory limit one sends $v^0 \to 0$  decompactifying the fourth dimension by sending to infinity the radius of the $4d/3d$ circle $r \sim \cV^{3/2} \to \infty$. Thus after the limit  all volumes of the base $B_3$ are expressed in terms of the string units $l_s$.
In the following  we omit the warp-factor $\mathscr{W}$ and thus $\mathscr{F}$ from the discussion. 

In section \ref{F-theoryOneLoop} we shortly comment on the  uplift of F-theory involving one-loop corrections resulting from integrating out massive KK-modes at one-loop in the circular reduction from four to three dimensions. As those results are not well studied in the literature we present an superficial discussion. 
Let us stress however, that as we are not able to fix all parameters  in the $3d,\, \cN=2$ coordinates the ambiguity of the "one-loop"  up-lift can be hidden in the following section in the uncertainty of the parameters and the generic conclusions of this work are expected to be unchanged.
In section \ref{topDivUplift} we then analyse the terms in the K\"ahler potential \eqref{Kahlerpot} and  K\"ahler metric \eqref{Kahlercord} surviving the  F-theory uplift. 
Finally, in section \eqref{4dSuggestionforKahlermetric} we then combine the conclusions of sections \ref{F-theoryOneLoop}  and \ref{topDivUplift} to discuss the $4d, \cN =1$   K\"ahler potential and K\"ahler metric. In particular we give a string theory  interpretation of the novel corrections and discuss the breaking of the no-scale structure and the $\alpha'^2$-modified scalar potential.

\subsection{The F-theory uplift  }
\label{F-theoryOneLoop}

In this section we review the supergravity perspective of the F-theory lift identifying the connection in-between the four and three-dimensional fields and their kinematic couplings \cite{Grimm:2010ks}.
Note that by compactifying a general four-dimensional, $\cN=1$ supergravity theory on a circle one matches the original four-dimensional K\"ahler potential  with the  three-dimensional K\"ahler potential $K$ or kinetic 
potential $\tilde K$. 
The resulting kinetic potential  arising in the $4d/3d$ circular  dimensional reduction takes the form
\beq \label{4d3dK}
   \tilde K (r , T_\alpha) = -\log(r^2) + K^{F}(T_\alpha)  \,.
\eeq
To match \eqref{4d3dK} with the natural three-dimensional multiplets
one may split  $L^i$ and $T_i$  such that 
\beq\label{def-R}
    L ^i = \big( \, L^0 \equiv R  \, , \,\ L^\alpha \, \big) \ , \qquad  T_i = (\, T_0 \, , \,T_\alpha\, )\ .
\eeq
One is then led to identify  that $R$ is given by $R = r^{-2}$, where $r$ is the radius of the $4d/3d$ circle  \cite{Grimm:2010ks}.
Furthermore, the fields $T_\alpha$ remain complex scalars in four dimensions  whilst  $T_0$ should 
be dualized already in three dimensions into vector multiplets with $(R,A^0)$ and then uplifted to four dimensions as  it arises from the four-dimensional metric. Note that one  computes  the dualized kinetic potential $\tilde K(R , \Re \, T_\alpha)$  by Legendre dualization as discussed in \ref{3dN2sugra}.
In  the F-theory limit one then identifies
\beq \label{LT_limit}
L^\alpha_b =  L^\alpha|_{\epsilon = 0} \ , \qquad T_\alpha^b = T_\alpha|_{\epsilon = 0}\,,
\eeq
where we denote the four-dimensional fields $L^\alpha_b $ and $ T_\alpha^b $ due to the fact that they correspond to  fields with couplings  related to the base $B_ 3$  representing the Calabi--Yau orientifold in the IIB picture, i.e.\;\;in the  F-theory limit.
Let us next review the classical analysis to determine $K^F(T_\alpha^b)$.  Evaluating the intersection numbers $\cK_{ijkl}$ for an elliptic 
fibration the  non-vanishing coupling is 
\begin{equation} \label{triplebase}
\cK_{0 \alpha \beta \gamma} = \cK^b_{\alpha \beta \gamma} \;\; ,\;\;\;
    \cK^b_{\alpha \beta \gamma} = \int_{B_3} \omega_\alpha \wedge \omega_\beta \wedge \omega_\gamma\ .
\end{equation} 
The kinetic potential  and coordinates take the following form for an elliptic fibration
\bea
   \tilde K (L^i) &=& \log (R ) -2 \log \Big(\cV_b+ \mathcal{O}(R) \Big)+4\ , \\
   \text{Re}\,T_\alpha &=&  \cK^b_\alpha
  + \mathcal{O}(R)\ , \qquad   \cV_b =  \frac{1}{3!} \cK_{\alpha \beta \gamma } v_b^\alpha v_b^\beta v_b^\gamma \ ,
\eea
or equivalently
\bea
   \tilde K (L^i) &=& \log (R ) + \log \Big(\frac{1}{3!} \cK^b_{\alpha \beta \gamma} L_b^\alpha L_b^\beta L_b^\gamma + \mathcal{O}(R) \Big)+4\ , \\
   \text{Re}\,T_\alpha &=&  \frac{1}{2!}\frac{ \cK^b_{\alpha \beta \gamma } L_b^\beta L_b^\gamma}{\hat \cV_b(L_b)} 
  + \mathcal{O}(R)\ , \qquad   \hat \cV_b(L^b) =  \frac{1}{3!} \cK_{\alpha \beta \gamma } L_b^\alpha L_b^\beta L_b^\gamma \ ,
\eea
where we have replaced the $L^\alpha$ with $L^\alpha_b$ by means of \eqref{LT_limit} and made use of the relation $\hat \cV^b(L_b) = (\cV^b)^{-2}$ .
Performing the Legendre transform in order to express everything in terms of $T_\alpha^{b}$
and comparing the result with \eqref{4d3dK}  by setting $R = r^{-2}$
in the limit $r \to \infty$ one encounters
\begin{equation} \label{KFresult}
    K^F (T_\alpha^b) = - 2  \log \big( \cV_b\big) = \log \big(\hat \cV_b(L_b)  \big)\ ,  \qquad  \text{Re}\,
    T^b_\alpha  = \cK^b_\alpha =  \frac{1}{2!}\frac{ \cK^b_{\alpha \beta \gamma } L_b^\beta L_b^\gamma}{\hat \cV_b(L_b)}\ ,
\end{equation}
where one has to solve $T_\alpha^b$ for $L^\alpha_b(T_\alpha^b)$ and insert the result into $K^F$.

Let us next comment on the case present in this work namely  where one encounters higher-order  $l_{\rm M}$-corrections to the three-dimensional fields. As suggested by the generic $4d/3d$ circular  reduction result and   one infers  for  the corrected the K\"ahler coordinates that
\beq \label{4dcorrectedKcoord}
  \Re   T_\alpha  \;\;\; \rightarrow  \;\;\;\;  \Re   T^b_\alpha  \;\; ,
  \eeq
where we  analyse $ \Re   T^b_\alpha  $ in the next section \ref{topDivUplift}. The  corrected K\"ahler potential \eqref{Kahlerpot}  can be re-written as
\beq
K =  - \log\big( R \, \big)   -2 \log \left( \cV_b \Big(1 +  \al^2   \tfrac{3}{2 \cV_b} \big((512 - \kappa_2) \,  \cZ^b_\al v_b^\al + \kappa_2  \cT^b_\al v_b^\al  \big) \Big)    + \mathcal{O}(R)\right) \;\;  ,
\eeq 
by making use of the sub-leading order of $\al^2$. Thus in the limit $r \to \infty$ one encounters
\beq
  K^F (T_\alpha^b) = - 2  \log \Big( \cV_b  + \al^2 \big( ( 768  - \tilde\kappa_2 ) \cZ^b_\al v_b^\al + \tilde\kappa_2  \cT^b_\al v_b^\al  \Big)  \;\; ,
\eeq
where is  $\cZ^b_\al$  the F-theory limit of $\cZ_\al$  derived in  in the following  section \ref{topDivUplift} and $\tilde \kappa_2 = \tfrac{3}{2}\kappa_2$. The identification of the dependence  $T_\alpha^b$ is implicit.

Let us close this section with remarks on  one-loop corrections to the F-theory limit resulting from integrating out massive KK-modes which is expected to modify the relation \eqref{4dcorrectedKcoord}. The $\log \cV$-correction to the K\"ahler coordinates \eqref{Kahlercord} is reminiscent of such a one loop correction.
To see this one is to preform a dimensional reduction of  a general $4d, \, \cN=1$ supergravity theory on the circle to three dimensions where massive KK-modes are integrated out at one-loop.  The case for pure supergravity is discussed in \cite{Tong:2014era} which yields the three-dimensional K\"ahler coordinates
\beq\label{gravoneLoop}
\Re T^{1-loop}_0 = \frac{2 \pi^2}{R} - \frac{7}{48} \log( R) \;\;.
\eeq
However, we are interested in a theory with additional chiral multiplets and vector multiplets  which will lead to a modification of the purely gravitational result \eqref{gravoneLoop}. We are not aware of  such a discussion in the literature and thus have no rigorous tool to argue for the up-lift of the $ \cZ_\alpha \log \cV$  correction in F-theory except the comments made in \cite{Grimm:2017pid}. Note that the main result of this work is obtained from the novel divisor integral modification of the K\"ahler potential and coordinates in  \eqref{Kahlerpot}  and  \eqref{Kahlercord}  thus the one-loop discussion is not expected to change these conclusions. 
Let us assume in the following that the  $\log \cV$-correction in the K\"ahler coordinates \eqref{Kahlercord}   is absorbed entirely by the F-theory uplift. This leads us to write 
\beq\label{correctKcoordf}
\Re T_i = \cK_ i + \al^2 \kappa_4 \cZ_i \log \cV \;\; ,
\eeq
where  for simplicity we only write the  logarithmic correction to the K\"ahler coordinates. Considering \eqref{correctKcoordf}  on the elliptically fibered Calabi--Yau fourfold  one finds
\ba\label{correctedKaehlcoordiantes}
\Re   T_0 &= \frac{ 1}{R} -\tfrac{  \al^2  \kappa_4}{3} \cZ^b_0 \log{R} -\tfrac{  \al^2 \kappa_4}{3} \cZ^b_0  \log\big((\cV^b{)}^{-3} + \mathcal{O}(R)\big) \; , \nn \\[0.2cm]
  \Re   T_\alpha &= \cK^b_{\alpha}  - \tfrac{  \al^2 \kappa_4}{3} \cZ^b_\al \log (R) -\tfrac{  \al^2 \kappa_4}{3} \cZ^b_\al \log \big((\cV^b){}^{-3}  + \mathcal{O}(R)\big) \;\; .
\ea
 The assumption that it is absorbed in the uplift immediately  leads us to a revision of \eqref{4dcorrectedKcoord} to
\ba \label{4dcorrectedKcoord2}
 \Re   T_0^{1-loop}     & \rightarrow   \frac{ 1}{R} - q_0 \log{R} +  \frac{1}{2}q_0 K^{4d} \; ,\nn \\[0.2cm]
  \Re   T_\alpha^{1-loop}   &  \rightarrow  \Re   T^{b \, tree}_\alpha     - q_\alpha \log{R} +  \frac{1}{2} q_\alpha K^{4d}  \;\; .
  \ea
  where  $ K^{4d}$ is the four-dimensional classical K\"ahler potential.
By  matching  \eqref{correctedKaehlcoordiantes} and \eqref{4dcorrectedKcoord2} one fixes the charges to $q_i = - \tfrac{ \al^2 \kappa_4}{3} \cZ^b_i + \mathcal{O}(R)$.  Note that  \eqref{4dcorrectedKcoord2} and \eqref{correctedKaehlcoordiantes} are very sensitive to a conspiration of factors. Thus at this stage as the one-loop F-theory uplift remains elusive  it cannot be excluded that a finite contribution in \eqref{correctedKaehlcoordiantes} may survive the F-theory uplift such as
\beq\label{onelopplogV}
 \Re   T^b_\alpha  = \cK^b_{\alpha}  +   \al^2 \kappa_4  \cZ^b_\al \log \big(\cV_b \big) \;\; .
\eeq
 Let us stress that different assumptions lead us to find  \eqref{4dcorrectedKcoord2}  and \eqref{onelopplogV} but a honest one-loop computation needs to be performed to decide their validity.  Note that  \eqref{4dcorrectedKcoord2} would imply that by integrating out massive Kaluza-Klein modes only the three-dimensional K\"ahler coordinates receive modifications whilst the K\"ahler potential remains uncorrected.

\subsection{Topological integrals on elliptic Calabi--Yau fourfolds}\label{topDivUplift}

In this section we discuss the F-theory uplift of the higher-order $l_ M$-corrections  appearing in \eqref{correctionTbeforeUplift} and \eqref{correctionZbeforeUplift} resulting in $\al'$-corrections.  For topological integrals  we can use adjunction formulae to express Chern-classes of $CY_4$  and the divisors $D_\alpha$ in terms of Chern-classes of the  base $B_3$. For details of the derivation of the  adjunction formulae see appendix \ref{adjunctoinChernSec}. One infers that 
\ba \label{finalChern}
 \tilde c_3 (D_\alpha)  = &  \;  c_3(B_3) -  c_1(B_3) \wedge c_2(B_3)  - 60 c^3_1(B_3) -60  c^{ 2}_1(B_3)  \wedge \omega_0 \nonumber -  \tilde c_2 (D_\alpha)\wedge \o_\alpha \;\; , \\[0.2cm]
 \tilde  c_2 (D_\alpha) = &  \;   c_2(B_3)  +11    c^{2}_1(B_3)  +12 c_1(B_3)  \wedge \omega_0 + \o_\alpha^2
  \nonumber \;\; , \\[0.2cm]
  \tilde  c_1 (D_\alpha) = &  - \o_\alpha \;\; ,
\ea 
where the $c_{i=1,2,3}(B_3)$ on the r.h.s. of these expressions denote the Chern classes of $B_3$ pulled-back to $CY_4$ restricted to $D_\alpha$. Note that the Poincare duals of the harmonic $(1,1)$-forms in  \eqref{finalChern} are given by $PD(\o_0) = B_3$, and $PD(\o_\alpha) = D_\alpha$.  We  choose to omit the pull-back map in expressions in this section for notational simplicity.
One furthermore finds that
\beq
 \omega_0^2= -  c_1(B_3)  \wedge \omega_0  \;\; .
\eeq
 Note that the new contribution to the K\"ahler coordinates $\cT_i$  is expressed as  integrals on the divisors $D_i$ where the K\"ahlerform is inherited from the ambient $CY_4$ and one thus may use the decomposition \eqref{KaehlerFormCY4} as well for $\tilde J$.  In the F-theory limit one finds  the scalings discussed at the beginning of this section to imply 
\beq\label{Ftheoryscaling}
v^\alpha \sim \epsilon^{-\tfrac{1}{2}}\; , \;\;\;\;\; \ v^0 \sim \epsilon \; \;\; \Rightarrow \;\;\;\;\; \cV_\alpha = \cK_\alpha \sim \epsilon^0 \;\;.
\eeq
Using \eqref{finalChern} and \eqref{Ftheoryscaling} one infers the contributions in \eqref{Kahlercord}  and \eqref{Kahlerpot} which survive the F-theory limit. 
 For the object defined as the third Chern-form of the Calabi--Yau fourfold \eqref{correctionZbeforeUplift} one finds in the limit 
\ba\label{liftedquant2}
   \cZ_\alpha  & \;\;\; \longrightarrow \;\;\ {\cZ^b_\alpha} = - 60 \,  (2 \pi)^2 \int_{D^b_\alpha}    c_1(B_3) \wedge  c_1(B_3) \;\;\; .
\ea
The leading order  contributions which are non vanishing in the limit must scale as  $\cT_\alpha \sim \mathcal{O}(\epsilon^0)$. The integrals  in \eqref{correctionZbeforeUplift} which  thus contribute are 
\ba \label{limitIntegral}
 &  \int_{D_\alpha} \tilde c_1 \wedge \tilde c_2   \;\;\; \quad  \longrightarrow \quad   \;\;\;  - 12 \int_{D^b_\alpha} c_1(B_3) \wedge \o^b_\alpha  \nn \;\;,  \\[0.2cm]
 & \;\;  \int_{D_\alpha} \tilde c_3  \quad \quad  \;\;  \quad  \longrightarrow \quad   \;\;\; -60 \int_{D^b_\alpha} c_1(B_3) \wedge c_1(B_3) - 12 \int_{D^b_\alpha} c_1(B_3) \wedge \o^b_\alpha  \nn \;\;,  \\[0.2cm]
 &  \int_{D_\alpha}  \; \tilde   \ast_6 ( H\tilde c_1  \wedge   \tilde J ) \wedge \tilde c_2\;\;  \longrightarrow   \ - \frac{ 12}{\cK_\alpha^b} \int_{D^b_\alpha}  \o^b_\alpha \wedge J^b \int_{D^b_\alpha} c_1(B_3) \wedge J^b  +  12 \int_{D^b_\alpha} c_1(B_3) \wedge \o^b_\alpha \;\; ,\nn \\[0.2cm]
& \frac{1}{\cK_\alpha}    \int_{D_\alpha}  \; \tilde c_1 \wedge   \tilde J ^2    \int_{D_\alpha}  \; \tilde c_2 \wedge   \tilde J  \quad  \;\;\; \longrightarrow  \;\;\;  \ - \frac{ 12}{\cK_\alpha^b} \int_{D^b_\alpha}  \o^b_\alpha \wedge J^b \int_{D^b_\alpha} c_1(B_3) \wedge J^b  \;\; ,
\ea
where we used  \eqref{integralsplit3} and  where $ D^b_\alpha$ are the divisors of the base such that their pre-image w.r.t.\,the projection $\pi: CY_4 \to B_3$ gives the  vertical divisors of the Calabi--Yau fourfold as $D_\alpha = \pi^{-1}( D^b_\alpha) $.\footnote{Note that in order to rewrite the integrals we note that e.g.
\beq
\int_{B_3} c_1(B_3) \wedge \o_\al^b \wedge \o_\al^b = \int_{D^b_\alpha}  c_1(B_3) \wedge \o_\al^b \;\;,
\eeq
where we again omit the pull-back map on $c_1(B_3)$ in the r.h.s.\,of the equality.
}
One thus infers the divisor integral contribution of the K\"ahler coordinates in the limit to take  the form \vspace{0.2 cm}
\ba \label{liftedquant1}
\cT_\alpha  \;\;\; \longrightarrow  \;\;\;  \cT_\alpha^b =   \cZ^b_ \alpha 
-  18(1 + \al_2) \, (2\pi)^2 \frac{1}{\cK_\alpha^b} \int_{D^b_\alpha}  \o^b_\alpha \wedge J^b \int_{D^b_\alpha} c_1(B_3) \wedge J^b  
\ea
with $\cK_\alpha^b  = \tfrac{1}{2!} \int_{B_3} \omega^b_\alpha \wedge J^{{\sm b} 2}$ the volume of the divisor $D^b_\alpha$ and the K\"ahler form $J^b = \o_\al^b v_b^\al$. For further use let us define
 \ba\label{W-novelCorrection}
 \mathcal{W}^b_\alpha \;\; := \;\; &  \frac{ (2\pi)^2 }{\cK_\alpha^b} \int_{D^b_\alpha}  \o^b_\alpha \wedge J^b \int_{D^b_\alpha} c_1(B_3) \wedge J^b   \;\; , \\
 \mathcal{U}^b_{\alpha}\;\;\, := \;\;&  (2\pi)^2  \int_{D^b_\alpha} c_1(B_3) \wedge \o^b_\al\;\; .\label{U-novelCorrection}
 \ea
The contribution \eqref{W-novelCorrection}  takes a special role as it depends on the K\"ahler form of the divisor and thus  is non-vanishing upon taking derivatives w.r.t.\,K\"ahler moduli fields. This will be of particular interest in the following sections.
 Note that the F-theory uplift absorbs two-derivatives along the fiber thus the resulting corrections are  of order $\alpha'^2$.  It would be interesting to establish a connection to the $\alpha'^2$-corrections to the K\"ahler potential predicted in the Heterotic string \cite{Anguelova:2010ed}.
 The $ \mathcal{U}^b_{\alpha}$-correction \eqref{U-novelCorrection} vanishes from  \eqref{liftedquant1}  due to a vanishing pre-factor. As one may find that our constraints imposed are too restrictive this correction may survive if an additional parameter freedom is somehow introduced in the present discussion of divisor integrals. In the following we thus as well comment on its potential origin and interpretation.

\vspace{0.4 cm}

Let us next comment on some special cases before  providing a  Type IIB string interpretation of the $ \al'$-corrections in \eqref{liftedquant2} and \eqref{liftedquant1}.
  Firstly,
for a trivial elliptic fibration, i.e.\;\;$CY_4= CY_3\times T^2$ with $CY_3$ a Calabi--Yau threefold, one infers that
 $c_{i}(CY_4)=c_{i}(CY_3),\, i=1,2,3$, in particular $c_1(CY_3) = 0$. Furthermore, the divisors relevant in the K\"ahler coordinates \eqref{Kahlercord}  are a direct product and obey $c_1(D^b_\al \times T ^2) = c_1(D^b_\al), \,c_2(D^b_\al \times T ^2) = c_2(D^b_\al)$ and $c_3(D^b_\al \times T ^2) = 0$, see appendix \ref{adjunctoinChernSec} for details. One  infers that in this case all corrections in \eqref{limitIntegral} and \eqref{liftedquant2} go to zero due to their scaling behavior in the limit $v^0 \to 0$ and thus the $\al'$-corrections  in the resulting $4d, \,\cN=2$ theory are absent.

Secondly, one may study other   $\cN=2$ F-theory vacua by taking 
${Y}_4={\rm K3}\times {\rm K3}$, a configuration discussed in \cite{GarciaEtxebarria:2012zm} 
with a focus on  $\al'$-corrections.
In this case $c_3({Y}_4)=0$ and thus the $ \cZ^b$-correction \eqref{liftedquant2} vanishes identically. The corrections resulting from the divisors \eqref{limitIntegral} vanish due to analogous arguments as in the above case. Concludingly, the $\alpha'$-corrections discussed in this work vanish in these $\cN = 2$ set-ups.

Finally, let us  stress that there are several additional $l_{\rm M}$-corrections 
to the fourfold volume surviving the F-theory limit. Let us again go back to the example of the product geometry $Y_4=X_3 \times T^2$, without $D7$-branes.
The $\alpha'$-corrections involving the Type IIB axio-dilaton $\tau$ have been computed by integrating out the whole tower of $T^2$ Kaluza-Klein modes 
of the 11d supergravity multiplet \cite{Green:1997as}, which results in
$ v^0{}^{-\tfrac{1}{2}}  \, \chi({CY}_3) \,  E_{3/2}(\tau,\bar\tau) $ with $ E_{3/2} $ the non-holomorphic Eisenstein series. Note that it obeys the correct scaling behavior to survive the F-theory limit. One  expects that the proper treatment of the KK-modes in a generic elliptic fibration is crucial to encounter the Euler-characteristic  $\alpha'^3$-correction \cite{Becker:2002nn}  to the $4d, \cN =1$ K\"ahler potential  inside the  F-theory framework.\footnote{An alternative approach was taken in \cite{Minasian:2015bxa}.}

\subsection{4d$,\, \cN=1 $ K\"ahler potential and coordinates}
\label{4dSuggestionforKahlermetric}

 The discussion of the uplift of the $\al'$-corrections in the previous sections \ref{F-theoryOneLoop} and \ref{topDivUplift}   enables us to infer the resulting $4d, \cN=1$ K\"ahler potential and coordinates. Let us use the dimensionless coefficients from now one, where all dimensionful quantities, e.g.\,\,$\al'$-corrections are expressed in terms of the string length $l_s$, we thus write
\beq
\al^2 \to  \frac{1}{3^2 \cdot 2^{13}} \;\;.
\eeq
One infers that
\beq\label{4dKaehler}
\cK^{4d, \cN=1} = - 2 \log\Big(\cV_b + \alpha^2 \big( ( 768  - \tilde \kappa_2) \,  \cZ^b_{ \; \alpha} v_b^ \alpha   +   \tilde \kappa_2 \cT^b_{ \; \alpha} v_b^ \alpha  \big)  \Big) 
\; ,
\eeq
and
\beq \label{4dCoord}
\Re T^b_\alpha = \cK^b_\alpha +  \alpha^2 \left(  \kappa_3  \frac{\cK^b_\alpha }{\cV_b} \cZ^v_\alpha v_b^\alpha  +  \kappa_5  \frac{\cK^b_\alpha }{\cV_b} \cT^v_\alpha v_b^\alpha   - 4 ( \kappa_3 + \kappa_5) \, \cT^b_\alpha \vspace{0.3 cm} \right) \;\; .
\eeq
Note that we have argued in \ref{F-theoryOneLoop} that the logarithmic  term  may be absorbed in the F-theory uplift as a one-loop correction and is thus not present in \eqref{4dCoord}. To verify this assumption is of great interest. One may however perform the  uplift of this term  which then  leads to a modification of \eqref{4dCoord} of the form
\beq\label{logaithmicCorrection}
\al^2 \,\kappa_4 \cZ^b_\alpha \log{\cV_b} \;\;,
\eeq
as suggested in \eqref{onelopplogV}.
 Note that \eqref{4dKaehler} and \eqref{4dCoord} depend on four unfixed parameters, due to the additional freedom in \eqref{liftedquant1}. Further studies are required to proof the existence of the  $\alpha'$-corrections in \eqref{4dKaehler} and \eqref{4dCoord} and especially \eqref{logaithmicCorrection}. Nevertheless, let us proceed by giving  a string theory interpretation of the  novel corrections to the four-dimensional K\"ahler potential and coordinates.

\paragraph
{String theory interpretation and  weak string-coupling limit. }

We follow the weak string-coupling limit  by 
Sen \cite{Sen:1996vd} which is is performed in the complex structure moduli space of $CY_4$ to give a 
weakly coupled description of F-theory in terms of Type IIB string theory on a Calabi--Yau threefold  
with an $O7$-plane and $D7$-branes. Where $CY_3$ is a double cover of the 
base $B_3$ branched along the $O7$-plane. Let us stress that the class of this branching locus is the pull-back of $c_1(B_3)$ to $CY_3$. 
In section \ref{topDivUplift} we considered the  topological divisor integrals on the geometries described by the smooth Weierstrass model i.e.\;non-Abelian singularities are absent. In this case Sen's limit  contains a single recombined $D7$-brane wrapping a divisor of class $8 c_1(B_3)$. This follows from the  seven-brane 
tadpole cancellation condition. 
As was noted in \cite{Collinucci:2008pf,Braun:2008ua} this $D7$-brane is of the characteristic Whitney-umbrella shape. It would be interesting to extend the study to geometries  with non-Abelian singularities analogously to \cite{Grimm:2013bha}.

The $\cZ^b_\al$-correction     \eqref{liftedquant2} was discussed extensively in \cite{Grimm:2013gma,Grimm:2013bha} and we refer the reader to this work for details. Let us mention here however, that in more generic geometries it morally  counts the number of self-intersections of stacks of $D7$-branes and the $O7$-plane. It should arise at tree-level in string theory and is of order $\alpha'^2$. 
In the geometry studied in this work this can be checked by identifying
\beq  
   \cV_{{\rm D7} \cap {\rm O7}} = 8 \int_{CY_3} c^2_1(B_3) \wedge J_b\ ,
\eeq
where we omitted the pull-back map from $B_3$ to its double cover $CY_3$
in the integrand. 
To give the string theory interpretation  one  identifies the string amplitude 
capturing it by considering the Einstein-Hilbert term of the  four-dimensional action in the string frame\footnote{The superscript $^s$ denotes quantities computed using the string frame metric. } 
\beq\label{StringFrameIIB}
S_{(4)}\supset\frac{1}{(2\pi)^7l_s^2 g_{\sm{ IIB}}^{2}}\int \Big(\cV^s_b -\tfrac{5 \pi^2}{2}g_{\sm{ IIB}} \cV_{{\rm D7} \cap {\rm O7}}^s  \Big) R^s_{\rm sc}  \ast^s_{4} 1 \;\; .
\eeq
Let us recall the  general formula for the Euler number of Riemann surfaces, possibly non-orientable and with boundaries, is
\beq
\chi(\Sigma)=2-2g-b-c\,,
\eeq
where $g,b,c$ denote the genus, the number of boundaries, and the number of cross caps, respectively. 
One thus infers  that  the  correction in \eqref{StringFrameIIB} arises from a string 
amplitude that involves the sum over two topologies, namely the disk $g=c=0, \,b=1$ and the projective 
plane $g=b=0,\, c=1$. 
These are tree-level amplitudes of the orientable open strings and 
non-orientable closed strings which is in agreement with the property that the correction is intrinsically $\cN=1$, i.e.\;\;its presence is constrained by having  $D7$-branes intersecting with an $O7$-plane.

Let us next give a string theory interpretation of  \eqref{liftedquant1}. 
In fact, at weak string coupling one infers that
\beq \label{O7D7-split2}
 \int_{D^b_\alpha} c_1(B_3) \wedge J^b   \sim  \big(\cV_{\rm D7 \cap D^b_\alpha} + 4 \cV_{\rm O7  \cap D^b_\alpha} \big)\,,
\eeq
where $\cV_{\rm D7}$ and $\cV_{\rm O7}$ are the volumes of the $D7$-brane and the $O7$-plane in $CY_3$, respectively.
Both volumes are in the Einstein frame and in units of $l_s$.  By tadpole cancellation one infers $\cV_{\rm D7}=8 \cV_{\rm O7}$.
It follows that
\beq \label{O7D7-split3}
 \int_{D^b_\alpha} c_1(B_3) \wedge \omega^b_\alpha  \;\;  \sim   \;\; \rm D7 \cap D^b_\alpha  \cap D^b_\alpha + 4 \,  \rm O7  \cap D^b_\alpha \cap D^b_\alpha,
\eeq
are the self-intersection curves of the base Divisors  $ D^b_\alpha$  intersected with the $D7$-brane and the $O7$-plane in $CY_3$, respectively. Lastly, the $\mathcal{W}^b_\alpha$-correction in \eqref{W-novelCorrection}  which is of order  $\alpha'^2$ and depends on  the volume the self intersection curve of $D^b_ \alpha$  \vspace{0.3 cm}
\beq\label{selfINtCuve}
 \int_{D^b_\alpha}  \o_\alpha \wedge J^b \;\; \sim  \;\; \cV_{ D^b_\alpha\cap D^b_\alpha} \;\; ,
\eeq
and furthermore \vspace{0.3 cm}
\beq \label{O7D7-split4}
 \int_{D^b_\alpha} c_1(B_3) \wedge J^b   \;\;  \sim  \;\;  \cV_{ \rm D7 \cap D^b_\alpha}   +  4 \,  \cV_{ \rm O7 \cap D^b_\alpha} \;\;  ,\vspace{0.3 cm}
\eeq 
the volume to the intersection curves of  the $D7$-branes and $O7$-planes with the base divisors $D^b_\alpha$. 
One concludes that \eqref{W-novelCorrection} is a product of the curve volumes  \eqref{selfINtCuve} and \eqref{O7D7-split4} weighted over the volume of the divisor $D^b_\al$.
It is of same order  in the string coupling as \eqref{StringFrameIIB} and is thus expected to arise equivalently   from a  tree-level amplitude of the orientable open string and non-orientable closed string amplitude in an orientifold  background.

\paragraph
{$4d$ Scalar Potential and no-scale condition. }
We next comment on the scalar potential resulting from  \eqref{4dKaehler} and \eqref{4dCoord}. We  assume that the complex structure moduli have been fixed and thus the superpotential remains to be a function of the K\"ahler moduli. The F-term scalar potential of a $4d, \, \cN=1$ theory is well known and adjusted to our case results in
\beq\label{4dFTerm}
V^{4d}_F = e^{K}\Big( K_{\alpha \beta} D^\alpha W \overline{D^\beta W} - 3 \big| W \big|^2  \Big) \;\; ,
\eeq
with the superpotential $W = W(\Re T_\alpha)$ and the K\"ahler covariant derivative given by
\beq
D^\alpha W =  \frac{\pd K}{\pd \Re T_\alpha}   W + \frac{\pd W }{\pd \Re T_\alpha}   \;\; ,  \;\;\; K_{\alpha \beta} = \big( \pd_{\Re T_\alpha} \pd_{\Re T_\beta} K \big)^{-1} \;\; .
\eeq
Let us next discuss the special case in which the superpotential is given generated by fluxes in F-theory \eqref{GVWsuper} and  non-perturbative effects are absent. We denote $W_0$ as the vacuum expectation value of the superpotential resulting after stabilizing the complex structure moduli. One then infers that for  the K\"ahler potential \eqref{4dKaehler} and K\"ahler coordinates  \eqref{4dCoord} the F-term scalar potential \eqref{4dFTerm} the resulting corrections at  order  $\alpha'^2$ vanish. This cancellation was observed in \cite{Cicoli:2007xp,Berg:2007wt}.  The terms in the K\"ahler coordinates \eqref{4dCoord} admit a functional structure which remarkably  never breaks the no-scale condition. Thus the sole contribution to the scalar potential at order $\alpha'^2$ may arise from the speculative logarithm term \eqref{logaithmicCorrection}  to be
\ba\label{no-scale4d}
V_F  \;\; \;= &  \;\;  - \al^2 \frac{3 |W_0|^2}{2 \, \cV_ b^3} \;   \kappa_4 \cZ^b_\alpha \,  v_ b^\alpha \;  \;\;.
\ea
Let us emphasize that this  is due to an assumption on the F-theory uplift. As the uplift of this one-loop term remains elusive note that the pre-factor of the $ \cZ^b_\alpha $-correction in \eqref{no-scale4d} might be subject to change and may vanish.
In the  further context of this work  however it is suggested that  the $\alpha'^2$-correction breaks the no-scale structure as seen from \eqref{no-scale4d} as it is interesting to study potential phenomenological consequences.\footnote{ Not at least to increase the interest in the tedious study of $\alpha'$-corrections.} 
Let us emphasize that the corrections  $\cZ^b_\alpha$-correction \eqref{no-scale4d}  is  of order $\alpha'^2$ and   thus leading with respect to the well known Euler-characteristic correction \cite{Becker:2002nn}.

  Let us close this section with two critical remarks. Firstly, the F-theory lift is performed by shrinking the fiber i.e.\,making the geometry singular and thus other higher-order corrections may become relevant. However, let us emphasize that all the corrections discussed in this work are of topological nature and thus are expected to be protected in the F-theory limit. Secondly,  let us stress that we did not aim to prove the integration into  $3d, \cN=2$  variables of the reduction result. However, we suggest an Ansatz for the K\"ahler coordinates and K\"ahler potential which allow to obtain all the higher-derivative couplings in the K\"ahler  metric obtained by dimensional reduction from the $l_{\rm M}^6$ eight-derivative couplings  to eleven-dimensional supergravity. This is a necessary but not sufficient step, and it thus remains to ultimately decide the faith  of the $\alpha'$-corrections  $ \mathcal{W}^b_\alpha $ and $\cZ^b_\alpha$. In particular the faith of the phenomenologically interesting correction to the scalar potential \eqref{no-scale4d}   is certainly not decided at this stage.

\subsection{Moduli Stabilisation}
\label{moduli-stabilisation}

In this section we comment on the vacuum structure of the potential generated by the novel conjectured $\alpha'$-correction \eqref{no-scale4d}. Furthermore, we study the interplay with the well known Euler-characteristic $\al '^3$-correction to the K\"ahler potential
\beq\label{BBHL}
 \xi =  - g_s^{{-\tfrac{3}{2}}}  \, (2\pi)^3 \,\frac{\zeta(3)}{4} \, \chi(CY_3) \;\; ,
 \eeq
with $\chi(CY_3)$ Euler-characteristic of $CY_3$. Note that it is of order $ \mathcal{O}(\alpha^{\prime3})$ and it depends on 
the Type IIB string coupling.\footnote{The correction is known to depend on the dilaton $e^{-\Phi}$. We assume that the dilaton is stabilized by the flux background and we thus encounter the string coupling constant  $g_s = \langle e^{\Phi}\rangle$.}  It is obtained from the parent $\cN=2$ theory arising  from compactification of Type IIB on Calabi--Yau orientifolds \cite{Becker:2002nn,Bonetti:2016dqh}.\footnote{To compute the correction to the scalar potential resulting from \eqref{BBHL} we use the  K\"ahler potential and coordinates obtained in \cite{Bonetti:2016dqh}.} Note that  Calabi-Yau threefold in IIB is the  double cover of the base $B_3$ branched along the $O7$-plane. Thus in particular we find that $\chi(CY_3) = \chi(B_3)$. As we discuss intrinsic $\cN=1$  vacua in this work we continue in the terminology of F-theory.  We comment on the potential F-theoretic origin of the correction \eqref{BBHL} in section  \ref{topDivUplift}. Note that the string coupling dependence of \eqref{BBHL} makes it parametrically relevant although being sub leading in $\al'$ compared to \eqref{no-scale4d}. 

Lastly let us close with a remark on the stability of the following scenarios in regard to higher-order corrections in $\al'$ and $g_s$ in the light of \cite{Dine:1985he}. The classical correction to the scalar potential vanishes due to the no-scale condition and thus the leading order $g_s$ and $\al'$-correction determine the vacuum. Higher-order $\alpha'$-corrections are parametrically under control as one stabilizes the internal space at large volumes. Moreover the string coupling constant $g_s$ may be achieved to be parametrically small thus  higher-order string loop corrections may be safely neglected. Details depend however on the relative pre-factors of the perturbative terms given by  topological quantities  of the internal space.



\paragraph{Extrema in the generic moduli case.} In this section we discuss a scenario in which all K\"ahler moduli might be stabilized in a non-supersymmetric anti-de Sitter minimum for manifolds with $\chi(B_3)< 0$. We argue for a model independent extremum and  provide a sufficient condition  for the existence of a local minimum in generic geometric backgrounds for K\"ahler cone coordinates $ v^\al_b > 0$  to be,
 \beq
 \langle \cK^b_\al \rangle  \langle \cK^b_\beta \rangle \, v_b^\alpha v_b^\beta \;\;\; >  \;\;  - \tfrac{1}{2}  \langle \cK^b_{\al \beta}\rangle \langle \cV^b \rangle  \, v_b^\alpha v_b^\beta   \;\; ,\;\;\;  \forall \,  v_b^\alpha \; . 
 \eeq 
However, to show that is a true local minimum further studies in explicit geometries are required.
 The stabilization is achieved by an interplay of the correction proportional to $\cZ^b_\al$  in \eqref{no-scale4d}  with the  $\alpha'^3$ Euler-characteristic correction \eqref{BBHL} to the K\"ahler potential \cite{Becker:2002nn,Bonetti:2016dqh}. 
To achieve positivity of the four-cycle volumes in the vacuum the $\al'$-corrections additionally needs to obey strict positivity and negativity conditions, i.e.\,\,the geometric background must be suitable. 
 Note that due to a similar potential all K\"ahler moduli may be stabilized for $\chi(B_3)> 0$ as discussed in \cite{Ciupke:2015msa}. Let us emphasize that the we do not require non-perturbative effects which are generically exponentially suppressed by the volume of the cycles.
In future work \cite{Weissenbacher:2019} we study  a modified scenario  by  additionally considering the $\alpha'^3 g_s^{- 3\, / \, 2}$-correction to the scalar potential discussd in \cite{Grimm:2017okk,Ciupke:2015msa}.
The resulting potential in the large volume limit then takes the form\footnote{We refer to the large volume limit to the regime at large volumes $\cV^b$ and weak string coupling  such that higher order $\alpha'$ and $g_s$-corrections can be neglected. }
\beq\label{no-scale4dBBHL}
V_F =   \frac{3  g_s |W_0|^2}{4\, \cV_ b^3} \; \Big(   \, \xi \,    +  \hat\kappa_4 \,   { \cZ^b_\alpha} \,  v_ b^\alpha \; \Big) \;\; .
\eeq
where $\hat\kappa_4 =- \tfrac{1}{64}$. We note that the functional structure is similar to the $\alpha'$-correction discussed in  \cite{Grimm:2017okk,Ciupke:2015msa}.\footnote{The overall factor  $g_s/2$ in \eqref{no-scale4dBBHL}  stems from the dilation dependence  of the K\"ahler potential.}   One finds the AdS vacua where all four-cycle volumes $\cK_\al^b$ are stabilized at 
\beq\label{vacuum_one}
\langle \cK_\al^b \rangle =  \, \Lambda^2 \, \cZ^b_\alpha  \;\; , \;\;\; \text{with} \;\;\; \Lambda =   \frac{9\, \xi}{8 \, |\hat\kappa_4|}  \,  \cdot \frac{1}{\cZ^b_\alpha \langle v^\al_0 \rangle }  \;\; ,
\eeq
where $ \langle v^\al_0 \rangle =\tfrac{1}{\Lambda}  \langle v^\al_b \rangle $ is the expectation value of the fields such that
\beq\label{defV0}
  \cK^b_{\al \beta \gamma} \langle v^\beta_0 \rangle  \langle v^\gamma_0 \rangle  =  \, \cZ^b_\alpha \;\; .
\eeq
This scaling is required  to ensure that
 $\cZ^b_\alpha   \langle v^\al_b \rangle =   9\, \xi \, / \,(8 \, |\hat\kappa_4|)$.  In other words this additional condition can always be satisfied  as  one concludes  from \eqref{defV0}  which fixes $ \langle v^\al_0 \rangle $ uniquely, and thus implies \eqref{vacuum_one}. One infers the volume in the extremum to be
 \beq\label{volumeVac}
\langle \cV^b \rangle=  \, \frac{3\,  \xi \, \Lambda^2}{8 \, |\hat\kappa_4|} \, \sim g_s^{-\tfrac{9}{2}}\;\;, 
 \eeq
and moreover that the value of the potential in the extremum takes the form
\beq\label{Vacuum}
\langle V^F \rangle =  - \frac{3 \, g_s\, \xi }{32 }\, \cdot\, \frac{|W_0|^2}{\langle \cV^b \rangle ^3} \, \sim  - g_s^{13}\, |W_0|^2 \;\; \; < 0 \;\; .
\eeq 
Note that since $\chi(B_3) < 0$ one  infers that $\xi > 0$. 
In the weakly coupled string regime $g_s < 1$  one  generically achieves a  large positive overall volume $\langle \cV^b\rangle > 0 $ in \eqref{volumeVac}.   Moreover, positivity of all four-cycles volumes $\langle \cK_\al^b \rangle> 0$   for $\cZ^b_\alpha  > 0$  for all $ \al =1,\dots,h^{1,1}(B_3)
$  in \eqref{vacuum_one} and \eqref{defV0}.\footnote{Note that the mechanism could also be applied for different sign of the pre-factor of the $\cZ^b_\al$-correction in \eqref{no-scale4dBBHL} and would then lead to  $\cZ^b_\alpha  > 0$  for all $ \al =1,\dots,h^{1,1}(B_3)
$ with opposite overall sign  in \eqref{vacuum_one} and \eqref{defV0}.}
From \eqref{Vacuum} one finds that one may achieve small values of $\langle V^F \rangle $ also for a moderately large $|W_0|$ due to the  strong string coupling suppression.
By analyzing the matrix of second derivatives in the extremum one infers
\beq\label{variation2}
 \Big<  \frac{ \pd^2\, V_F}{\pd v_b^\al \pd v^\beta_b}   \Big>  \;\; =  \frac{3 \, g_s\, |W_0|^2 \Lambda^2}{4\, \langle \cV^b \rangle^5}\Big(\gamma_1\cK^b_{\al \beta}  + \gamma_2\cZ^b_{\al}  \cZ^b_{\beta}   \Big)\;\;\;\; , \;\; \gamma_1 = \tfrac{9}{64\, |\hat\kappa_4|}  \xi^2 \;\;\; ,\;\;\; \gamma_2 = \tfrac{3\,}{4}  \xi \Lambda^2 \;\; ,
\eeq
where one concludes that $\gamma_1 >  0$ and $\gamma_2 > 0$. The matrix  $\gamma_2\cZ^b_{\al}  \cZ^b_{\beta}$ is positive semi-define, however it was argued  in \cite{Candelas:1990pi} that $ \cK_{\al \beta}^b$ is of signature $(1,h^{1,1}(B_3))$, i.e.\,it exhibits one positive eigenvalue  in the direction of the vector $\langle v^\al_ b \rangle$. 
 Thus to argue for a local minimum  one needs to  analyse \eqref{variation2} in explicit models. One may rewrite \eqref{variation2} to be in the form
 \beq\label{variation3}
 \Big<  \frac{ \pd^2\, V_F}{\pd v_b^\al \pd v^\beta_b}   \Big> = \frac{9\,g_s\, |W_0|^2 \xi}{16 \, \langle \cV^b \rangle ^4}\Big( \tfrac{1}{\langle \cV^b \rangle }\langle \cK^b_\al \rangle  \langle \cK^b_\beta \rangle + \tfrac{1}{2}  \langle \cK^b_{\al \beta} \rangle  \Big) \;\; ,
 \eeq
 from which one infers a sufficient condition on the geometry for positive semi-definiteness of \eqref{variation3} and thus for the existence of a local minimum to be
 \beq\label{conditionVac1}
 \langle \cK^b_\al \rangle  \langle \cK^b_\beta \rangle \, v_b^\alpha v_b^\beta \;\;\; >  \;\;  - \tfrac{1}{2}  \langle \cK^b_{\al \beta}\rangle \langle \cV^b \rangle  \, v_b^\alpha v_b^\beta   \;\; ,\;\;\;  \forall \,  v_b^\alpha \; . 
 \eeq 
 Note that in this paragraph we have assumed that the self-intersection numbers are vanishing to argue for the vanishing of the $\mathcal{W}^b_\al$ in the scalar potential  \eqref{no-scale4dBBHL}. Thus \eqref{conditionVac1} is automatically satisfied by the  non-vanishing of all  four-cycle volumes in the vacuum. Thus one encounters a local minimum for those geometries.
 Let us next compare the gravitino mass with  the string  and Kaluza-Klein scale \cite{Conlon:2005ki} for which one finds that
\beq
m_S \sim \langle \cV^b \rangle^{-\tfrac{1}{2}} \;\;\;\;\ ,\;\;\;\;\; m_{KK}  \sim \langle\cV^b\rangle^{-\tfrac{2}{3}}  \;\;\;\;\ ,\;\;\;\;\; m_{3/2} \sim \frac{|W_ 0|}{\langle \cV^b \rangle} \;\; .
\eeq
Thus one infers by using \eqref{volumeVac}  that
\beq\label{hirachy}
\frac{ m_{3/2}}{ m_S} \sim  \frac{|W_0| }{2 |\Lambda|}  \sqrt{\frac{3}{\xi}} \sim g_s^{\tfrac{9}{4}} |W_0| \;\;, \;\;\; \frac{ m_{3/2}}{ m_{KK}} \sim \, g_s^{\tfrac{3}{2}} |W_0| \;\;, \;\;\; \frac{ m_{KK}}{ m_{S}} \sim \, g_s^{\tfrac{3}{4}} \;\;.
\eeq
Thus for a weakly coupled string regime the hierarchies in \eqref{hirachy} can be satisfied accordingly.
Let us conclude that this mechanism might lead to a stabilization for all four-cycles for geometric backgrounds with $\chi(B_3)< 0$. This is achieved solely by an interplay of the Euler-Characteristic $\alpha'^3$-correction \cite{Becker:2002nn} with the $\alpha^2$-correction \cite{Grimm:2013gma,Grimm:2013bha}. As the volume can be stabilized at sufficiently large values higher-order $\alpha'$-corrections are under control i.e.\,\,the vacuum may  not be shifted.

Let us close this section with some remarks concerning  the recent conjecture by  \cite{Obied:2018sgi} which in particular implies the absence of  local  de Sitter  extrema in any  controlled string theory set-up. Note that the discussion in this section can be performed analogously for $\chi(B_3)>  0$ which then leads to a de Sitter extremum  as seen by equation \eqref{Vacuum} with $\xi < 0$. To achieve  a positive overall volume $\cV^b > 0$ in \eqref{volumeVac} and positivity of all four-cycles volumes $\cK_\al^b> 0$  one infers that  
 $\cZ^b_\alpha  < 0$  for all $ \al =1,\dots,h^{1,1}(B_3)
$ and the opposite overall   sign choice  in \eqref{vacuum_one} and \eqref{defV0} which as well constitutes  a solution.
It would be interesting to study explicit geometries where $\cZ^b_\al$ takes values such that a de Sitter extremum is obtained. Let us close this section by emphasizing that  the scenario in this section might thus suffice as the starting point for a concrete counter example to the conjecture \cite{Obied:2018sgi}.



\section{Conclusions}

In this work we established a connection in between  eight-derivative $l_{\rm M}^6$-couplings in eleven-dimensional supergravity i.e.\,the low wave length limit of M-theory, and $\alpha'$-corrections to the K\"ahler potential and K\"ahler coordinates of four-dimensional $\cN=1$ supergravity.  The derivation relies on the M/F-theory duality.  In particular we argue for two novel corrections  to the K\"ahler coordinates  and potential at order  $\alpha'^2$. Noteworthy one of them breaks the no-scale structure. However, we are not able to  ultimately determine the faith of the proposed correction as a  more complete analysis of the $3d, \cN=2$ variables needs to be performed.
This work constitutes the foundation for such a  future study. We provide the completion of the eleven-dimensional $G^2 R^3$ and $(\nabla G)^2 R^2$-sectors  relevant four Calabi--Yau fourfold reductions. We suggest that it would be  of great interest to match our proposal against 5-point  and 6-point scattering amplitudes.  Furthermore, we provide the reduction result of the $G^2 R^3$ and $(\nabla G)^2 R^2$-sectors  for  Calabi--Yau fourfolds with an arbitrary number of K\"ahler moduli. 

One of the main achievements presented is the establishment  of a divisor integral basis for the  three-dimensional K\"ahler coordinates at higher-order in $l_{\rm M}$. This  allows  us to derive the non-topological higher-derivative couplings  obtained in the dimensional reduction from  the novel Ansatz for the K\"ahler potential and K\"ahler coordinates.
We suggest that in order to prove the integration into the proposed $3d,\, \cN =2$ variables additional  non-trivial identifies relating the higher-derivative building blocks are required. Then this amounts to fixing the remaining parameters in our Ansatz. 
We are able to fix several parameters by ensuring compatibility  with the one-modulus case in which the K\"ahler potential and K\"ahler coordinates can be determined exactly as no non-trivial  higher-derivative couplings appear in the K\"ahler metric.

To connect the $l_{\rm M}$-corrections in the three-dimensional K\"ahler coordinates  and K\"ahler potential  to the $\alpha'^2$-corrections in the  $4d, \cN= 1$ theory we employ the classical well understood F-theory uplift. Although it is expected that a one-loop modification of the F-theory lift is needed we argue that in particular the novel $\text{log}(\cV)$- contribution to the K\"ahler potential and coordinates is expected to remain untouched by such an extension. It would be interesting to perform a dimensional reduction of a generic $4d, \cN=1$ supergravity theory in particular  with vector and chiral multiplets where the Kaluza Klein-modes are integrated out at one-loop extending the work of \cite{Tong:2014era}. 
The  novel divisor integral contribution in four-dimensions  is of order $\alpha'^2$. Let us stress that the  ultimate faith of the novel $\alpha'$-corrections to the scalar potential  shall be decided in a forthcoming work as the present result  admits one free parameter. 
Let us continue with a critical remark. The F-theory lift is performed by shrinking the fiber of the Calabi--Yau fourfold, i.e.\,the geometry becomes singular in this process. In this limit  other higher-order UV-corrections may become relevant and modify the uplift. 
However, the corrections discussed in this work are of topological nature and  are thus expected to be protected in the F-theory limit. 

Although the  resulting $\al'^2$-corrected scalar potential arisies from a conjectured correction to the K\"ahler coordinates it is of interest to study possible scenarios to obtain stable vacua. We discuss a  scenario in which the   $\mathcal{Z}^b_\alpha$-correction at order $\alpha'^2$ interplays with the $\alpha'^3$ Euler-characteristic correction  to achieve a  non-supersymmetric  anti-de Sitter minimum for geometric backgrounds with $\chi(B_ 3)< 0$. Moreover   constraints on the topological quantities of the geometric backgrounds are derived such that a minimum may be obtained. It would be  of great interest to realize our constraints in explicit examples of elliptically fibered Calabi-Yau fourfolds. Furthermore, we note that  the scenarios provide a model independent de Sitter extremum for geometric backgrounds with $\chi(B_ 3)> 0$.  One may extend the present analysis  \cite{Weissenbacher:2019}  by additionally considering the $\alpha'^3$-correction to the scalar potential discussd in \cite{Grimm:2017okk,Ciupke:2015msa}.
Lastly let us point the reader to an obvious extension of the present work. Our analysis of geometries does not allow for no-Abelian singularities, i.e.\,\,no non-Abelian four-dimensional gauge fields are present. It would be highly desirable to  analyse the uplift of the   K\"ahler potential and K\"ahler coordinates for such backgrounds.

\vspace{1.5 cm}

\noindent{\bf Acknowledgements.}

\noindent 
In particular, I am grateful to Andreas Braun, Michele Cicoli, Thomas Grimm and Raffaele Savelli for enlightening discussions and comments on the draft.  Moreover, I would like to take this opportunity to express my gratitude  to Sharon Law for being a beacon of light and for her constant  kind support.
 Lastly,  many thanks to the string theory group at the Seoul National University  for its hospitality during my stay, where parts of this work were initiated. This work was supported by the WPI program of Japan.

\appendix

\section{\bf Conventions, definitions, and identities} \label{Conv_appendix}

In this work we denote the eleven-dimensional space indices by capital Latin letters $M,N = 0,\ldots,10$ and
the external  ones by  $\mu,\nu = 0,1,2$, and the internal complex ones by $m,n,p=1,...,4$ and $\bar m, \bar n,\bar p =1, \ldots,4$. The metric signature of the eleven-dimensional space  is $(-,+,\dots,+)$.
Furthermore, the convention for the totally 
anti-symmetric tensor in Lorentzian space in an orthonormal frame is $\epsilon_{012...10} = \epsilon_{012}=+1$. 
The epsilon tensor in $d$ dimensions then satisfies
\ba
\epsilon^{R_1\cdots R_p N_{1 }\ldots N_{d-p}}\epsilon_{R_1 \ldots R_p M_{1} \ldots M_{d-p}} &= (-1)^s (d-p)! p! 
\delta^{N_{1}}{}_{[M_{1}} \ldots \delta^{N_{d-p}}{}_{M_{d-p}]} \,, 
\ea
where  $s=0$ if the metric has Riemannian signature and $s=1$ for a Lorentzian metric.
We adopt the following conventions for the Christoffel symbols and Riemann tensor 
\ba
\G^R{}_{M N} & = \fr12 g^{RS} ( \pa_{M} g_{N S} + \pa_N g_{M S} - \pa_S g_{M N}  ) \, , &
R_{M N} & = R^R{}_{M R N} \, , \nonumber\\
R^{M}{}_{N R S} &= \pa_R \G^M{}_{S N}  - \pa_{S} \G^M{}_{R N} + \G^M{}_{R  T} \G^T{}_{S N} - \G^M{}_{ST} \G^T{}_{R N} \,, &
R & = R_{M N} g^{M N} \, , 
\ea
with equivalent definitions on the internal and external spaces. Written in components, the first and second  Bianchi identity are
\bea\label{Bainchiid}
{R^O}_{PMN} + {R^O}_{MNP}+{R^O}_{NPM} & = & 0 \nonumber\\
(\nabla_L R)^O{}_{PMN} + (\nabla_M R)^O{}_{PNL} + (\nabla_N R)^O{}_{PLM} & = & 0 \;\;\; .
\eea
Differential p-forms are expanded in a basis of differential one-forms as
\beq
\Lambda = \frac{1}{p!} \Lambda_{M_1\dots M_p} dx^{M_1}\wedge \dots \we dx^{M_p} \;\; .
\eeq
The wedge product between a $p$-form $\Lambda^{(p)}$ and a $q$-form $\Lambda^{(q)}$ is given by
\beq
(\Lambda^{(p)} \we \Lambda^{(q)})_{M_1 \dots M_{p+q}} = \frac{(p+q)!}{p!q!} \Lambda^{(p)}_{[M_1 \dots M_p} \Lambda^{(q)}_{M_1 \dots M_q]} \;\; .
\eeq
Furthermore, the exterior derivative on a $p$-form  $\Lambda$ results in
\beq
 ( d \Lambda)_{N M_1\dots M_p} = (p+1) \pa_{[N}\Lambda_{ M_1\dots M_p]} \;\;,
\eeq
while the Hodge star of   $p$-form  $\Lambda$ in $d$ real coordinates is given by
\beq
(\ast_d \Lambda)_{N_1 \dots N_{d-p}} = \frac{1}{p!} \Lambda^{M_1 \dots M_p}\epsilon_{M_1 \dots M_p N_1\dots N_{d-p}} \;\; .
\eeq
Moreover, 
\beq\label{idwestar}
 \Lambda^{(1)} \we \ast \Lambda^{(2)} = \frac{1}{p!}\Lambda^{(1)}_{M_1\dots M_p} \Lambda^{(2)}{}^{M_1\dots M_p} \ast_1 \;\; ,
\eeq
which holds  for two arbitrary $p$-forms $\Lambda^{(1)}$ and $\Lambda^{(2)}$. 

Let us next define the intersection numbers,  where $\{ \o_i \}$ are harmonic w.r.t.~to the Calabi- Yau metric $g_{m \bar n}$
\ba\label{IN1}
\cK_{i j k  l}  &= \int_{X} \o_i \we   \o_j \we   \o_k  \we   \o_l   \, ,& 
\cK_{i j}  & = \cK_{i j k l } v^l \,, & 
\cK_{i j  }  & = \fr12 \cK_{i j k l }v^k  v^l,  & \nonumber \\
\cK_{i  }  & = \fr1{3!} \cK_{i j k l } v^j v^k v^l  \, , & 
\cV  & = \fr1{4!} \cK_{i j k l } v^i v^j v^k v^l \, . & 
\ea 
Let us review  well known identities such as
\beq
\int \o_i \wedge \ast_8 \o_j = - \cK_{ij} + \frac{1}{\cV} \cK_i \cK_j \;\; .
\eeq
Let us note that the intersection numbers obey the properties
\ba \label{intersectionids}
\cK_{i} v^i &= 4 \cV \;\;\; , \;\;\; \cK_{ij} v^j = 3 \cK_{j} \;\;\; ,\;\; \cK_{ijk} v^k =  2 \cK_{ij} &\nn \\
\cK_{ijkl} v^l &=  \cK_{ijk}  \;\;\; , \;\;\; \cK^{ik} \cK_{jk} = \delta_j^i  \;\;\; , \;\;\; \cK^{ik} \cK_{k} = \tfrac{1}{3} v^i& \nn \\
\Big(\frac{\pd}{\pd v^k}\cK^{ij}\Big)\cK_{j} &= -\tfrac{2}{3} \delta_{k}^i \;\; ,\;\;\;   \Big(\frac{\pd}{\pd v^k}\cK^{ij}\Big)\cK_{jl} = - \cK^{i j}\cK_{k j l}    \;\;\; ,&
\ea
with the inverse intersection matrix  $\cK^{ij}$.
The intersection numbers for the K\"ahler base are given by
\ba\label{IN2_6d}
\cK^b_{\al \beta \gamma}  &= \int_{B_3} \o_\al  \we   \o_\beta \we   \o_\gamma  \  \, ,& 
\cK_{\al^b  \beta }  & = \cK^b_{\al \beta \gamma}v_b^\gamma \,, & 
\cK^b_{\al }  & = \fr12 \cK^b_{\al \beta \gamma} v_b^\beta v_b^\gamma,  & \nonumber \\
\cV^b  & = \fr1{3!} \cK^b_{\al \beta \gamma}v_b^\al v_b^\beta v_b^\gamma \, . & 
\
\ea 
One may show that for a six-dimensional K\"ahler  manifold
\beq\label{integralsplit2}
\ast_6(\o^b_\al \wedge J^b) = \frac{\cK^b_\al}{\cV^b} J^b - \omega^b_\al\;\; .
\eeq
with intersection numbers defined analogously to  \eqref{IN1}. In particular, this implies the analogous relation 
\beq\label{integralsplit3}
\int_{D_\alpha}  \; \tilde   \ast_6 ( H\tilde c_1  \wedge   \tilde J ) \wedge \tilde c_2\;\;   = \frac{1}{\cK_\alpha}    \int_{D_\alpha}  \; \tilde c_1 \wedge   \tilde J ^2    \int_{D_\alpha}  \; \tilde c_2 \wedge   \tilde J   -  \int_{D_\alpha} \tilde c_1 \wedge \tilde c_2  \;\;,
\eeq
which holds due to the harmonicity of $H \tilde c_1(D_\alpha)$.

 We define the curvature two-form for Hermitian manifolds to be
  \begin{equation}\label{curvtwo}
 {\cR^m}_n  =  {{R^m}_n }_{ r \bar s} dz^ r \wedge d\bar{z}^\bar{s}\;\;,
  \end{equation}
 and 
 \bea \label{defR3}
 \Tr{\cR}\;\;&  =& {{R^ m }_ m }_{ r \bar{s}}dz^ r \wedge d\bar{z}^{\bar{s}} \;,\nn \\[0.2cm]
 \Tr{\cR^2} &= & {{R^{ m }}_{n }}_{ r \bar{s}} {{R^{n }}_{ m }}_{ r_1 \bar{s}_1}dz^{ r}
 \wedge d\bar{z}^{\bar{s}}\wedge dz^{ r_1} \wedge d\bar{z}^{\bar{s}_1} \;,\nn  \\[0.2cm]
 \Tr{\cR^3} &=& {{R^{ m }}_{n }}_{ r \bar{s}}  R^{n }{}_{n _1  r_1 \bar{s}_1}
 {{R^{n _1}}_{ m }}_{ r_2 \bar{s}_2}dz^{ r} \wedge d\bar{z}^{\bar{s}}\wedge dz^{ r_1} \wedge d\bar{z}^{\bar{s}_1}\wedge dz^{ r_2} \wedge d\bar{z}^{\bar{s}_2} \; .
 \eea
 The Chern forms can be expressed in terms of the curvature two-form as
\begin{align}  \label{Chernclasses}
 c_1 &=  i \Tr{ \mathcal{R}} \nonumber\;, \\[0.2cm]
 c_2 &=  \frac{1}{2}\left( \Tr{\cR^2} -  (\Tr{\cR})^2 \right)\;, \\[0.2cm]
 c_3 &= \frac{1}{3}c_1c_2 + \frac{1}{3} c_1 \wedge \Tr \cR^2 - \frac{i}{3} \Tr \cR^3 \;,\nonumber \\[0.2cm]
 c_4 &= \frac{1}{24} \left( c_1^4 - 6c_1^2 \Tr\cR^2 - 8i c_1 \Tr\cR^3\right) + \frac{1}{8}((\Tr\cR^2)^2 - 2 \Tr\cR^4)\,. \nonumber 
\end{align}
The Chern classes of a $n$ complex-dimensional Calabi-Yau manifold $CY_n$ reduce to
\beq\label{chern34}
c_3 (CY_{n \geq 3}) =  -\frac{i}{3}  \Tr{\cR^3} \;\; \text{and} \;\; c_4 (CY_{n \geq 4}) = \frac{1}{8}((\Tr\cR^2)^2 - 2 \Tr\cR^4)\;,
\eeq
with $\Tr \cR^4$ defined analogous as in \eqref{defR3}. 
Let us next define a set of higher-derivative building blocks  identified  in \cite{Grimm:2014efa} as
\ba \label{def-Zmmmm}
Z_{m \bar m n \bar n} = \fr{1}{4!} \e^\tbzero_{ m \bar m m_1 \bar m_1 m_2 \bar m_2 m_3 \bar m_3}  \e^\tbzero_{ n \bar n n_1 \bar n_1 n_2 \bar n_2 n_3 \bar n_3} R^\tbzero{}^{\bar m_1 m_1 \bar n_1 n_1} R^\tbzero{}^{\bar m_2 m_2 \bar n_2 n_2} R^\tbzero{}^{\bar m_3 m_3 \bar n_3 n_3}\ ,
\ea
and 
\ba \label{def-Yij}
Y_{i j m \bar n} = \fr{1}{4!} \e^\tbzero_{ m \bar m m_1 \bar m_1 m_2 \bar m_2 m_3 \bar m_3}  \e^\tbzero_{ n \bar n n_1 \bar n_1 n_2 \bar n_2 n_3 \bar n_3} &\na^\tbzero{}^{n} \o^\tbzero_i{}^{\bar m_1 m_1} \na^\tbzero{}^{\bar m} \o^\tbzero_j{}^{\bar n_1 n_1}  \nn \\ &\times R^\tbzero{}^{\bar m_2 m_2 \bar n_2 n_2} R^\tbzero{}^{\bar m_3 m_3 \bar n_3 n_3}\ .
\ea
It turns out that the tensor $Z_{m \bar m n \bar n}$ given in \eqref{def-Zmmmm} plays 
a central role in the following and is related 
to the key topological quantities on $Y_4$. It satisfies the identities
\ba
Z_{m \bar m n \bar n} &= Z_{n \bar m m \bar n} = Z_{m \bar n n \bar m}\ , &
\na^\tbzero{}^{m} Z_{m \bar m n \bar n} &= \na^\tbzero{}^{\bar m} Z_{m \bar m n \bar n}  = 0\ .
\ea
It is related to the third Chern-form $c_3^\tbzero{}$ via
\ba \label{def-Z_plain}
Z_{m \bar m} &= i 2 Z_{m \bar m n}{}^n = (2\pi)^3 \fr12 (*^\tbzero c_3^\tbzero{})_{m \bar m}\, , \nn
\ea
\vspace{-0.5cm}
\ba 
Z &=  i 2 Z_{m}{}^m  =(2 \pi)^3 *^\tbzero( J^\tbzero{} \we c_3^\tbzero{})\, ,  & 
(2 \pi)^3*^\tbzero (c_3^\tbzero{} \wedge \o^\tbzero_i ) & = - 2 Z_{m \bar n}\o^\tbzero_i{}^{\bar n m}\, , 
 \ea
and yields the fourth Chern-form $c^\tbzero_4$ by contraction with the Riemann tensor as
\ba 
Z_{m \bar m n \bar n} R^\tbzero{}^{\bar m m \bar n n} &= (2 \pi)^4*^\tbzero c^\tbzero_4\ .
\ea
We note that $Y_{i j m \bar n} $ is also related to $Z_{m \bar m n \bar n}$ upon integration as
\ba
 \int_{Y_4} Y_{ij}{}_m{}^m *^\tbzero 1= - \fr16 \int_{Y_4} (  i Z_{m \bar n} \o^\tbzero_{i}{}^{\bar r m} \o^\tbzero_{j}{}^{\bar n}{}_{\bar r } + 2  Z_{m \bar n r \bar s} \o_i^\tbzero{}^{\bar n m} \o_j^\tbzero{}^{\bar s r} ) *^\tbzero1 \, , 
\ea
where the right hand side represents the same linear combination that will be relevant in \ref{secCYDivisorINt}. Let us for further use define
\beq\label{def-Yint}
\mathcal{Y}_{ij} :=  \int_{Y_4} Y_{ij}{}_m{}^m *^\tbzero 1 \;\;.
\eeq
Lastly in this work we encounter a  new (2,2)-form object  
\beq\label{defOmega}
\Omega_{ij}  =    R^\tbzero_{m \bar n r \bar s} \omega^\tbzero_{i}{}^{r}{}_{t}\omega^\tbzero_{j}{}^{\bar s}{}_{\bar u} \;\; dz^m \wedge dz^t \wedge d\bar z^{\bar n} \wedge d \bar z^{\bar u} \;\; .
\eeq

\subsection{Divisor integrals in terms of $CY_4$ integrals} \label{secDivisorintegralsDetails}

We define an arbitrary  basis of higher-derivative $ (1,1)$ -forms convenient for the computations in this work
\bea\label{basisCalabiYauRiemann}
X_1  &=&   R_{m}{}^{ m_2}{}_{m_5}{}^{n_2}  R_{m_2}{}^{n_3}{}_{n_2}{}^{n_4}    R_{n_3 \bar m n_4}{}^{m_5}     \;\;  d z^m \wedge d \bar z^{\bar m}\nonumber  \\ \nonumber
X_2  &=&  R_{m}{}^{m_2}{}_{m_5}{}^{n_2} 
 R_{m_2 \bar m}{}_{n_3}{}^{n_4}    R_{n_2}{}^{m_5}{}_{n_4}{}^{n_3}  \;\; d z^m \wedge d \bar z^{\bar m} \\ \nonumber
X_3  &=&  R_{m \bar m}{}_{m_2}{}^{m_5}    R_{m_5}{}^{n_2}{}_{n_3}{}^{n_4}    R_{n_2}{}^{m_2}{}_{n_4}{}^{n_3}   \;\;   d z^m \wedge d \bar z^{\bar m} \nonumber \\ 
X_4  &=&   g_{m \bar m} R_{m_1}{}^{ m_2}{}_{m_5}{}^{n_2}  R_{m_2}{}^{n_3}{}_{n_2}{}^{n_4}    R_{n_3}{}^{ m_1}{}_{ n_4}{}^{m_5}     \;\;  d z^m \wedge d \bar z^{\bar m}\nonumber  \\ 
X_5  &=&  g_{m \bar m} R_{m_1}{}^{m_2}{}_{m_5}{}^{n_2} 
 R_{m_2}{}^{m_1}{}_{n_3}{}^{n_4}    R_{n_2}{}^{m_5}{}_{n_4}{}^{n_3}  \;\; d z^m \wedge d \bar z^{\bar m} 
\eea

These $(1,1)$-forms can be expressed as integrals on Calabi--Yau  fourfolds which admit an interpretation as integrals on divisors $D_i$  of a Calabi--Yau fourfold as
\beq\label{basisCalabiYauRiemann2}
\int_{CY_4} \big( \ast_8 X_{k=1,..,5}  \big)\wedge \omega_i = \int_{D_i}\ast_8 X_{k=1,..,5}  \;\; ,
\eeq
where the r.h.s.\,is  to be seen as pulled back to the divisor. Let us now recall the fact \cite{Kobayashi} that any complex sub-manifold of a K\"ahler manifold $M$ is itself K\"ahler with induced metric  and K\"ahler form $g, J$ of M. Thus in particular we find for the Divisors  $i : D_i \xhookrightarrow{ } CY_4$  the K\"ahler metric and form $^\ast i g$ and  $^\ast i J$, respectively, which are  pulled back from the Calabi--Yau  fourfold. One may thus as well restrict Riemann tensors on the Calabi--Yau fourfold to divisors $D_ i$ expressed by the induced metric which generically obeys $c_1(D_i) \neq 0$.  In particular contractions of the Riemann tensors which do not vanish on the Calabi--Yau manifold due to the Calabi--Yau conditions may be pulled back to the divisors and expressed in terms of Riemann tensors in terms of the induced metric on $D_i$. Note that  the $(1,1)$-forms in \eqref{basisCalabiYauRiemann} expressed as integrals on divisors \eqref{basisCalabiYauRiemann2} in  are of this form.
By

We may write the K\"ahler coordinates  as \eqref{divisorInt} in terms as the new basis on $CY_4$ in the following way if the coefficients obey the following relations
\bea\label{coeffIntegralB3toY4appendix}
\alpha_5  &=&  - \tfrac{1}{8}\alpha_1 +  \tfrac{1}{24} \alpha_3 + \tfrac{1}{4} \alpha_4\; ,\nonumber \\
 \alpha_6  &=&  \tfrac{1}{2}\alpha_2 + \tfrac{1}{2}\alpha_3 \; ,\nonumber \\
   \alpha_7  &=& \alpha_2 + \alpha_3 \; ,\nonumber \\
   \alpha_8  &=&  \tfrac{1}{2} \alpha_1 - \tfrac{1}{3}\alpha_3 - \alpha_4 \; , \nonumber \\
    \alpha_9  &=& - \alpha_1 + \tfrac{1}{6}\alpha_3 \; .
\eea
one then infers that
\ba\label{paramteriseT}
\mathcal{T}_i =  -\frac{i}{3 } \int_{CY_4} \omega_i \wedge \ast_8\Big(  & (\alpha_3 + 3 \gamma_2 + 6 \gamma_3) X_1 +  (\alpha_3 + 3 \gamma_1 - 12 \gamma_3) X_2+  3( \gamma_3 - \alpha_3 )  X_3  \nn\\ 
&+
3 (- \gamma_2 + \gamma_4 + \gamma_5) X_4 - 3 ( \gamma_1 + 2 \gamma_4 + 2 \gamma_5)  X_5  \Big)\;\; ,
\ea
where $X_{i=1,2,3,4,5}$ are defined in \eqref{basisCalabiYauRiemann}, and where the freedom in the real parameters  $\gamma_1, \dots,\gamma_5$ results due to total derivatives which take different form on the divisors integrals and Calabi--Yau fourfold integrals, respectively.
The simplest choice in this work for coefficients $\gamma_1, \gamma_2,\gamma_3$ defines  the higher-derivative $(3,3)$-form to be 
\beq\label{definitionOfT}
 \mathcal{X} = -\frac{i}{3 }\ast_8\big(X_1 +X_2-X_3  \big) \;\; ,
\eeq
and thus the K\"ahler coordinate modification  is
\beq
\label{tiasCY4intApp}
\cT_i =  - \frac{i }{3 }  \int_{CY_4} \omega_i \wedge \ast_8 \big(X_1 +X_2-X_3  \big) \;\;\; .
\eeq
Note that $\alpha_3 = 1$ which is in agreement with the divisor integral one-modulus limit.\footnote{The coefficients in \eqref{paramteriseT} are chosen as $ \gamma_1 = 8/3, \,   \gamma_2 =  - 4/3,\,   \gamma_3 = 2/3, \, \gamma_5 = -4/3 -  \gamma_4$.} Note that the choice of fixing $\alpha_1$ does not limit the Ansatz for the K\"ahler coordinates as it amounts only to an overall coefficients which is anyway taken into account for in \eqref{Kahlercord}. One may easily show that thus
\beq \label{propXYApp}
\mathcal{T}_i v^i =  \int_{CY_4} J \wedge \mathcal{X} =  \cZ \;\; .
\eeq
and from this property \eqref{propXYApp} that
\ba\label{propXYApp2}
\Big(\frac{\pd}{\pd v^j}\mathcal{T}_i \Big) v^i  & =  - \cT_j + \cZ_j \nn \\
\Big(\frac{\pd^2}{\pd v^k\pd v^j}\mathcal{T}_i \Big) v^i  &=  - \Big(\frac{\pd}{\pd v^j}\mathcal{T}_{k} \Big)   -\Big(\frac{\pd}{\pd v^k}\mathcal{T}_{j} \Big)   \;\; .
\ea

Let us comment on \eqref{definitionOfT}. the combination of basis elements $X_{i=1,2,3,4,5}$ is a choice compatible with the match to six-dimensional divisor integrals.
In section \ref{variation} we  discuss the variation of $\mathcal{T}_i$ w.r.t.\,to the K\"ahler deformations.

As the matching of the  correction to the K\"ahler coordinates  in terms of  $CY_4$ integrals  to the  divisor integral expression is not unique,  let us close this section on remarks other possible choices of $\gamma_1, \gamma_2,\gamma_3$.   Due to \eqref{propXYApp} the Ansatz \eqref{Kahlerpot} and \eqref{Kahlercord} cannot depend separately on $\mathcal{T}_i v^i$. It is interesting to study the possible where \eqref{Kahlerpot} is modified by this expression as well and  \eqref{Kahlercord} by $\tfrac{1}{\cV}\cK_i \mathcal{T}_j v^j$. Let us close this section by discussing a caveat to the Ansatz in this work namely that our choice for $\cT_i$ \eqref{definitionOfT} may be rewritten by splitting integrals using the harmonicity of $\o_i$
\beq\label{eqcaveat}
\cT_i = \tfrac{1}{3} \big( - \cZ_i + \tfrac{1}{\cV}\cK_i \cZ \big)\;\; .
\eeq
Let us emphasize that the insights of this work is that the higher-derivative structures derived in dimensional Calabi--Yau fourfold reductions for $h^{1,1} > 1$ can be obtained by variation of $\cT_i$ before applying the integral split \eqref{eqcaveat} which suggests an interpretation in terms of divisor integrals. One infers that by imposing \eqref{eqcaveat} first the Ansatz for the K\"ahler coordinates \eqref{Kahlercord} does not carry any new information, i.e.\;\;those to steps seem not to commute. However, by choosing a more involved combination for the correction to the K\"ahler coordinate in terms of Calabi--Yau fourfold quantities in  \eqref{paramteriseT} this caveat can be prevented as then no analogous relation for \eqref{eqcaveat} holds. Generically we expect the form $ \cT_i + \cT^0_i $ where in the one-modulus limit $ \cT_i  \to \tilde \cZ$ and $ \cT^0_i  \to 0$. This suggests that one might need to extend the basis \eqref{basisCalabiYauRiemann} to also contain terms with explicit covariant derivatives such as e.g.\,$ \sim (\nabla R)^2$. Moreover, one may  not expect to  capture the information  of topological quantities of divisors entirely  by local covariant integral densities on the entire space but  may need to include  additional global obstructions to succeed in the matching.


\subsection{3d K\"ahler coordinates as topological divisor integrals }
\label{appendixsecTopDivInt}

In this section we argue that the  Ansatz for the K\"ahler coordinates \eqref{divisorInt} may  be rewritten in terms of topological integrals by fixing the coefficients in the Ansatz.
Any closed form on such as $\tilde c_1$ may be written in terms of its harmonic part plus a double exact contribution
\beq
\tilde c_1 = H\tilde c_1 + \partial\bar{\partial} \lambda \;\; ,
\eeq
where $\lambda$ is a function on the divisor. From the closure of $\tilde c_1$ we infer that
\beq 
\nabla_{[m}\tilde R_{n] \bar n r}{}^r = 0 \;\; .
\eeq
But equivalently one may use that
\ba
\nabla_{m}\tilde R_{n}{}^n{}_{r}{}^r = \nabla_{m} \nabla_{n}\nabla^n \lambda  \;\;, \nonumber \\
\nabla^{m}\tilde R_{n}{}^n{}_{r}{}^r = \nabla^{m} \nabla_{n}\nabla^n \lambda  \;\;,  \nonumber \\
\nabla_{m}\tilde R_{n}{}^m{}_{r}{}^r = \nabla_{m} \nabla_{n}\nabla^m \lambda    \;\;, \nonumber \\
 \nabla^{m}\tilde R_{m}{}^n{}_{r}{}^r =  \nabla^{m} \nabla_{n}\nabla^m \lambda   \;\;.
\ea
Using the above set of equations  one may show that the Ansatz for the K\"ahler coordinates  \eqref{divisorInt} can be written as
\ba\label{newTi1app}
\cT_i =  &\;\;\;  \alpha_1  \int_{D_i} \tilde c_1 \wedge  \tilde c_1 \wedge  \tilde c_1 + \alpha_2    \int_{D_i}  \tilde c_1 \wedge  \tilde c_2 \ + \alpha_3   \int_{D_i} \tilde c_3
 +  \frac{\alpha_4}{\cK_i}    \int_{D_i} \tilde c_1 \wedge \tilde J^2   \int_{D_i}  \; \tilde c_1 \wedge  \tilde c_1 \wedge  \tilde J   \nonumber  \\[0.2 cm]
 &+  \frac{\alpha_5}{\cK_i^2}   \int_{D_i}  \; \tilde c_1 \wedge   \tilde J ^2  \int_{D_i}  \; \tilde c_1 \wedge   \tilde J ^2 
 \int_{D_i}  \; \tilde c_1 \wedge   \tilde J ^2   + \frac{\alpha_6}{\cK_i}    \int_{D_i}  \; \tilde c_1 \wedge   \tilde J ^2    \int_{D_i}  \; \tilde c_2 \wedge   \tilde J  \nonumber  \\[0.2 cm] 
& + 2\alpha_6  \int_{D_i}  \; \tilde   \ast_6 ( H\tilde c_1  \wedge   \tilde J ) \wedge \tilde c_2
   -  \big( 2\alpha_4 + 8 \alpha_5 \big)  \int_{D_i}  \; \tilde \ast_6 (H\tilde c_1  \wedge \tilde J) \wedge  \tilde c_1  \wedge   \tilde c_1 \   \nonumber \\[0.2 cm]
   &-\frac{4\alpha_5}{\cK_i} \int_{D_i} \tilde c_1 \wedge \tilde J^2 \int_{D_i}  \tilde c_1 \wedge \ast_6 H \tilde c_1   \;\; ,
\ea
where $\cK_i$ denotes the volume of the divisor $D_i$.
Note that in order to obtain  \eqref{newTi1} one fixes the coefficients such that
\ba
\alpha_7  = 2 \alpha_6  \;\; , \;\;\; 
\alpha_8 =  2 \alpha_4 + 8  \alpha_5  \;\; , \;\;\;
\alpha_9 = - 4 \alpha_5 \;\;.
\ea
Additionally requiring that we can write $\cT_i$ as integrals on the Calabi--Yau fourfold i.e.\;\;the constraints \eqref{coeffIntegralB3toY4} then imposes
\ba \vspace{0,3 cm}\label{finalconstraintsapp}
\quad & \alpha_1 = \tfrac{1}{6} \;\; , \;\;\; \;\;\quad\quad
\quad  \alpha_3 = 1 \;\; , \;\;\; \;\;\quad\ \;\;\;\;\;
  \alpha_4 = -\tfrac{1}{12}  \;\; ,\;\;\;\; \;\;\;
  \alpha_5 = 0   \;\; , \;\;\;\; & \quad\nonumber \\[0.2cm]
 \quad &  \alpha_6 = \tfrac{1}{2} +\tfrac{1}{2} \alpha_2 \;\;\;,\;\;\;\;
     \alpha_7 = 1  +  \alpha_2  \;\; , \;\;\;\; \;\;\;\;
     \alpha_8 = -\tfrac{1}{6}  \;\;\; ,\;\;\;\;  \;\;\;
      \alpha_9 = 0 \;\;. & \;\; \vspace{0,3 cm}
\ea
Note that this coordinate \eqref{finalconstraintsapp} depends on the free parameter $\al_2$. It would be interesting to determine it by imposing some other constraint. 

\subsection{Variation w.r.t.\,K\"ahler moduli fields}
\label{variation}
To compute the variation of covariant integral densities such as \eqref{Texpr} w.r.t.\,K\"ahler moduli fields we deform the Calabi--Yau fourfold metric $g_{m \bar n}$ in complex coordinates by
\beq\label{varmetric}
g_{m \bar n} \to g_{m \bar n} + i  \delta v^i \o_{i m \bar n} \;\; \text{and} \;\;\; g^{ \bar n m} \to g^{  \bar n m} - i  \delta v^i \o_i^{  \bar n m} \;\; .
\eeq
The determinant of the metric  subject to \eqref{varmetric} derives to
\beq
\sqrt{- g} \to \sqrt{- g} +   i \sqrt{- g} \, v^i \o_{i m}{}^{m} \;\; .
\eeq
Note that we are only interested in linear deformations here thus we need to expand the expression to $\mathcal{O}(\delta v^i)$.
The Riemann tensors variation compute to
\beq
R_{m \bar m n \bar n}  \to R_{m \bar m n \bar n} + i \delta v^i \nabla_m \nabla_{\bar m} \o_{i \bar n n}  
+ \tfrac{i}{2}\delta v^i R_{m \bar n n}{}^{r} \o_{i r\bar m}
 + \tfrac{i}{2}\delta v^i R_{m \bar m  n}{}^{r} \o_{i r \bar n} \;\; .
\eeq
To evaluate the variation of higher-derivative object a computer algebra package such as xTensor \cite{Nutma:2013zea} is highly desirable.
One may employ its power  to generate a complete set of Shouten identities, Bianchi identities and total derivatives to show that the variation of \eqref{tiasCY4intApp}  can be written as 

 \beq \label{variationTi2}
 \frac{\partial}{\partial v^j} \cT_i =- \frac{3}{\cV}   \cK_{j} \cT_{i} + \frac{5}{\cV}   \cK_{i} \cT_{j} + 3\, \cT_{ij} + \,  \Lambda_{ij}  \;\; ,
 \eeq
 where
 \beq
 \cT_{ij} =   \int_{CY_4} \ast_8 \big( \o_i \wedge \o_j \wedge J \big) \wedge \mathcal{X}  \;\; ,
 \eeq
 and
 \ba\label{encounter002}
 \Lambda_{ij} =   & \;\;   4 i \int_{CY_4} Z_{m \bar n}  \omega_{i}{}^{\bar n s} \omega_{j s}{}^m  \ast 1 - 6 i \int_{CY_4} Z_{m \bar n}  \omega_{j}{}^{\bar m n} \omega_{i s}{}^s  \ast 1  \nn \\
 = & \;\;  4 i \int_{CY_4} Z_{m \bar n}  \omega_{i}{}^{\bar n s} \omega_{j s}{}^m  \ast 1 +  \frac{3}{\cV}\cZ_j \cK_i \;\; .
 \ea
Let us stress that in order to compute \eqref{variationTi} we make extensive use of the computer algebra package \cite{Nutma:2013zea}, and a non-publicly self-developed extension for complex manifolds and tools to perform the above computation.  By using the relation 
 \ba
\mathcal{Y}_{ij}= - \fr16 \int_{Y_4} (  i Z_{m \bar n} \o_{i}{}^{\bar r m} \o_{j}{}^{\bar n}{}_{\bar r } + 2  Z_{m \bar n r \bar s} \o_i {}^{\bar n m} \o_j{}^{\bar s r} ) *1 \, , 
\ea
We note in section \eqref{secDivisorintegralsDetails} that the we are not able to fix $\cT_i$ precisely in this work. Thus let us present here the variation of a different possible choice of the parameter freedom in \eqref{paramteriseT} which one may show then leads to analogous expression as \eqref{variationTi2}. It is intriguing to note that one can obtain also the novel higher-derivative structure in \eqref{new-reductioResult} by variation of the alternative K\"ahler coordinates
 \ba\label{variationTi3}
 \frac{\partial}{\partial v^j} \cT^{alt}_i  \supset \;\;     \int_{CY_4} c_2 \wedge \Omega_{ij}  +   \int_{CY_4} c_2 \wedge J \wedge   \Omega^{1}_{ij}  +     \int_{CY_4} c_2 \wedge J \wedge \Omega^{2}_{ij}  
   \ea
 with $\Omega_{ij}$ defined in \eqref{defOmega} and with the $(1,1)$-forms 
 \beq
  \Omega^{1}_{ij \, m \bar n}  :=  \big (\nabla_m \nabla_{\bar n} \o_{i r \bar s} \big)  \o_j{}^{\bar s r }\;\;\;\, \;\;\;\;    \Omega^{2}_{ij \, m \bar n}  :=  \big(\nabla_r \nabla^r \o_{i m \bar s}  \big) \o_{j }{}^{\bar s }{}_{\bar n}\;\;\; .
 \eeq
Note that the second Chern-form $c_2$ appears in this case \eqref{variationTi3}  in particular in the combination as in \eqref{new-reductioResult}.   Note that \eqref{variationTi3} is of schematic form and we do not specify the factors in this work.
 
 \paragraph{Warp-factor and the K\"ahler potential.}
Let us  next review the integration of the warp factor into a K\"ahler potential following \cite{Grimm:2015mua}. To begin with, let us reduce our Ansatz \eqref{Kahlercord} and \eqref{Kahlerpot} to the warp factor related quantities which gives
\bea \label{K-AnsatzApp}
   K &=& -3 \log \bigg( \cV + 4 \al^2 \mathscr{W}_i v^i\bigg) 
\eea
 We therefore suggest that they take the form 
\ba \label{PotentialAndCoordinates}
\Re T_i  = \ \cK_i + \al^2\Big(  \mathscr{F}_i + 3 \mathscr{W}_i \Big) 
\ea
where $D_i$ are $h^{1,1}(Y_4)$ divisors of $Y_4$ that span the homology $H_2(Y_4,\mathbb{R})$.
The six-form $F_6$ in this expression is a function of degrees of freedom associated with the internal space metric. It is constrained by a relation to the fourth Chern form $c_4$ such that $F_6$ determines the non harmonic part of $c_4$ as 
\beq \label{def-F6}
   c_4 = H c_4 + i \pa \bar \pa F_6\ .
\eeq 
Note that \eqref{def-F6} leaves the harmonic and exact part of $F_6$ unfixed and 
we will discuss constraints on these pieces in more detail below. The justification 
of the first term in Re$T_i$ is simpler. Remarkably, this definition of 
the K\"{a}hler coordinates as $D_i$ integrals will help us to obtain the couplings 
$\int e^{3 \alpha^2 W^\tbtwo} J \wedge J \wedge \o_i \wedge \o_j$, which, as we stressed in our previous work \cite{Grimm:2014efa},
cannot be obtained as $v^i$-derivatives of the considered $CY_4$-integrals. 
In order to evaluate the derivatives of $T_i$ with respect to $v^i$ and to make contact with 
the K\"{a}hler metric found in the reduction result \eqref{actionvs10}, we have to rewrite the integrals over $D_i$ 
into integrals over $CY_4$. Due to the appearance of the warp-factor and the non-closed form 
$F_6$ in \eqref{PotentialAndCoordinates} this is not straightforward. In particular, one cannot 
simply use Poincar\'e duality and write $T_i$ as an integral over $CY_4$ with inserted $\omega_i$. 
Of course, it is always possible to write $T_i$ as a $CY_4$ integral 
when inserting a delta-current localized on $D_i$, i.e.
\beq \label{T_idelta}
\R T_i  =  \int_{CY_4} \Big( \fr{1}{3!} e^{3 \al^2  W^\tbtwo} J  \we J  \we J    + 1536 \al^2  F_6 \Big)  \we \delta_i\ ,
\eeq
where $\d_i$ is the (1,1)-form delta-current that restricts to the divisor $D_i$. 
Appropriately extending the notion of cohomology to include currents \cite{GriffithsHarris}, we 
can now ask how much $\delta_i$ differs from the harmonic form $\omega_i$ in the same class.
In fact, any current $\d_i$ is related to the harmonic element of the same class $\o_i$ 
by a doubly exact piece as 
\ba
\d_i = \o_i   + i \pa   \bar \pa   \l_i\ .
\label{delToOm}
\ea
This equation should be viewed as relating currents. Importantly, as we assume 
$D_i$ and hence $\delta_i$ to be $v^i$-independent, the $v^i$ dependence of the harmonic 
form $\o_i$ and the current $\l_i$ has to cancel such that $\partial_j \o_i =  - i\pa   \bar \pa  \pa_j  \l_i$. 
Importantly, once we determine $\partial_j \R T_j$ we can express the result as $Y_4$-integrals without 
invoking currents. We therefore need to understand how each part of $T_i$ 
varies under a change of moduli. This will also fix the numerical factor in front of $F_6$ in \eqref{PotentialAndCoordinates}.

In order to take derivatives of $T_i$ we first use the fact that $D_i$ and hence $\d_i$ 
are independent of the moduli $v^i$, which implies 
\beq \label{eqdelta}
\partial_j \R T_i = \int_{Y_4} \Big( \fr{1}{2} e^{3 \al^2  W^\tbtwo} \omega_j  \we J  \we J    +  \fr{1}{2} \alpha^2 \partial_jW^\tbtwo J  \we J  \we J  +  1536 \al^2  \pa_j F_6 \Big)  \we \d_i\ .
\eeq
We next claim that we can replace $\d_i$ with $\omega_i$ such that finally 
\beq \label{eqomega}
\partial_j \R T_i = \fr{1}{2}   \int_{Y_4} e^{3 \al^2  W^\tbtwo} \omega_i  \we \omega_j \we J  \we J    +  \fr{1}{2} \alpha^2  \int_{Y_4} \partial_jW^\tbtwo \o_i \we J  \we J  \we J  +  1536 \al^2  \int_{Y_4} \o_i\we  \pa_j F_6  \ .
\eeq
Note that by using \eqref{delToOm} the two expressions \eqref{eqdelta} and \eqref{eqomega} only differ by a
term involving $\pa   \bar \pa   \l_i$. By partial integration this term is proportional to  
\bea \label{extraterm}
&&\int_{Y_4} \l_i \pa \bar \pa \Big( \fr{1}{2} e^{3 \al^2  W^\tbtwo} \omega_j  \we J  \we J    +  \fr{1}{2} \alpha^2 \partial_jW^\tbtwo J  \we J  \we J  +  1536 \al^2  \pa_j F_6 \Big)  \\
&=& \int_{Y_4} \l_i  \Big( \fr{1}{2} \pa \bar \pa (e^{3 \al^2  W^\tbtwo}) \omega_j  \we J  \we J    +  \fr{1}{2} \alpha^2 \pa \bar \pa ( \partial_jW^\tbtwo ) J  \we J  \we J  +  1536 \al^2  \pa \bar \pa \pa_j F_6 \Big) \nn \ . 
\eea
It is now straightforward to see that the terms multiplying $\l_i$ are simply the $\partial_j$ derivative of the 
warp-factor equation \eqref{eq:warp}. One first writes \eqref{eq:warp} as
\beq \label{warprewrite}
   d^\dagger d e^{ 3 \alpha^2 W^\tbtwo} *_8 1 -  \alpha^2 Q_8  = - \frac{1}{3} i \pa \bar \pa  (e^{ 3 \alpha^2 W^\tbtwo} ) \wedge J\wedge J \wedge J - \alpha^2 Q_8\ .
\eeq
Then one takes the $v^j$-derivative of \eqref{warprewrite} by using 
the fact that $Q_8$ is given via 
\beq
Q_8 = -\tfrac{1}{2}  \, G^\tbo \wedge G^\tbo -3^2 2^{13}\, X^\tbzero_8\;\;,
\eeq
which can easily be inferred by comparison to
\eqref{eq:warp} and \eqref{def-F6}. The moduli dependence of $Q_8$ only arises from the 
term involving $F_6$, i.e.~one has $\pa_i Q_8  = 3072 \, i \, \pa \bar \pa \pa_i F_{6} $. Hence one finds exactly 
the terms in  \eqref{extraterm} such that this $\l_i$ dependent part of the $T_i$ variation vanishes due to the warp-factor equation \eqref{eq:warp}. 
The final expression \eqref{eqomega} is  then written
as
\ba
\pa_j \R T_i &   =  \fr12 \int_{Y_4} e^{3 \al^2 W^\tbtwo} \o_i \we \o_j \we J \we J   
 +  3  \al^2 \cK_i \mathscr{W}_j 
 +1536  \al^2 \int_{Y_4}  \o_i  \we \pa_j F_6   \ .
\label{dTbydvfromAnsatz}
\ea
Evaluating \eqref{KpotfromAnsatz} effective action will depend  on the quantities
\ba \label{deriv-results}
\int_{Y_4}  \o_i \we \pa_j F_6 | \;\;\; \text{and } \;\;\;
\int_{Y_4}  J  \we \pa_i \pa_j F_6 | \;\; .
 \ea
in order for the results to match the reduction result those terms need to interact with the higher-derivative building blocks. One may use the freedom in the definition  \eqref{def-F6} to accomplish this task. A concise match with the reduction result is beyond the scope of this work.

\section{\bf Higher-derivatives and F-theory} \label{HigherFtheory}
\subsection{11d higher-derivative Terms} \label{HigherDerivativeTerms}
The terms $ t_8  t_8  R^4$ and $ t_8  t_8  G^2  R^3$ in  require the definition
\bea \label{def-t8}
&&  t_8^{N_1\dots N_8}=   \\
&&  \fr{1}{16} \big( -  2 \left(     g^{ N_1 N_3  }  g^{  N_2  N_4  }  g^{ N_5   N_7  }  g^{ N_6 N_8  } 
 +   g^{ N_1 N_5  }  g^{ N_2 N_6  }  g^{ N_3   N_7  }  g^{  N_4   N_8   }
 +    g^{ N_1 N_7  }  g^{ N_2 N_8  }  g^{ N_3   N_5  }  g^{  N_4 N_6   }  \right) \nonumber \\
 && \quad +\quad
 8 \left(    g^{  N_2     N_3   }  g^{ N_4    N_5  }  g^{ N_6    N_7  }  g^{ N_8   N_1   } 
  +  g^{  N_2     N_5   }  g^{ N_6    N_3  }  g^{ N_4    N_7  }  g^{ N_8   N_1   } 
  +     g^{  N_2     N_5   }  g^{ N_6    N_7  }  g^{ N_8    N_3  }  g^{ N_4  N_1   } 
\right) \nonumber \\ \nn
&&  \quad-\quad (N_1 \lra N_2) -( N_3 \lra N_4) - (N_5 \lra N_6) - (N_7 \lra N_8) \big) \,. 
\eea
Let us now discuss the various eight-derivative couplings in in more detail. We recall the definition
\ba\label{defX8}
    X_8 = \fr{1}{192} \Big( \Tr  \cR^4 - \fr14 (\Tr  \cR^2)^2 \Big) \ ,
\ea
where $ \cR$ is the eleven-dimensional curvature two-from $   \cR^{M}_{\ \ N} = \frac12    R^{M}_{\ \ NPQ} dx^P \wedge dx^Q$,
and  
\ba\label{deft8t8R4}
    \e_{11}    \e_{11}    R^4 &=  \epsilon^{R_1 R_2 R_3  M_1\ldots M_{8} }  \epsilon_{R_1 R_2 R_3 N_1 \ldots N_{8}}    R^{N_1 N_2}{}_{M_1 M_2}    R^{N_3 N_4}{}_{M_3 M_4}     R^{N_5 N_6}{}_{M_5 M_6}    R^{N_7 N_8}{}_{M_7 M_8} \ , \nn \\[.1cm]
    t_8    t_8    R^4 & =    t_{8}^{  M_1 \dots M_8}    t_{8 \, N_1  \dots N_8}        R^{N_1 N_2}{}_{M_1 M_2}    R^{N_3 N_4}{}_{M_3 M_4}     R^{N_5 N_6}{}_{M_5 M_6}    R^{N_7 N_8}{}_{M_7 M_8} \ , 
\ea
where $\e_{11}$ is the eleven-dimensional totally anti-symmetric epsilon tensor and 
$t_8$ is given explicitly in \eqref{def-t8}.
Using $\e_{11}$ and $t_8$ the explicit form for the terms in section \ref{sec-HIghDerM} are precisely
 given by 
\ba
    \epsilon_{11}    \epsilon_{11}    G^2    R^3 &=    \epsilon^{R M_1 \ldots M_{10} }    \epsilon_{R N_1 \ldots N_{10}}     G^{N_1 N_2}{}_{M_1 M_2}    G^{N_3 N_4}{}_{M_3 M_4}   \nn \\
     & \quad \quad \quad \quad \quad \quad \quad \quad \;\; \times    R^{N_5 N_6}{}_{M_5 M_6}    R^{N_7 N_8}{}_{M_7 M_8}    R^{N_9 N_{10}}{}_{M_9 M_{10}} \,, \nn \\[0.2cm] 
 \label{eq:ttGR} 
    t_8    t_8    G^2    R^3  & =    t_8^{ M_1  \dots M_8}     t_{8}{}_{  N_1 \dots N_8}     G^{N_1}{}_{M_1}{}_{R_1 R_2}    G^{N_2}{}_{M_2}{}^{R_1 R_2}     R^{N_3 N_4}{}_{M_3 M_4}     R^{N_5 N_6}{}_{M_5 M_6}    R^{N_7 N_8}{}_{M_7 M_8} \, .
\ea
Finally, we need to introduce the tensor $   s_{18}^{N_1 \ldots N_{18}} $, 
however its precise form not known. Significant parts of it 
may be fixed following \cite{Peeters:2005tb}. We argue for an extension in \ref{sec-HIghDerM} of this work.
In order to express the kwon parts we use the basis $B_i,\ i=1,...,24$ of \cite{Peeters:2005tb}, 
that labels all unrelated index contractions in $   s_{18} (   \na    G)^2    R^2 $.
The basis $\{B_i\}$  is explicitly given in  section \ref{section-basisG2R3}. 
The result can then be expressed in terms of a four-point amplitude contribution $\cA$ 
and a linear combination of six contributions $\mathcal{S}_{i=1,\dots,6}$ which do not affect the 4-point amplitude as
\ba \label{s18term_exp}
    s_{18} (   \na    G)^2    R^2 =    s_{18}^{N_1 \ldots N_{18}}    R_{N_1\ldots N_{4}}    R_{N_5\ldots N_{8}}    \na_{N_9}    G_{N_{10} \ldots N_{13} }    \na_{N_{14}}      G_{N_{15} \ldots N_{18}} = \cA + \sum_n a_n \mathcal{S}_n \, . 
\ea
The combinations $\cA$ and $\mathcal{S}_n$ are then given in terms of the basis elements as
\ba
\cA &= -24B_{5} -48B_{8} -24B_{10} -6B_{12} -12B_{13} +12B_{14}
 +8B_{16} -4 B_{20} \nn \\
 &\;\;\;\;\; + B_{22} + 4B_{23} + B_{24}  \,, \nn \\
\mathcal{S}_1 &= 48B_1 + 48B_2 - 48B_3 + 36B_4 + 96B_6 + 48B_7 - 48B_8 +
96B_{10} \nn \\
&\quad\quad + 12B_{12} +24B_{13} -12B_{14} + 8B_{15} + 8B_{16} - 16B_{17} + 6B_{19} + 2B_{22} + B_{24}\,,\nn \\
\mathcal{S}_2 &=-48B_1 -48B_2 -24B_4 -24B_5 +48B_6 -48B_8 -24B_9
-72B_{10}  \nn \\
 &\;\;\;\;\; -24B_{13} +24B_{14} -B_{22} +4B_{23}\,,\nn \\
\mathcal{S}_3 &= 12B_1 + 12B_2 - 24B_3 + 9B_4 +48B_6 + 24B_7 - 24B_8
 + 24B_{10}
 \nn \\
& \quad \quad
+ 6B_{12} 
 + 6B_{13}  + 4B_{15} - 4B_{17} + 3B_{19} + 2B_{21}\,,\nn \\
\mathcal{S}_4 &= 12B_1 + 12B_2 - 12B_3 + 9B_4 +24B_6 + 12B_7 - 12B_8 + 24B_{10}
+ 3B_{12} \nn \\
 &\;\;\;\;\;+ 6B_{13} + 4B_{15} - 4B_{17} + 2B_{20}\,,\nn \\
\mathcal{S}_5 &= 4B_{3} -8B_{6} -4B_7 + 4B_8 -B_{12} -2B_{14} + 4B_{18}\,,\nn \\
\mathcal{S}_6 &= B_4 + 2B_{11} \,.
\ea
Note that $\mathcal{S}_3$ to $\mathcal{S}_6$ vanish both on the considered Calabi-Yau fourfold  background solution.

\subsection{Adjunction of Chern-classes}
\label{adjunctoinChernSec}

Let us next discuss the adjunction of Chern-classes of divisors on an elliptically  fibered Calabi--Yau fourfold $CY_4$ which is a hyper-surface in a $\mathbb{P}_{321}$ bundle of the K\"ahler base $B_3$ denoted by $\mathbb{P}_{321}(\mathcal{L})$given by the vanishing locus of the Weierstrass equation
\beq
y^2 −(x^3 +f x z^4 +g z^6) = 0 \;\; ,
\eeq
with $f,g$ holomorphic sections of $\mathcal{L}^4$ and $\mathcal{L}^6$, respectively. The $SL(2,\mathbb{Z})$ line bundle $\mathcal{L}$ over $B$ together with the choice of $f,g$ defines the elliptic fibration.
One may show that the first Chern class is given by
\beq
c_1(CY_3) = c_1(B_3) - c_1(\mathcal{L}) \;\;,
\eeq
where the r.h.s\, is pulled back to $CY3$. Then the total Chern class is given by
\beq
c(\mathbb{P}_{321}(\mathcal{L})) = c(B_3)(1 + 2 \o_0 +2 c_1(B_3)))(1 + 3\o_0 + 3 c_1(B_3)))(1 + \o_0)
\eeq
were $\o_0$ is the harmonic $(1,1)$-form such that $PD(\o_0) = B$.\footnote{We are using abuse of notation in the following using $\o_0$ and $c_{1,2,3}$ in the context of a concrete representative of the class as well as the class itself. }
  Using adjunction formulae  for the 
  \beq
c(CY_4)= \frac{c(\mathbb{P}_{321}(\mathcal{L}))}{ (1+ \mathcal{L})}
\eeq
with 
\beq
\mathcal{L} = 6 \o_0 + 6 c_1(B_3)
\eeq
  one then derives
\ba 
c_3 (CY_4) = &  \;  c_3(B_3) -  c_1(B_3) \wedge c_2(B_3)  - 60 c^3_1(B_3) -60  c^{ 2}_1(B_3)  \wedge \omega_0 \nonumber \\[0.2cm]
  c_2 (CY_4) = &  \;   c_2(B_3)  +11    c^{2}_1(B_3)  +12 c_1(B_3)  \wedge \omega_0 
  \nonumber \\[0.2cm]
   c_1 (CY_4) = &  0
\ea
and furthermore
\beq
 \omega_0^2= -  c_1(B_3)  \wedge \omega_0  \;\; .
\eeq
where the $c_{i=1,2,3}(B_3)$ on the r.h.s. of these expressions denote the Chern classes of $B$ pulled-back to $CY_4$.

One may next iterate the adjunction formulae to find The Chern-forms of the  vertical divisors  $D_\alpha$ of the Calabi--Yau fourfold which are pullbacks of divisors of the base $D^b_\alpha$. Thus we denote the class of such divisors via its representatives of harmonic $(1,1)$-forms $\omega_\alpha$, $\alpha=1,...,h^{1,1}$. Thus one may use adjunction to write
  \beq
c(D_\alpha)= \frac{c(\mathbb{P}_{321}(\mathcal{L}))}{ (1+ \mathcal{L}) (1+ \o_\alpha)} \;\; ,
\eeq
with which  one then derives
\ba \label{finalChernapp}
c_3 (D_\alpha) = &  \;  c_3(B_3) -  c_1(B_3) \wedge c_2(B_3)  - 60 c^3_1(B_3) -60  c^{ 2}_1(B_3)  \wedge \omega_0 \nonumber -  c_2 (D_\alpha)\wedge \o_\alpha \\[0.2cm]
  c_2 (D_\alpha) = &  \;   c_2(B_3)  +11    c^{2}_1(B_3)  +12 c_1(B_3)  \wedge \omega_0 + \o_\alpha^2
  \nonumber \\[0.2cm]
   c_1 (D_\alpha) = &  - \o_\alpha  \;\;\; .
\ea
where $c_{i=1,2,3} (B_3) $ on the r.h.s\,of the above equality  are pulled back to the divisor $D_\al$, which amounts to a simply restriction to the subspace $D_\al \subset CY_4$.
In particular we find that the self intersection of divisors $[D_\alpha] \cdot [D_\alpha] $ is generically non-vanishing.

Let us close this section by analyzing the case where the Calabi--Yau fourfold is a direct product manifold e.g.\,\,$CY_4 = CY_3 \times T^2$ or $CY_4 = K3 \times K3$. The Chern-character on product spaces $X = Y \times Z$ obeys $c(X) = c(Y) c(Z)$. Thus we find for the Chern-forms 
\ba \label{finalChernProductSpace}
c_3 (X) = & \; c_1 (Y)\wedge c_2 (Z) + c_2 (Y)\wedge c_1 (Z) +  c_3 (Y) +      c_3(Z) \; , \nn\\[0.2 cm]
  c_2 (X) = & \;  c_1 (Y)\wedge c_1 (Z) +  c_2 (Y) +      c_2 (Z)
  \nonumber \; , \\[0.2 cm]
   c_1 (X) = &  \;     c_1 (Y) +      c_1 (Z)   \;\;\; .
\ea
Furthermore, on may apply adjunction to compute the Chern-forms of $CY_3$ in therms of Chern-forms Divisors $D^b_\alpha$ pulled back to $CY_3$ which results in 
 \beq\label{CY3T2chern}
 c_1(D^b_\alpha) = \o_\alpha^b \;\; ,\;\;\;   c_2(D^b_\alpha) = c_2(CY_3) \;\; \;\; ,
 \eeq
where we have used the Calabi--Yau condition $c_1(CY_3) = 0$. Divisors inside $CY_4 = CY_3 \times T^2$ wrapping the torus are as well a direct product of $D^b_\al \times T^2$. Thus by combing \eqref{finalChernProductSpace} and \eqref{CY3T2chern} one can straightforwardly infer their Chern-forms.

\subsection{Basis of the  $G^2R^3$  and  $(\nabla G)^2R^2$-sector}\label{section-basisG2R3}
\paragraph{Basis of the  $G^2R^3$-sector.} The complete eleven-dimensional $G^2R^3$ terms may be written in terms of the basis \cite{Grimm:2017okk}  The basis for the potentially relevant eight-derivative terms involving the four-form field strength is
\bea
\mathcal{B}_1 & ={{{{G}}}}_{M_5}{}^{M_7M_8M_9} \,{G}_{M_6M_7M_8M_9} \,  {R}_{MM_2}{}^{M_4M_5}\, {R}^{MM_1M_2M_3} \, {R}_{M_1M_3M_4}{}^{M_6}\\[0.2cm]
\mathcal{B}_2 & ={G}_{M_4M_6}{}^{M_8M_9} \, G_{M_5M_7M_8M_9}\, R_{MM_2}{}^{M_4M_5}\, R^{MM_1M_2M_3}\, R_{M_1M_3}{}^{M_6M_7}\nonumber \\[0.2cm]
\mathcal{B}_3 & =G_{M_4M_5}{}^{M_8M_9} \, G_{M_6M_7M_8M_9} \,R_{MM_2}{}^{M_4M_5} \,R^{MM_1M_2M_3}\, R_{M_1M_3}{}^{M_6M_7}\nonumber\\[0.2cm]
\mathcal{B}_4 & =G_{M_6M_7M_8M_9} \, G^{M_6M_7M_8M_9}\,R_{MM_2}{}^{M_4M_5}\,R^{MM_1M_2M_3} \, R_{M_1M_4M_3M_5}\nonumber\\[0.2cm]
\mathcal{B}_5 & =G_{M_6M_7M_8M_9}\,  G_4^{M_6M_7M_8M_9}\, R_{M}{}^{M_4}{}_{M_2}{}^{M_5}\, R^{MM_1M_2M_3}\, R_{M_1M_4M_3M_5}\nonumber\\[0.2cm]
\mathcal{B}_6 & =G_{M_5}{}^{M_7M_8M_9}\,  G_{M_6M_7M_8M_9} \, R_{MM_2}{}^{M_4M_5} \, R^{MM_1M_2M_3} \, R_{M_1M_4M_3}{}^{M_6}\nonumber\\[0.2cm]
\mathcal{B}_7 & =G_{M_5}{}^{M_7M_8M_9} \, G_{M_6M_7M_8M_9}\, R_{M}{}^{M_4}{}_{M_2}{}^{M_5}\, R^{MM_1M_2M_3}\, R_{M_1M_4M_3}{}^{M_6}\nonumber\\[0.2cm]
\mathcal{B}_8 & =G_{M_3M_6}{}^{M_8M_9} \, G_{M_5M_7M_8M_9}\, R_{MM_2}{}^{M_4M_5}\, R^{MM_1M_2M_3}\, R_{M_1M_4}{}^{M_6M_7}\nonumber\\[0.2cm]
\mathcal{B}_9 & =G_{M_3M_5}{}^{M_8M_9} \, G_{M_6M_7M_8M_9}\, R_{MM_2}{}^{M_4M_5}\, R^{MM_1M_2M_3} \, R_{M_1M_4}{}^{M_6M_7}\nonumber\\[0.2cm]
\mathcal{B}_{10}&=G_{M_3M_6}{}^{M_8M_9} \, G_{M_5M_7M_8M_9}\, R_{M}{}^{M_4}{}_{M_2}{}^{M_5}\, R^{MM_1M_2M_3} \, R_{M_1M_4}{}^{M_6M_7}\nonumber\\[0.2cm]
\mathcal{B}_{11}&=G_{M_3M_5}{}^{M_8M_9}\,  G_{M_6M_7M_8M_9} \, R_{M}{}^{M_4}{}_{M_2}{}^{M_5} \, R^{MM_1M_2M_3}\, R_{M_1M_4}{}^{M_6M_7}\nonumber\\[0.2cm]
\mathcal{B}_{12}&=G_{M_4M_7}{}^{M_8M_9} \, G_{M_5M_6M_8M_9}\, R_{M}{}^{M_4}{}_{M_2}{}^{M_5} \, R^{MM_1M_2M_3} \, R_{M_1}{}^{M_6}{}_{M_3}{}^{M_7}\nonumber\\[0.2cm]
\mathcal{B}_{13}&=G_{M_3M_7}{}^{M_8M_9}\,  G_{M_5M_6M_8M_9}\, R_{MM_2}{}^{M_4M_5}\, R^{MM_1M_2M_3} \, R_{M_1}{}^{M_6}{}_{M_4}{}^{M_7}\nonumber\\[0.2cm]
\mathcal{B}_{14}&=G_{M_3M_7}{}^{M_8M_9} \, G_{M_5M_6M_8M_9} \, R_{M}{}^{M_4}{}_{M_2}{}^{M_5}\, R^{MM_1M_2M_3} \, R_{M_1}{}^{M_6}{}_{M_4}{}^{M_7}\nonumber\\[0.2cm]
\mathcal{B}_{15}&=G_{M_5}{}^{M_7M_8M_9} \, G_{M_6M_7M_8M_9}\, R_{MM_1M_2}{}^{M_4}\, R^{MM_1M_2M_3} \, R_{M_3}{}^{M_5}{}_{M_4}{}^{M_6}\nonumber\\[0.2cm]
\mathcal{B}_{16}&=G_{M_4M_6}{}^{M_8M_9} \, G_{M_5M_7M_8M_9}\, R_{MM_1M_2}{}^{M_4} \, R^{MM_1M_2M_3}\, R_{M_3}{}^{M_5M_6M_7}\nonumber\\[0.2cm]
\mathcal{B}_{17}&=G_{M_4M_6}{}^{M_8M_9} \, G_{M_5M_7M_8M_9} \, R_{MM_1M_2M_3} \, R^{MM_1M_2M_3}\, R^{M_4M_5M_6M_7}.  \nonumber
\eea

\paragraph{Basis of the $(\nabla G)^2R^2$-sector.}
The complete eleven-dimensional $(\nabla G)^2R^2$ terms may be written in terms of the basis \cite{Peeters:2005tb}. In order to discuss the term $ s_{18}$ appearing in \eqref{newAnsatzG2R3} and \eqref{s18term_exp} we introduce the basis 
\bea
B_{1}  &=    R_{M_{1}M_{2} M_{3}M_{4}}   R_{M_{5}M_{6}M_{7}M_{8}}               \na^{M_{5}}  G^{M_{1}M_{7}M_{8}}{}_{M_{9}}      \na^{M_{3}}  G^{M_{2}M_{4}M_{6}M_{9}} \,,  \vspace{0.2cm} \nonumber \\[0.2cm]
B_{2}  &=    R_{M_{1}M_{2}M_{3}M_{4}}   R_{M_{5}M_{6}M_{7}M_{8}}               \na^{M_{5}}  G^{M_{1}M_{3}M_{7}}{}_{M_{9}}      \na^{M_{8}}  G^{M_{2}M_{4}M_{6}M_{9}} \,, \vspace{0.2cm}\nonumber \\[0.2cm]
B_{3}  &=    R_{M_{1}M_{2}M_{3}M_{4}}   R_{M_{5}M_{6}M_{7}M_{8}}               \na^{M_{5}}  G^{M_{1}M_{3}M_{7}}{}_{M_{9}}      \na^{M_{6}}  G^{M_{2}M_{4}M_{8}M_{9}} \, \vspace{0.2cm}\nonumber \\[0.2cm]
B_{4}  &=    R_{M_{1}M_{2}M_{3}M_{4}}   R_{M_{5}M_{6}M_{7}M_{8}}               \na_{M_{9}}  G^{M_{3}M_{4}M_{7}M_{8}}           \na^{M_{6}}  G^{M_{9}M_{1}M_{2}M_{5}} \, \vspace{0.2cm}\nonumber \\[0.2cm]
B_{5}  &=    R_{M_{1}M_{2}M_{3}M_{4}}   R_{M_{5}M_{6}M_{7}}{}^{M_{4}}          \na^{M_{1}}  G^{M_{2}M_{3}}{}_{M_{8}M_{9}}      \na^{M_{5}}  G^{M_{6}M_{7}M_{8}M_{9}} \, ,\vspace{0.2cm}\nonumber \\[0.2cm]
B_{6}  &=    R_{M_{1}M_{2}M_{3}M_{4}}   R_{M_{5}M_{6}M_{7}}{}^{M_{4}}          \na^{M_{1}}  G^{M_{2}M_{5}}{}_{M_{8}M_{9}}      \na^{M_{3}}  G^{M_{6}M_{7}M_{8}M_{9}} \,\vspace{0.2cm},\nonumber  \\[0.2cm]
B_{7}  &=    R_{M_{1}M_{2}M_{3}M_{4}}   R_{M_{5}M_{6}M_{7}}{}^{M_{4}}          \na^{M_{1}}  G^{M_{2}M_{5}}{}_{M_{8}M_{9}}      \na^{M_{7}}  G^{M_{3}M_{6}M_{8}M_{9}} \, \vspace{0.2cm}\nonumber \\[0.2cm]
B_{8}  &=    R_{M_{1}M_{2}M_{3}M_{4}}   R_{M_{5}M_{6}M_{7}}{}^{M_{4}}          \na^{M_{1}}  G^{M_{3}M_{5}}{}_{M_{8}M_{9}}      \na^{M_{2}}  G^{M_{6}M_{7}M_{8}M_{9}} \,, \vspace{0.2cm}\nonumber \\[0.2cm]
B_{9}  &=    R_{M_{1}M_{2}M_{3}M_{4}}   R_{M_{5}M_{6}M_{7}}{}^{M_{4}}          \na^{M_{1}}  G^{M_{3}M_{5}}{}_{M_{8}M_{9}}      \na^{M_{6}}  G^{M_{2}M_{7}M_{8}M_{9}} \,, \vspace{0.2cm}\nonumber \\[0.2cm]
B_{10} &=   R_{M_{1}M_{2}M_{3}M_{4}}   R_{M_{5}M_{6}M_{7}}{}^{M_{4}}          \na_{M_{9}}  G^{M_{3}M_{5}M_{7}M_{8}}           \na^{M_{9}}  G^{M_{1}M_{2}M_{6}M_{8}} \, ,\vspace{0.2cm}\nonumber \\[0.2cm]
B_{11} &=   R_{M_{1}M_{2}M_{3}M_{4}}   R_{M_{5}M_{6}M_{7}}{}^{M_{4}}          \na_{M_{8}}  G^{M_{1}M_{2}M_{6}}{}_{M_{9}}      \na^{M_{9}}  G^{M_{3}M_{5}M_{7}M_{8}}  \,,\vspace{0.2cm}\nonumber \\[0.2cm]
B_{12} &=   R_{M_{1}M_{2}M_{3}M_{4}}   R_{M_{5}M_{6}M_{7}}{}^{M_{4}}          \na^{M_{3}}  G^{M_{5}M_{6}}{}_{M_{8}M_{9}}      \na^{M_{7}}  G^{M_{2}M_{1}M_{8}M_{9}} \, ,  \vspace{0.2cm}
\nonumber
\\ [0.2cm]
B_{13} &=   R_{M_{1}M_{2}M_{3}M_{4}}   R_{M_{5}}{}^{M_{1}}{}_{M_{6}}{}^{M_{3}}   \na_{M_{9}}  G^{M_{2}M_{6}}{}_{M_{7}M_{8}}   \na^{M_{9}}  G^{M_{4}M_{5}M_{7}M_{8}} \,, \vspace{0.2cm}\nn \\[0.2cm]
B_{14} &=   R_{M_{1}M_{2}M_{3}M_{4}}   R_{M_{5}}{}^{M_{1}}{}_{M_{6}}{}^{M_{3}}   \na_{M_{9}}  G^{M_{2}M_{4}}{}_{M_{7}M_{8}}   \na^{M_{9}}  G^{M_{5}M_{6}M_{7}M_{8}} \,, \vspace{0.2cm}\nn \\[0.2cm]
B_{15} &=   R_{M_{1}M_{2}M_{3}M_{4}}   R_{M_{5}}{}^{M_{1}}{}_{M_{6}}{}^{M_{3}}   \na^{M_{2}}  G^{M_{6}}{}_{M_{7}M_{8}M_{9}}   \na^{M_{5}}  G^{M_{4}M_{7}M_{8}M_{9}} \,,\vspace{0.2cm}\nn \\[0.2cm]
B_{16} &=   R_{M_{1}M_{2}M_{3}M_{4}}   R_{M_{5}}{}^{M_{1}}{}_{M_{6}}{}^{M_{3}}   \na^{M_{2}}  G^{M_{4}}{}_{M_{7}M_{8}M_{9}}   \na^{M_{5}}  G^{M_{6}M_{7}M_{8}M_{9}} \, ,\vspace{0.2cm},\nn \\[0.2cm]
B_{17} &=   R_{M_{1}M_{2}M_{3}M_{4}}   R_{M_{5}}{}^{M_{1}}{}_{M_{6}}{}^{M_{3}}   \na^{M_{2}}  G^{M_{5}}{}_{M_{7}M_{8}M_{9}}   \na^{M_{4}}  G^{M_{6}M_{7}M_{8}M_{9}} \, ,\vspace{0.2cm}\nn \\[0.2cm]
B_{18} &=   R_{M_{1}M_{2}M_{3}M_{4}}   R_{M_{5}}{}^{M_{1}}{}_{M_{6}}{}^{M_{3}}   \na_{M_{9}}  G^{M_{5}M_{6}}{}_{M_{7}M_{8}}   \na^{M_{4}}  G^{M_{2}M_{7}M_{8}M_{9}} \, \, ,\vspace{0.2cm}\nn \\[0.2cm]
B_{19} &=   R_{M_{1}M_{2}M_{3}M_{4}}   R_{M_{5}M_{6}}{}^{M_{3}M_{4}}          \na_{M_{9}}  G^{M_{1}M_{5}}{}_{M_{7}M_{8}}      \na^{M_{9}}  G^{M_{2}M_{6}M_{7}M_{8}} \, \, ,\vspace{0.2cm}\nn \\[0.2cm]
B_{20} &=   R_{M_{1}M_{2}M_{3}M_{4}}   R_{M_{5}M_{6}}{}^{M_{3}M_{4}}          \na^{M_{1}}  G^{M_{5}}{}_{M_{7}M_{8}M_{9}}      \na^{M_{2}}  G^{M_{6}M_{7}M_{8}M_{9}} \, ,\vspace{0.2cm}\nn \\[0.2cm]
B_{21} &=   R_{M_{1}M_{2}M_{3}M_{4}}   R_{M_{5}M_{6}}{}^{M_{3}M_{4}}          \na^{M_{1}}  G^{M_{5}}{}_{M_{7}M_{8}M_{9}}      \na^{M_{6}}  G^{M_{2}M_{7}M_{8}M_{9}} \vspace{0.2cm}\nn  \, ,\\[0.2cm]
B_{22} &=   R_{M_{1}M_{2}M_{3}M_{4}}   R_{M_{5}}{}^{M_{1}M_{3}M_{4}}          \na^{M_{2}}  G_{M_{6}M_{7}M_{8}M_{9}}           \na^{M_{5}}  G^{M_{6}M_{7}M_{8}M_{9}} \ \, ,\vspace{0.2cm}\nn \\[0.2cm]
B_{23} &=   R_{M_{1}M_{2}M_{3}M_{4}}   R_{M_{5}}{}^{M_{1}M_{3}M_{4}}          \na_{M_{9}}  G^{M_{2}}{}_{M_{6}M_{7}M_{8}}      \na^{M_{9}}  G^{M_{5}M_{6}M_{7}M_{8}} \, , \nn \\[0.2cm]
B_{24} &=   R_{M_{1}M_{2}M_{3}M_{4}}   R^{M_{1}M_{2}M_{3}M_{4}}               \na_{M_{5}}  G_{M_{6}M_{7}M_{8}M_{9}}           \na^{M_{6}}  G^{M_{5}M_{7}M_{8}M_{9}} \, .
\eea
The contributions to $ s_{18} (\nabla  G)^2  R^2 $ are then formed from the linear combinations described in  \eqref{s18term_exp}.
We write the eleven-dimensional action as
\be
2 \kappa_{11}^2 \,  S^{\rm{extra, \,gen}}=\alpha^2 \, \int_{M_{11}}\sum_{i=1}^{17}  C_{i} \, \mathcal{B}_i \,   \ast 1 \, + \sum_{i=1}^{24} C_{i+17} \, B_i \,   \ast 1 \, 
\ee
with real parameters $C_1, \dots, C_{41}$ which are fixed by the reduction on a Calabi--Yau threefold and   compatibility with $5d,\, \cN_2$ super symmetry to
\be\nonumber
\begin{array}{cclccl}
             C_5&=&-\tfr{1}{2}  C_4  \, ,           &   \qquad   \qquad      C_7 &=& -C_1 -\tfrac{1}{2} C_6   \, ,                    \\[0.5cm]
             C_{9}&=& 4  C_3    \, ,           &   \qquad   \qquad 
                   C_{10} &=&  -3 C_1 - 2 C_2 - 8 C_3 - 18 C_4 - \tfrac{3}{2} C_6 - C_8 \, ,       \\[0.5cm]
                           C_{11} &=&  -4 C_3 \, ,                       & \qquad    \qquad  C_{12}&=& 4 C_3  \, ,                      \\[0.5cm]
                       C_{13} &=& -8 C_3 \, ,           &   \qquad   \qquad      C_{14} &=& -6 C_1 - 2 C_2 - 12 C_3 - 36 C_4 - 3 C_6 - C_8   \, ,           \\[0.5cm]
                              C_{15}&=&   \tfrac{1}{3 }C_2 + 3 C_4\, ,         &   \qquad   \qquad     
                   C_{16} &=& -2 C_2 - 4 C_3 \, ,    \\[0.5cm]
                           C_{17} &=&  \tfrac{1}{4} C_2   \, ,                       & \qquad    \qquad  C_{25}&=&  2 C_{22} - C_{24} \, ,                      \\[0.5cm]     
                         
\end{array}
\ee
\vspace{-0.5 cm}
\be
\begin{array}{cclccl}
\;\;\;  \;\;\;     C_{29} &=&  \tfrac{1}{4}C_{22} + \tfrac{1}{4}C_{23} - \tfrac{1}{4} C_{24} + \tfrac{1}{4} C_{26} \, ,       &&&   \qquad   \qquad     \qquad       \\[0.5cm]     
           \;\;\;   \;\;\;                     C_{32}&=& -C_1 - \tfrac{1}{3} C_{22}  - \tfrac{4}{3} C_3 + \tfrac{2}{3} C_{30} - 6 C_4 - \tfrac{1}{2} C_6  \, ,       \qquad   \qquad           \qquad  &&&           \\[0.5cm]     
           
 \;\;\;  \;\;\;     C_{33} &=& C_1 - \tfrac{1}{3} C_{22} + \tfrac{4}{3}  C_3 + 6 C_4 + \tfrac{1}{2} C_6    \, ,       \qquad   \qquad           \qquad  &&&           \\[0.5cm]      
 \;\;\;  \;\;\;     C_{37} &=&  - C_1 - \tfrac{4}{3}C_3  - \tfrac{1}{3} C_{31} - \tfrac{1}{2}C_{34} -\tfrac{1}{6} C_{35} - \tfrac{2}{3} C_{36} - 
   6 C_4 -  \tfrac{1}{2} C_6   \, ,       \qquad   \qquad           \qquad  &&&           \\[0.5cm]     
 \;\;\;  \;\;\;    C_{38} &=& \tfrac{1}{3} C_{30} + \tfrac{1}{3} C_{31} + \tfrac{1}{2}C_{34} +\tfrac{1}{6} C_{35} + \tfrac{2}{3} C_{36}  \, ,       \qquad   \qquad           \qquad  &&&           \\[0.5cm]     
  \;\;\; \;\;\;    C_{39} &=&  \tfrac{1}{4} C_1  - \tfrac{1}{24} C_{22} +\tfrac{1}{3} C_3 + \tfrac{3}{2} C_4 + \tfrac{1}{8} C_6  \, ,       \qquad   \qquad           \qquad  &&&           \\[0.5cm]     
   \;\;\;  \;\;\;    C_{40} &=&  \tfrac{1}{2} C_1 +  \tfrac{2}{3}  C_3 +\tfrac{1}{3} C_{31} +\tfrac{1}{6} C_{35} + 3 C_4 +\tfrac{1}{4} C_6  \, ,       \qquad   \qquad           \qquad  &&&           \\[0.5cm]      
  \;\;\;  \;\;\;   C_{41} &=&  \tfrac{1}{4}C_1 +\tfrac{1}{3} C_3 + \tfrac{1}{12}C_{31} +\tfrac{1}{24}C_{35} + \tfrac{3}{2} C_4 + \tfrac{1}{8} C_6             \, ,       \qquad   \qquad           \qquad  &&&            
\end{array}\label{parameters}
\ee
We then check compatibility of the novel induces $H^2 R^3$ terms making use of the IIA - Heterotic duality. Compactifying type IIA on $K3$ is dual to the Heterotic string on $\mathbb{T}^4$. One  finds that additionally
\beq
C_2 = 0\;\; , \;\;\;  C_1 = - \tfr{1}{6} \big(8 C_3 + 2 C_{31} + C_{35} + 36 C_4 + 3 C_6 \big) \;\; ,
\eeq
where more details can be found  in section \ref{CY3checkM}. 
\nocite{*}
\bibliographystyle{utcaps}
\newpage
\bibliography{references}
\end{document}